\documentclass[twocolumn,astrosymb,twocolappendix]{aastex631}

\newcommand{\sbunit}{\mathrm{mag\ arcsec}^{-2}}
\newcommand{\sbcen}{\mu_{0}(g)}
\newcommand{\sbeff}{\overline{\mu}_{\mathrm{eff}}(g)}
\newcommand{\sbeffr}{\overline{\mu}_{\mathrm{eff}}(r)}

\newcommand{\code}[1]{\texttt{#1}}
\newcommand{\sersic}{S\'ersic}
\usepackage{CJKutf8}
\usepackage{bm}
\usepackage{appendix}
\usepackage{amsmath,amssymb}
\usepackage{soul}

\defcitealias{Greco2018}{G18}

\shorttitle{Mass--size outliers among MW satellites}
\shortauthors{Li et al.}
\graphicspath{{./}{figures/}}

\begin{document}
\begin{CJK*}{UTF8}{gbsn}

\title{Beyond Ultra-diffuse Galaxies. I. Mass--Size Outliers among the Satellites of Milky Way Analogs}

\correspondingauthor{Jiaxuan Li}
\author[0000-0001-9592-4190]{Jiaxuan Li (李嘉轩)}
\affiliation{Department of Astrophysical Sciences, 4 Ivy Lane, Princeton University, Princeton, NJ 08544, USA}
\email{jiaxuanl@princeton.edu}

\author[0000-0002-5612-3427]{Jenny E. Greene}
\affiliation{Department of Astrophysical Sciences, 4 Ivy Lane, Princeton University, Princeton, NJ 08544, USA}

\author[0000-0003-4970-2874]{Johnny P. Greco}
\affiliation{Department of Astrophysical Sciences, 4 Ivy Lane, Princeton University, Princeton, NJ 08544, USA}
\affiliation{Center for Cosmology and AstroParticle Physics (CCAPP), The Ohio State University, Columbus, OH 43210, USA}

\author[0000-0003-1385-7591]{Song Huang (黄崧)}
\affiliation{Department of Astrophysical Sciences, 4 Ivy Lane, Princeton University, Princeton, NJ 08544, USA}
\affiliation{Department of Astronomy and Tsinghua Center for Astrophysics, Tsinghua University, Beijing 100084, China}

\author[0000-0002-8873-5065]{Peter Melchior}
\affiliation{Department of Astrophysical Sciences, 4 Ivy Lane, Princeton University, Princeton, NJ 08544, USA}
\affiliation{Center for Statistics \& Machine Learning, Princeton University, Princeton, NJ 08544, USA}

\author[0000-0002-1691-8217]{Rachael Beaton}
\affiliation{Department of Astrophysical Sciences, 4 Ivy Lane, Princeton University, Princeton, NJ 08544, USA}
\author[0000-0002-2991-9251]{Kirsten Casey}
\affiliation{Center for Cosmology and AstroParticle Physics (CCAPP), The Ohio State University, Columbus, OH 43210, USA}
\author[0000-0002-1841-2252]{Shany Danieli}
\altaffiliation{NASA Hubble Fellow}
\affiliation{Department of Astrophysical Sciences, 4 Ivy Lane, Princeton University, Princeton, NJ 08544, USA}
\author[0000-0003-4700-663X]{Andy Goulding}
\affiliation{Department of Astrophysical Sciences, 4 Ivy Lane, Princeton University, Princeton, NJ 08544, USA}
\author[0000-0002-2704-5028]{Remy Joseph}
\affiliation{Department of Astrophysical Sciences, 4 Ivy Lane, Princeton University, Princeton, NJ 08544, USA}
\affiliation{Oskar Klein Centre for Cosmoparticle Physics, Department of Physics, Stockholm University, Stockholm SE-106 91, Sweden}
\author[0000-0002-0332-177X]{Erin Kado-Fong}
\affiliation{Department of Astrophysical Sciences, 4 Ivy Lane, Princeton University, Princeton, NJ 08544, USA}

\author[0000-0002-1418-3309]{Ji Hoon Kim}
\affiliation{Astronomy Program, Department of Physics and Astronomy, Seoul National University, 1 Gwanak-ro, Gwanak-gu, Seoul 08826, Republic of Korea}
\affiliation{SNU Astronomy Research Center, Seoul National University, 1 Gwanak-ro, Gwanak-gu, Seoul 08826, Republic of Korea}

\author{Lauren A. MacArthur}
\affiliation{Department of Astrophysical Sciences, 4 Ivy Lane, Princeton University, Princeton, NJ 08544, USA}

\begin{abstract}
Large diffuse galaxies are hard to find, but understanding the environments where they live, their numbers, and ultimately their origins, is of intense interest and importance for galaxy formation and evolution. Using Subaru's Hyper Suprime-Cam Strategic Survey Program, we perform a systematic search for low surface brightness galaxies and present novel and effective methods for detecting and modeling them. As a case study, we surveyed 922 Milky Way analogs in the nearby Universe ($0.01 < z < 0.04$) and built a large sample of satellite galaxies that are outliers in the mass--size relation. These ``ultra-puffy'' galaxies (UPGs), defined to be $1.5\sigma$ above the average mass--size relation, represent the tail of the satellite size distribution. We find that each MW analog hosts $N_{\rm UPG} = 0.31\pm 0.05$ ultra-puffy galaxies on average, which is consistent with but slightly lower than the observed abundance at this halo mass in the Local Volume. We also construct a sample of ultra-diffuse galaxies (UDGs) in MW analogs and find an abundance of $N_{\rm UDG} = 0.44\pm0.05$ per host. With literature results, we confirm that the UDG abundance scales with the host halo mass following a sublinear power law. We argue that our definition for ultra-puffy galaxies, which is based on the mass--size relation, is more physically motivated than the common definition of ultra-diffuse galaxies, which depends on the surface brightness and size cuts and thus yields different surface mass density cuts for quenched and star-forming galaxies. 
\end{abstract}


\section{Introduction} \label{sec:intro}
In a variety of galaxies in the Universe, there is an interesting subset of galaxies with low surface brightness \citep[dubbed low surface brightness galaxies or LSBGs, e.g.,][]{Sandage1984,Caldwell1987,Impey1988,McGaugh1995,Dalcanton1997a}. Recently, new excitement has been generated by the discovery of a large population of ultra-diffuse galaxies (UDGs) in the Coma cluster \citep{vanDokkum2015}, accompanied by many other works on detecting UDGs in clusters \citep[e.g.,][]{Koda2015,Mihos2015,Yagi2016,vdBurg2016,vdBurg2017,Lee2017,ManceraPina2018,Zaritsky2019,Danieli2019}, groups \citep[e.g.,][]{Roman2017b,SAGA-II,Roman2021,CarlstenELVES2022}, and the field \citep[e.g.,][]{Leisman2017,Greco2018,Roman2019,Prole2019,Tanoglidis2021,Kadowaki2021}. There have been a variety of studies of these diffuse galaxy systems, ranging from their dark matter content \citep[e.g.,][]{Mowla2017,vanDokkum2018,vanDokkum2019,Danieli2019DF2,Wasserman2019,ManceraPina2019b,Keim2022,ManceraPina2022,Kong2022}, to globular cluster populations \citep[e.g.,][]{vanDokkum2017,Somalwar2020,Forbes2020,Danieli2022,Gannon2022,vanDokkum2022GC,Gannon2022}, and stellar populations \citep[e.g.,][]{Gu2018,Ferre-Mateu2018,Pandya2018,Villaume2022}. We refer to LSBGs as objects selected purely based on surface brightness without distance information. With known distances, people define UDGs to be galaxies with large sizes ($r_e > 1.5$ kpc) and low surface brightnesses ($\sbcen > 24.0\ \sbunit$), and thus UDGs are outliers in size and surface brightness compared to normal dwarf galaxies. A central question about LSBGs, including UDGs, is what causes them to be outliers from the average mass--size relation of dwarf galaxies.

Various mechanisms have been proposed to explain the presence of large, diffuse galaxies (as mass--size outliers) across different environments. In the field (i.e, in isolation), their large sizes might be a result of stellar feedback \citep{DiCintio2017,Chan2018}, early galaxy mergers \citep{Wright2021}, passing into and out of a more massive halo (so-called backsplash satellites; \citealt{Benavides2021}), or of inhabiting halos with higher spins \citep{Dalcanton1997,Amorisco2016,Liao2019,ManceraPina2020,Benavides2022,Kong2022}. In groups and clusters, their large sizes may ensue from tidal heating near pericenter \citep{Jiang2019}, via adiabatic expansion due to mass loss \citep{Tremmel2020}, or even via a bullet-dwarf collision \citep{vandokkum2022Nat,vanDokkum2022GC}. Despite theoretical advances, it is still an open question how mass--size outliers like UDGs are formed. It is also unclear whether mass--size outliers belong to a distinct class of galaxies or are just a natural extension of dwarf systems to lower surface brightness or larger sizes \citep[e.g.,][]{Greene2022}. A straightforward way to answer these questions is to compare mass--size outliers with ``normal'' low-mass galaxies, which have been studied extensively. 

Our best understanding of the low mass galaxy regime comes from the satellite system of our Milky Way (MW) and the Local Group \citep[LG; e.g.,][]{McConnachie2012,Simon2019}. With the advent of deep sky surveys and dedicated spectroscopic programs, we are now in a position to also map the satellite systems of MW analogs (typically defined as galaxies having similar stellar mass or halo mass to MW) in the Local Volume and nearby Universe. The Satellites Around Galactic Analogs survey (SAGA; \citealt{SAGA-I,SAGA-II}) is designed to spectroscopically confirm the satellites of 100 MW analogs at 20--40 Mpc. The Exploration of Local VolumE Satellites survey (ELVES; \citealt{ELVES-I,ELVES-II,CarlstenELVES2022}) uses deep imaging data to identify satellite candidates of MW-like hosts in the Local Volume (LV; $D<12$ Mpc) and estimate surface brightness fluctuation distances from images to confirm their association with their hosts. Having these satellites of MW analogs as a reference frame, we can contrast them with the mass--size outliers in the same environment to understand the formation of large diffuse galaxies. However, there are only $\sim 40$ confirmed UDGs associated with MW analogs in the literature \citep{Roman2017b,Cohen2018,SAGA-II,CarlstenELVES2022,Nashimoto2022,Karunakaran2022b}. Therefore, a larger sample of mass--size outliers associated with MW-like hosts is needed to conduct such a comparison between the mass--size outliers and the normal satellites.

The goal of this paper is to find and study large diffuse satellite galaxies in the context of MW analogs. Leveraging the depth and wide sky coverage of Hyper Suprime-Cam (HSC) Subaru Strategic Program \citep{Aihara2018} imaging data, we perform a search for LSBGs using novel methods in detection, false-positive identification, and modeling. We further build a large sample of mass--size outliers that are likely satellites of MW-mass hosts in the nearby Universe. We propose a new definition of mass--size outliers, dubbed ``\textit{ultra-puffy galaxies}'' (UPGs), based on the observed mass--size relation of satellites of MW analogs in the LV using the ELVES sample. This concept is less affected by the galaxy color and provides a robust avenue to study the mass--size outlier population. 
We calculate the abundances of mass--size outliers in MW analogs and compare them with their abundances in other environments. Our results seek to shed light on the role of the environment on the formation of large diffuse dwarf galaxies.

The layout of this paper is as follows. Section \ref{sec:data} describes the data used in this work. In Section \ref{sec:lsbg_search}, we describe the LSBG search in HSC data, including source detection (\S \ref{sec:detection}), deblending (\S \ref{sec:deblending}), modeling (\S \ref{sec:modeling}), completeness, and uncertainty (\S\ref{sec:comp_meas}). In Section \ref{sec:sample_construction}, we propose a new definition of large and diffuse galaxies (UPGs) and present our UDG and UPG samples. In Section \ref{sec:results}, we compare the UDGs and UPGs on the mass--size plane. We calculate their abundances and compare them with the literature (\S \ref{sec:n_udg}). Based on these results, we further argue that we should move beyond UDGs and adopt the UPG definition in Section \ref{sec:discussion}. Section \ref{sec:summary} presents a summary of this work and prospects for the future. In a companion paper \citep{Li2023}, we discuss the size distributions, spatial distributions, and quenching of UDGs and UPGs. 

In this work, we refer to mass--size outliers (including both UDGs and UPGs) as galaxies with large physical sizes compared to normal dwarf galaxies. We adopt a flat $\Lambda$CDM cosmology from \citet{Planck15} with $\Omega_{\rm m}= 0.307$ and $H_0 = 67.7\ $km s$^{-1}$ Mpc$^{-1}$. We use the AB system \citep{Oke1983} for magnitudes. The stellar mass used in this work is based on a \citet{Chabrier2003} initial mass function.

\section{Data} \label{sec:data}
\subsection{The Hyper Suprime-Cam Subaru Strategic Program}
The Hyper Suprime-Cam Subaru Strategic Program (HSC-SSP, hereafter the HSC survey; \citealt{Aihara2018})\footnote{\url{https://hsc-release.mtk.nao.ac.jp/doc/}} is an optical imaging survey using the 8.2 m Subaru telescope and the Hyper Suprime-Camera \citep{Miyazaki2012, Miyazaki2018}. The \texttt{Wide} layer is designed to cover $\sim 1000\ \rm{deg}^{2}$ of the sky in five broad bands ($grizy$), reaching a depth of $g=26.6$ mag, $r=26.2$ mag and $i=26.2$ mag ($5\sigma$ point-source detection). HSC data are processed using  \code{hscPipe}\footnote{\url{https://hsc.mtk.nao.ac.jp/pipedoc_e/}} \citep{Bosch2018}, which is a customized version of the Vera Rubin Observatory Legacy Survey of Space and Time (LSST) pipeline \citep{LSST-pipeline}\footnote{\url{https://pipelines.lsst.io/}}.

In this work, we use the \code{Wide} layer of the co-added data from the Public Data Release 2 (PDR2, also known as \code{S18A}, \citealt{Aihara2018}) of the HSC survey. It covers $\sim 300\ \rm{deg}^2$ in all five bands, which is 1.5 times larger than the dataset (\code{S16A}) analyzed in \citet{Greco2018}. One of the key improvements made in PDR2 is the sky background subtraction. Compared with previous data releases, PDR2 adopted a full focal plane sky subtraction algorithm to overcome the oversubtraction of the local sky background around bright objects \citep{Aihara2018,Li2021}. The unprecedented depth and careful sky subtraction make PDR2 an ideal dataset to study low surface brightness galaxies. In this work we take the point-spread function (PSF) models in \code{hscPipe} \citep{Bosch2018}, where the Point Spread Function Extractor \citep[\texttt{PSFEx};][]{Bertin2011} takes a star catalog and generates the PSF model.

\subsection{NASA-Sloan Atlas}
We use the NASA-Sloan Atlas (NSA\footnote{\url{http://nsatlas.org}}, \citealt{Blanton2005,Blanton2011}) to select galaxies that are analogous to our Milky Way based on their stellar mass (see \S\ref{sec:match}); then we match the LSBGs to these MW analogs. The NSA catalog provides various physical properties of galaxies in the nearby Universe as derived from the Sloan Digital Sky Survey \citep[SDSS;][]{York2000}. We use the most recent version of the NSA catalog (\code{v1\_0\_1}\footnote{\url{https://www.sdss.org/dr13/manga/manga-target-selection/nsa/}}) which contains about $640,000$ galaxies out to $z < 0.15$. This version also updates the aperture photometry to elliptical Petrosian photometry, which is considered to be more reliable than the photometry used in the older versions. In this paper, we use the stellar masses derived from the elliptical Petrosian photometry using \code{kcorrect v4\_2} \citep{Blanton2007}. The redshifts of the galaxies in the NSA come from several spectroscopic surveys, \ion{H}{1} gas surveys, or direct distance measurements. 

\section{LSBG Search in HSC PDR2}\label{sec:lsbg_search}

Continuing the work of \citet[][hereafter \citetalias{Greco2018}]{Greco2018}, we conducted a systematic search for low surface brightness galaxies in the HSC PDR2 data, which covers $\sim 1.5$ times the sky area of \citetalias{Greco2018}. The schematic of the searching method in this paper remains similar to that in \citetalias{Greco2018}, but we do improve the search completeness and purity by adding several new steps, which are highlighted in this section. We outline the major steps below and refer the readers who are already familiar with the algorithms in LSBG searches to \S \ref{sec:sample_construction} for the mass--size outlier sample construction.

\begin{enumerate}
    \item Source detection (\S \ref{sec:detection}): We run \href{https://www.astromatic.net/software/sextractor/}{\code{SourceExtractor}} \citep{Bertin1996} on the co-added images after removing bright extended sources. Then we apply an initial size and color cut based on the output catalog. 
    \item Deblending (\S \ref{sec:deblending}): We remove false positives that are not likely to be LSBGs by running \href{https://pmelchior.github.io/scarlet/}{\code{scarlet}} \citep{Melchior2018}. We use the \code{scarlet} models to define color--size--morphology--surface brightness cuts. This step removes roughly 98\% of false positives.
    \item Modeling (\S \ref{sec:modeling}): We fit a parametric model to the LSBGs to estimate their properties, including their sizes, total magnitudes, and average surface brightnesses. 
    \item Completeness and measurement biases and uncertainties (\S \ref{sec:comp_meas}) are characterized by injecting mock \sersic{} galaxies into images and recovering them following the procedures described above. 
\end{enumerate}
The number of objects remaining after each step is shown in Table \ref{tab:steps_flow}. Our source detection pipeline \code{hugs}\footnote{\url{https://github.com/johnnygreco/hugs}} and the deblending and modeling code \code{kuaizi}\footnote{\url{https://github.com/AstroJacobLi/kuaizi}} are open-source and available online.

\begin{deluxetable*}{lcc}
\tablecaption{LSBG Search Steps and Number of Remaining Objects After Each Step}
\tablewidth{20cm}
\label{tab:steps_flow}
\tablehead{
\colhead{Process} & \colhead{Description} &
\colhead{Remaining Objects}
}
\startdata
Initial detection & \S\ref{sec:detection} & 86,002 \\
Matching with MW analogs at $0.01 < z < 0.04$ & \S\ref{sec:match} & 10,579 \\
Deblending &\S\ref{sec:deblending} & 2673\\
Spergel profile modeling & \S\ref{sec:modeling}, \S\ref{sec:match} & 2510\\
Completeness & \S\ref{sec:comp_meas},\S\ref{ap:comp_meas_unc} & 2510 \\
Measurement error and uncertainty & \S\ref{sec:comp_meas},\S\ref{ap:comp_meas_unc} & 2510\\
Mass--size outlier selection & \S\ref{sec:sample} & 337 UPGs, 412 UDGs
\enddata
\end{deluxetable*}

\subsection{Source Detection}\label{sec:detection}
\citetalias{Greco2018} performed a search for LSBGs in the first $\sim 200$ deg$^2$ of the HSC survey and uncovered 781 LSBGs. We continued the work in \citetalias{Greco2018} and extended the search to the HSC PDR2 data, which cover $\sim 300\ \rm{deg}^{2}$ and have sky subtraction better suited for LSB studies compared to prior data releases. We follow the same method for source detection as in \citetalias{Greco2018} but make several updates to accommodate the PDR2 data. These updates are guided by both our understanding of PDR2 data and completeness tests using mock galaxies. We summarize the main steps of the LSBG search method here and refer the interested reader to \citetalias{Greco2018} and Appendix \ref{ap:detection} for more details. 

We start by replacing the bright sources and their associated LSB outskirts with sky noise. Bright objects and their diffuse light are identified using surface brightness thresholding. We run an additional detection using \href{https://sep.readthedocs.io/en/v1.1.x}{\code{sep}} \citep{Barbary2016} to remove small compact sources and noisy peaks and replace their footprints with sky noise. Then we use \code{SourceExtractor} to detect sources on the ``cleaned'' images with a low detection threshold. Taking the output catalog from \code{SourceExtractor}, we apply an initial selection based on the size, color, and peakiness of objects. 

After these steps, we have an initial sample that contains 86,002 LSBG candidates. We obtain 4 times more sources per square degree ($\sim 280\ \mathrm{deg}^{-2}$) than \citetalias{Greco2018} mainly because HSC PDR2 has sky subtraction suited for finding LSB features. Our detection here is more inclusive of larger objects, and our size cut is less restrictive for small objects. Among these objects, there are still many false positives, including shredded galaxy outskirts, tidal features, Galactic cirrus, and blended compact sources. Therefore, we perform an extra ``deblending'' step to further remove false positives.

\subsection{Deblending}\label{sec:deblending}

\begin{figure*}
	\vbox{ 
		\centering
		\includegraphics[width=1\linewidth]{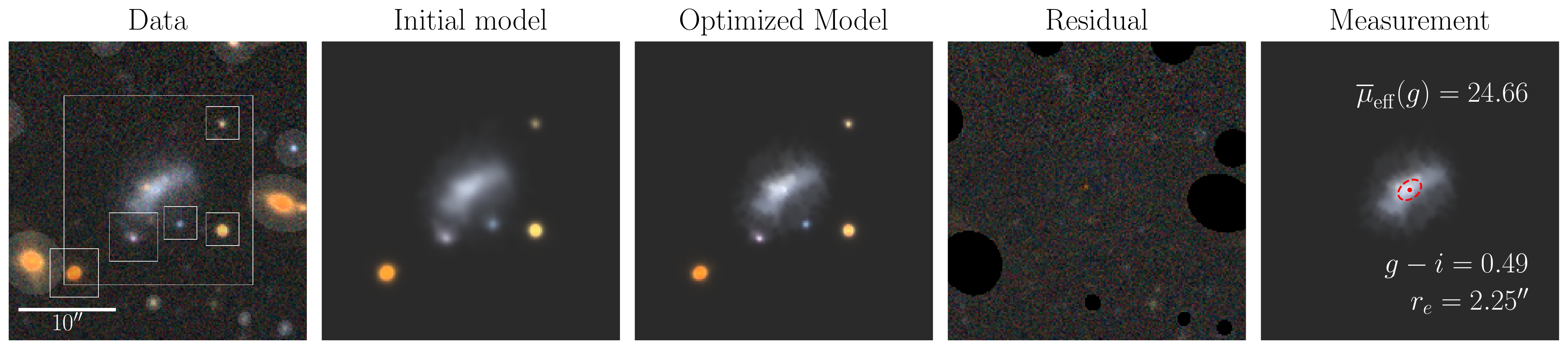}
		\includegraphics[width=1\linewidth]{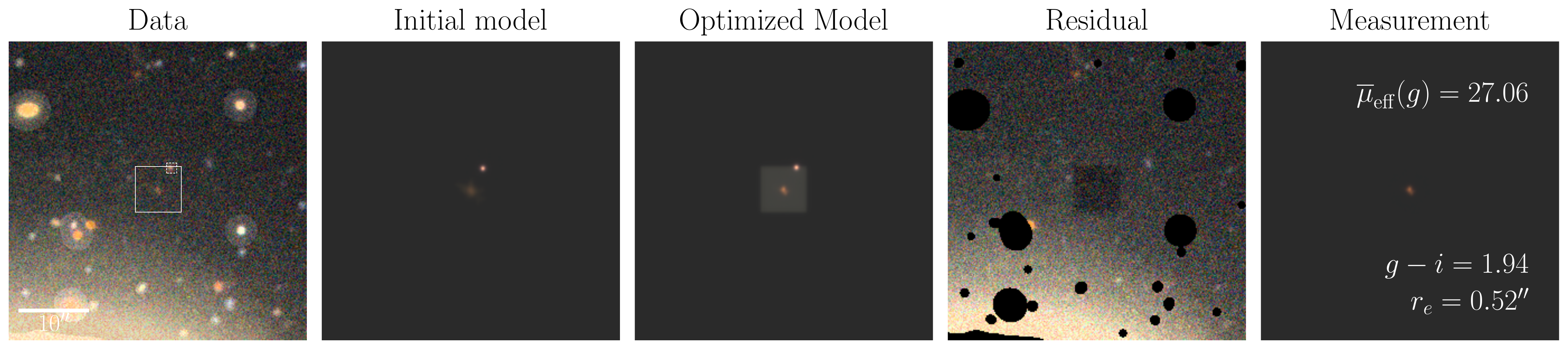}
	}
	\caption{Demonstration of the deblending step described in \S\ref{sec:deblending} and \S\ref{ap:deblending}. Here we show two objects from our initial sample: the one shown in the top panels is a blue LSBG, whereas the one shown in the bottom is a false positive, which is a blend between a high-$z$ galaxy and galaxy outskirts. The first columns show the $griz$ color-composite images with the bounding boxes overlaid. The objects covered by the gray shades are masked during fitting. The remaining columns show the initial model, the optimized model (PSF convolved), the residual image, and the optimized model of the target galaxy only. The red dashed circle in the rightmost column denotes the measured half-light radius.
	}
	\label{fig:vanilla_scarlet_demo}
\end{figure*}

Removing false positives from the initial sample while retaining high completeness is one of the major challenges in LSBG searches \citep[e.g.,][]{vanDokkum2015,Koda2015,Yagi2016,Greco2018,SAGA-I,Zaritsky2019,Zaritsky2021,Tanoglidis2021,Zaritsky2022}. A common type of false positives occurs when point-like sources are blended with diffuse light from background galaxies or the LSB outskirts of stars and galaxies. Our initial detection cannot distinguish them from real LSBGs. In order to identify and remove these objects from our sample, we perform nonparametric modeling of all LSBG candidates using \href{https://pmelchior.github.io/scarlet/}{\code{scarlet}} \citep{Melchior2018} and design an effective metric to remove false positives.

\subsubsection{\code{Scarlet} fitting}
\code{Scarlet} is a deblending and modeling tool designed for multiband and multiresolution imaging data. It utilizes color and morphology information to separate blended objects and model objects in a nonparametric fashion. In the following, we briefly summarize how \code{scarlet} works, and we refer the interested readers to \citet{Melchior2018,Melchior2021} and the online documentation\footnote{\url{https://pmelchior.github.io/scarlet/}} for more details. 

In \code{scarlet}, each source $k$ in the cutout is described by a morphology image $S_k$ and a spectral energy distribution (SED) vector $A_k$. The multi-band images are represented as $Y$. The goal of the modeling is to minimize the objective function $f = \frac{1}{2} ||Y - P \ast (\sum_k A_k^\top \times S_k)||^{2}$ under certain constraints, where $P$ is the PSF and $*$ is convolution. The morphology image of each source is limited by a bounding box. We assume two constraints when running \code{scarlet}: all sources have positive fluxes (positivity constraint), and the light profiles of all sources monotonically decrease from the center to the outskirts (monotonicity constraint). The monotonicity constraint is oversimplified for well-resolved galaxies with complicated structures, but for objects in our sample, this assumption still holds for most cases and provides an effective and robust way to deblend overlapping sources. We refer to this modeling method as \textit{vanilla} \code{scarlet}. 


We use vanilla \code{scarlet} to model the LSBG candidate, and the structural and morphological parameters are then used to exclude false positives from our LSBG sample. In the following, we briefly describe how we detect peaks on the image to initialize and optimize \code{scarlet} models. We refer the readers to Appendix \ref{ap:deblending} for a more detailed description. 

The deblending step is designed to model the sources in the vicinity of the LSBG candidate. Therefore, we run \code{sep} on the $griz$-combined image to detect objects (i.e., peaks) around the LSBG candidate. Objects that are close to the LSBG candidate are modeled using \code{scarlet} models. The models are initialized based on the smoothed image to capture the LSB outskirts of the target galaxy. Then the models are optimized using the adaptive proximal gradient method \citep{Melchior2019}. Typically, convergence is achieved after $\sim 50$ steps of optimization, and the whole modeling process takes about 40s for a typical LSBG.

The deblending procedures are demonstrated in Figure \ref{fig:vanilla_scarlet_demo} with two distinct examples. The one shown in the top panels is a blue LSBG, whereas the one shown on the bottom is a high-$z$ galaxy blended with the outskirts of a nearby galaxy, which falls in our initial sample as a false positive. 

We note that the optimized model of the blue LSBG captures its color and morphology quite well. For the false-positive case, the optimized model has a notable bright background because we model the sky together with all other sources. The model for the target itself is actually very red and compact. It passes our initial selection because the galaxy outskirts are shredded by \code{SourceExtractor} and happen to have a large size and similar color to our LSB galaxies. However, once modeled using \code{scarlet}, it becomes clear that this is just a false-positive detection and should be removed. A rubric based on the \code{scarlet} model is therefore needed to help us identify and exclude the false positives.

\subsubsection{False-positive Removal}\label{sec:non-par-measurement}

After running vanilla \code{scarlet} for LSBG candidates, the target object is successfully deblended from nearby sources. Unlike other parametric modeling methods, the nonparametric \code{scarlet} model is flexible enough to adequately represent galaxies with complex structures. However, the nonparametric nature of the code means that we have to make additional measurements to quantify the size and shape of the \code{scarlet} model. As shown in the rightmost column of Figure \ref{fig:vanilla_scarlet_demo}, we isolate the model of the target object and analyze it using \href{https://statmorph.readthedocs.io/en/latest/}{\code{statmorph}} \citep{statmorph}. The purpose of this analysis is to extract the structural and morphological parameters of the target object. We further use these diagnostics to remove false positives. 

\code{Statmorph} calculates non-parametric morphological and structural parameters, including effective radius, concentration-asymmetry-smoothness statistics \citep[CAS;][]{Conselice2003}, and Gini-M20 \citep{Abraham2003,Lotz2004}. The measurement requires an image, a variance map, and a PSF. For our purpose, the ``image'' is just the \code{scarlet} model convolved with the observed PSF. The variance map and PSF are taken from HSC cutouts. We also force the sky level in \code{Statmorph} to be zero because the sky has already been fit when running vanilla \code{scarlet}. 

One of the most important diagnostics is the effective radius of the object. In this paper, we refer to the effective radius as the circularized effective radius $r_{e}$, defined as the $r_{e} = r_{\rm eff,sma} \sqrt{b/a}$ where $r_{\rm eff,sma}$ is the effective radius measured along the semi-major axis of the aligned elliptical isophotes, and $b/a$ is the axis ratio of the isophotes. \footnote{We note that some authors use $r_{\rm eff, sma}$ as the effective radius. We do not find consensus in the literature regarding the most ideal definition of the effective radius.} In \code{statmorph}, the effective radius is calculated using elliptical apertures, where the ellipticity $\varepsilon$ is determined by calculating the second moment of the image. The total magnitudes are simply calculated by summing up the model flux in each band, and colors are defined using total magnitudes. A Galactic extinction correction is not applied at this point. The central surface brightness $\mu_0$ is measured by linearly extrapolating the surface brightness profile to $r\to 0$ using the \code{scipy} interpolation function\footnote{\url{https://docs.scipy.org/doc/scipy/reference/generated/scipy.interpolate.interp1d.html}} \citep{scipy}. We define the average surface brightness within the effective radius as 
\begin{equation}\label{eq:mu_eff}
    \overline{\mu}_{\rm eff} = m + 2.5 \log_{10}(2 \pi r_e^2),
\end{equation}
where $m$ is the total magnitude. 

We also use the Gini-M20 and CAS statistics. The Gini coefficient and M20 statistics \citep{Abraham2003,Lotz2004} quantify how concentrated/extended the flux distribution is across the image. The CAS statistics characterize the concentration $C$, asymmetry $A$, and smoothness $S$ of the light distribution of the object. We refer the reader to \citet{statmorph} for more details on their definitions and implementations.

To better guide us on how to use these diagnostics, we visually inspect a subset of LSBG candidates in our initial sample (\S \ref{sec:detection}) and use the classification results to help construct the metrics. More specifically, we randomly select 5000 LSBG candidates that are matched with a MW-like host at $0.01 < z < 0.04$ from the NSA catalog (see \S\ref{sec:match} for details). We run vanilla \code{scarlet} for all of them and measure the structural and morphological parameters as described above. Then we do visual inspections of the $griz$ color-composite images with 0\farcm5 on a side. False positives, including tidal streams, galaxy outskirts, blends, and other artifacts (dubbed \code{junk}), are identified during the visual inspection by coauthors J.E.G., J.G., S.H., R.B., K.C., A.G., and E.K-F. Each object has been inspected by at least two people. In the end, an object is classified as \code{junk} if the number of votes as \code{junk} is larger than the number of votes as \code{nonjunk}. In total, there are 1661 objects classified as \code{junk}. The visual inspection is summarized in Table \ref{tab:deblending_vis}.

We first apply the following selection cuts to remove false positives and background galaxies. They are motivated based on the knowledge of the color distribution of LSBGs \citep[e.g.,][]{SAGA-I,Greco2018,Zaritsky2019,Tanoglidis2021} and survey completeness in size and surface brightness (\S\ref{sec:comp_meas}). 
\begin{itemize}
    \item Color:
    \[0.0 < g-i < 1.2,\quad |(g-r) - 0.7\cdot (g-i)| < 0.25\]
    \item Size: \[1.8 \arcsec < r_e < 12 \arcsec\]
    \item Surface brightness: \[\mu_0(g) > 22.5,\quad 23.0 < \sbeff < 27.5\ \sbunit.\]
\end{itemize}
The color cuts here are more restrictive than the one used in the initial detection (Sec \ref{sec:detection}) to further remove junk and high redshift galaxies. We not only remove objects with small sizes but also exclude objects with large sizes and very low surface brightness because they are mostly spurious objects.
After such cuts, there are 200 \code{junk} and 1407 \code{non-junk} remaining. The color-size-surface brightness cuts remove 88\% of \code{junk} and 58\% of \code{nonjunk}. We note that although we lose a lot of \code{nonjunk} objects, most of them are small, red objects, which are not likely to be LSBGs of our interest.

\begin{figure}
    \centering
    \includegraphics[width=1\linewidth]{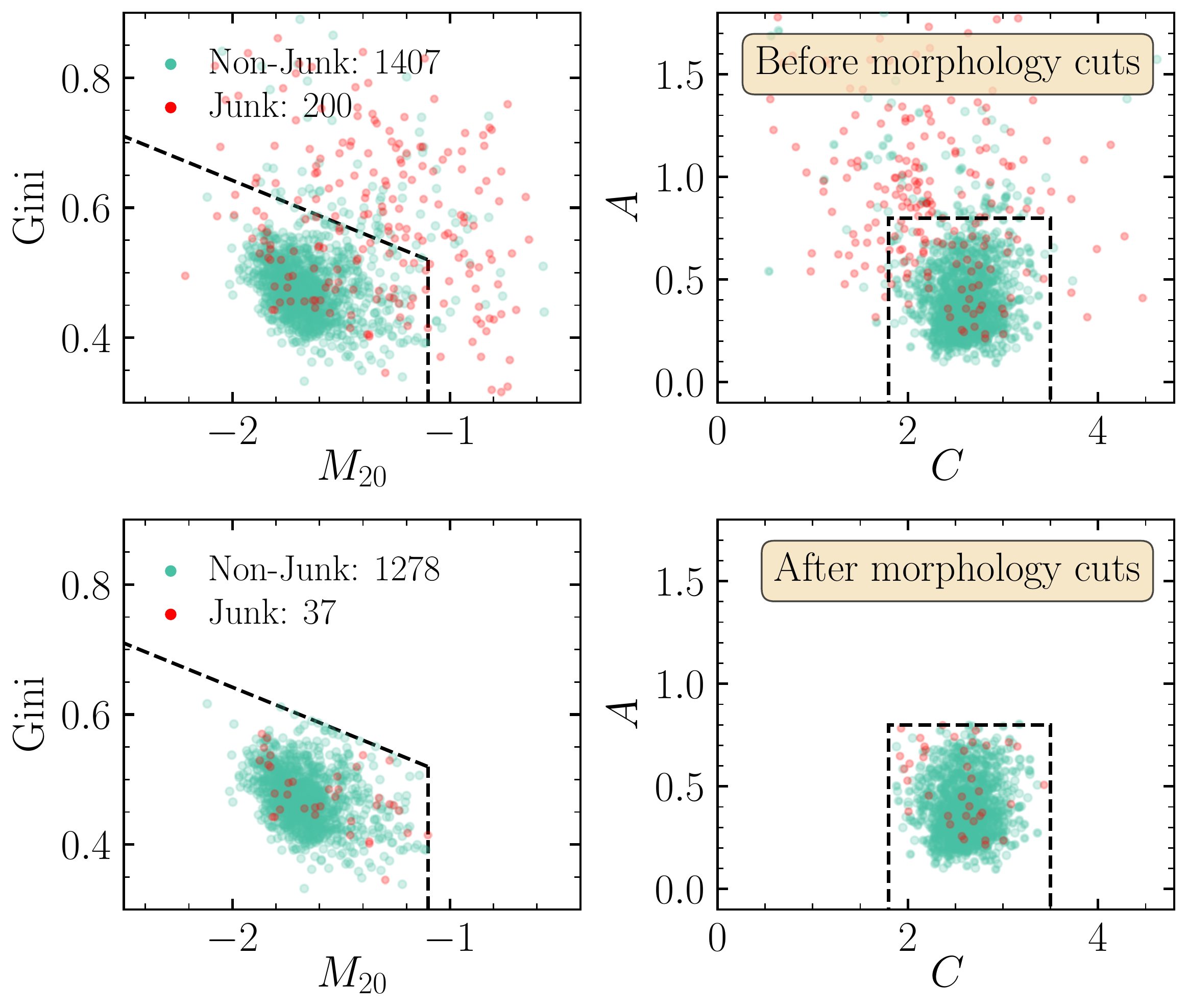}
    \caption{Distribution of \code{scarlet} morphological measurements for the LSBG candidates from our initial sample. The sample after the color--size--surface brightness cuts is shown in the top panels. We further remove false positives (\code{junk}; highlighted in red) by selecting their morphological parameters (dashed boxes). Morphology cuts (bottom panels) remove another 82\% of \code{junk}. In total, 98\% of false positives are removed in the deblending step.
	}
	\label{fig:deblending_cuts}
\end{figure}

We further use these visual inspection results to inform how to use morphological diagnostics. In the top panels of Figure \ref{fig:deblending_cuts}, we show the distributions of \code{junk} (red) and \code{nonjunk} (green) in the parameter space spanned by four morphological parameters. There is still a significant amount of \code{junk} with high $M_{20}$, Gini coefficient, and asymmetry. Therefore, we add another set of selection cuts to further remove \code{junk}:
\begin{itemize}
    \item Morphology: 
    \begin{gather*}
        \varepsilon < 0.65,\quad \mathrm{Gini} < 0.70,\quad M_{20} < -1.1,\\
        \mathrm{Gini} < -0.136\cdot M_{20} + 0.37,\\
        1.8 < C < 3.5,\quad A < 0.8.
    \end{gather*}
\end{itemize}
The slanted demarcation line on the Gini-$M_{20}$ diagram is motivated by the line used in \citet{Lotz2008} to separate merging galaxies from normal galaxies. The bottom panels in Figure \ref{fig:deblending_cuts} show the objects after applying the morphology selections, which remove another 82\% of \code{junk} while only removing 9\% of \code{nonjunk}. Such morphology cuts (shown as dashed lines in Figure \ref{fig:deblending_cuts}) effectively help us remove false positives that are not likely to be LSBGs of interest. 

In this way, we obtain an LSBG sample with high purity, where 98\% of the false positives are removed (see Table \ref{tab:deblending_vis}). Therefore, our empirical selection based on the nonparametric measurements on the scarlet models successfully removes most of the false positives and a large fraction of small red galaxies (most likely background galaxies) in our initial sample. Furthermore, the completeness of the real LSBG detection remains high. In \S\ref{sec:comp_meas}, we characterize the completeness of this ``deblending'' step by injecting mock \sersic{} galaxies, and we achieve $\sim80\%$ completeness at $\sbeff = 27.0\ \sbunit$ and $>50\%$ completeness at $\sbeff = 27.5\ \sbunit$ (Fig. \ref{fig:completeness}). We emphasize that the vanilla \code{scarlet} modeling and nonparametric measurements are not designed to measure the structural properties of the galaxies. They are used only as a diagnostic tool to remove false positives. We perform more detailed modeling in \S\ref{sec:modeling} to measure galaxy properties for science. 

\begin{deluxetable}{lccc}
\tablecaption{Number of Objects Remaining after Each Deblending Cut}
\tabletypesize{\footnotesize}
\label{tab:deblending_vis}
\tablehead{
\colhead{Process} & \colhead{\# \code{junk}} &
\colhead{\# \code{nonjunk}} & \colhead{\# Total}
}
\startdata
Visual inspection & 1661 & 3339 & 5000 \\
Color--size--surface brightness cuts & 200 & 1407 & 1607 \\
Morphology cuts & 37 & 1278 & 1305\\
\enddata
\end{deluxetable}

Many other works have used machine-learning (ML) algorithms to classify LSBG candidates and exclude spurious detection. Among others, \citet{Tanoglidis2021} take the output catalog from \code{SourceExtractor} and use the Support Vector Machine algorithm to classify objects in their initial sample and reduce the number of objects that are visually inspected. \citet{Zaritsky2019,Zaritsky2021,Zaritsky2022} use convolutional neural networks to classify LSBGs into binary classes based on candidate images. These endeavors certainly help reduce the human labor of visually inspecting tens of thousands of objects. However, our measurement-based method is more intuitive, reproducible, and transferable compared with many ML methods. We will explore ML methods and compare them with our deblending cuts in future work. Future work will also utilize information about Milky Way dust to exclude spurious detection of Galactic cirrus \citep[e.g.,][]{Zaritsky2021,Zaritsky2022}.


\subsection{Modeling}\label{sec:modeling}

Although the vanilla \code{scarlet} modeling provides useful information to help us remove false positives, it does not necessarily give us reliable estimations of galaxy size, color, total magnitude, etc. In fact, the vanilla \code{scarlet} model systematically underestimates the size and total flux of LSB objects. 
This is because nonparametric modeling cannot capture the faint outskirts of LSBGs very well. When doing parametric modeling such as \sersic{} fitting, we effectively apply radial averaging within elliptical annuli and boost the signal-to-noise ratio by binning pixels. In this case, pixels within the same isophotal annulus are assumed to have the same intensity and thus are strongly correlated. However, the nonparametric nature of vanilla \code{scarlet} only assumes monotonicity, which imposes quite weak correlations among pixels. Consequently, it is hard for nonparametric modeling to probe very LSB features. Furthermore, the monotonicity constraint stops the model from growing in certain directions if there is another source along this direction.\footnote{This issue can be seen in the third column of Figure \ref{fig:vanilla_scarlet_demo} where the model of the target object never exceeds the other objects next to it.} As a result, the nonparametric model often does not capture the very LSB outskirts of LSBG, thus biasing the measurements. In the following, we explore a novel method to perform robust parametric modeling and measurement for LSBGs.

A traditional way of doing parametric modeling is to mask out contaminants based on the detection segmentation map and fit a model to the masked image. However, such fitting results are very sensitive to the masking scheme and sky background \citep[e.g.,][]{Greco2018}. A possible solution to this problem is to simultaneously model all the objects in the cutout and the sky using parametric models \citep[e.g.,][]{Lang2016,Dey2019,Liu2022}. In this work, we combine the advantage of parametric modeling with the power of deblending in \code{scarlet}. To be specific, we follow the spirit of deblending as described in \S\ref{sec:deblending}, but replace the nonparametric model for the target galaxy with a parametric model. In this way, the LSB outskirts of LSBGs can be better captured with the parametric model, and the impact of contaminants is minimized because they are modeled simultaneously in all bands with nonparametric models. The parametric model for the target object can also extend to the whole scene without artificially truncating if it encounters a neighboring object. 

In this work, we use the Spergel surface brightness profile \citep{Spergel2010} to model the LSBGs (see Appendix \ref{ap:spergel} for details). The Spergel profile is motivated by having a simple analytical expression in Fourier space, making it easy to convolve with a PSF. Similar to the \sersic{} index, the parameter $\nu$ in Equation \eqref{eq:spergel} (the ``Spergel index'') controls the concentration of the light profile. As shown in Appendix \ref{ap:spergel}, the Spergel profile approximates the \sersic{} profile very well over the range of \sersic{} indices that are relevant to the study of LSBGs.

While other objects are initialized in the same way as in the deblending step (Section \ref{sec:deblending}), we initialize the Spergel model for the target object differently. First, we initialize a vanilla \code{scarlet} model for the target object, and we measure the effective radius $r_e$, total flux, and shape of the scarlet model. The size of the bounding box is also updated to be the maximum between 250 pixels and $10\, r_e$. For the target, we still require a positivity constraint, and the monotonicity is automatically satisfied by the Spergel profile. After optimization, we take the $r_0$ in Equation \eqref{eq:spergel} as the circularized half-light radius $r_e$ and take $L_0$ as the total flux. The average surface brightness $\overline{\mu}_{\rm eff}$ is calculated in the same way as in Equation \eqref{eq:mu_eff}. The Spergel modeling results are used for studying the properties of LSBGs. 
In the following section, we assess the quality of the Spergel modeling by injecting mock \sersic{} galaxies and comparing the recovered properties with the truth. 

\subsection{Completeness and Measurement Uncertainty}\label{sec:comp_meas}
\begin{figure*}
	\vbox{ 
		\centering
		\includegraphics[width=1\linewidth]{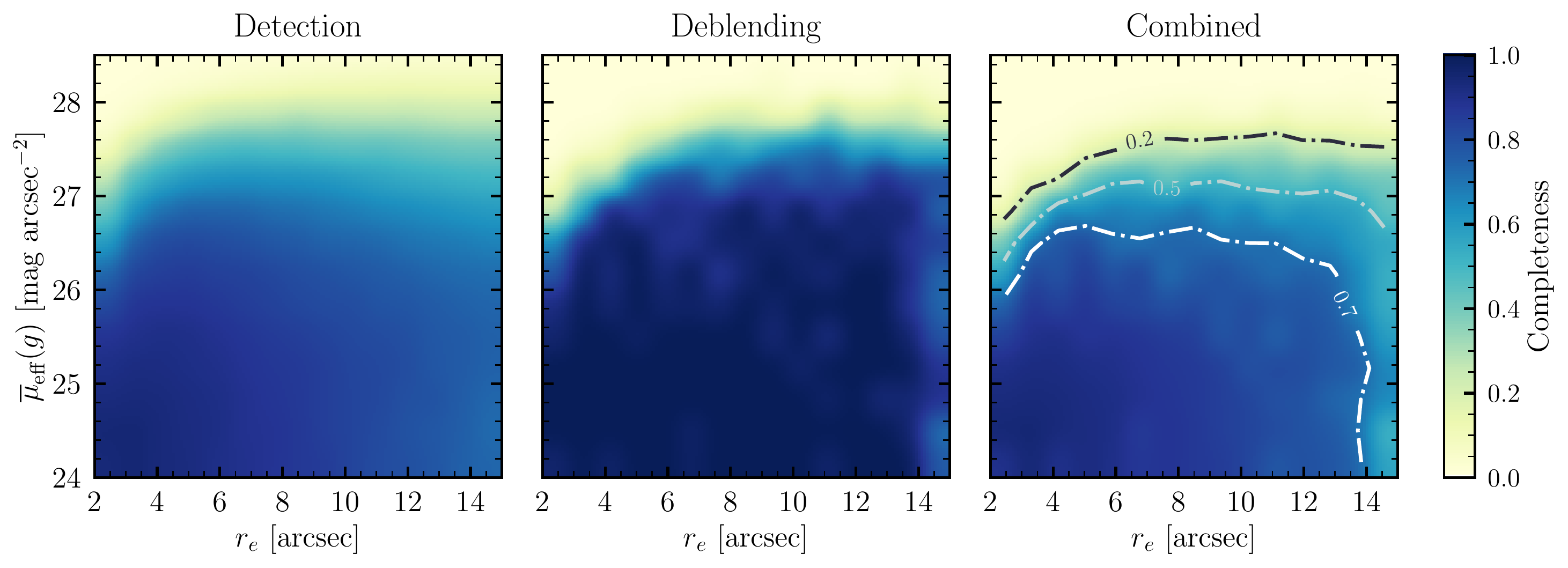}
	}
	\caption{Completeness of our LSBG search as a function of the effective radius $r_e$ and the average surface brightness $\sbeff$. The overall completeness (right panel) comprises the detection completeness (left panel) and deblending completeness (middle panel). The dashed lines in the right panel highlight the 70\%, 50\%, and 20\% completeness contours. The completeness drops as we go to fainter surface brightness and smaller size. For the size range of $3\arcsec < r_e < 14\arcsec$, we are $>70\%$ complete to $\sbeff < 26.5\ \sbunit$ and $>50\%$ complete to $\sbeff \leqslant 27.0\ \sbunit$. 
	}
	\label{fig:completeness}
\end{figure*}

The completeness of the LSBG search is important not only for understanding our search efficiency and improving the method in the future but also for deriving the completeness-corrected results for science purposes. Similar to many other works \citep[e.g.,][]{vdBurg2017,Zaritsky2021,CarlstenELVES2022,Greene2022}, we derive the completeness by simulating mock LSBGs and recovering them from the image. Using these mock galaxies, we also characterize the measurement bias and uncertainty by comparing the measured properties with the truth. We present the main steps below and describe details in Appendix \ref{ap:comp_meas_unc}.

A real LSBG can be missed in the detection step (\S \ref{sec:detection}) or excluded in the deblending step (\S \ref{sec:deblending}). Thus, the overall completeness combines the detection and deblending steps. We perform a large suite of image simulations to derive completeness. We inject single-\sersic{} light profiles \citep{Sersic1963} into the co-added images and try to recover them using the detection method (\S\ref{sec:detection}), model them using vanilla \code{scarlet} (\S\ref{sec:deblending}), and apply the deblending cuts. The mock \sersic{} galaxies span a wide range in size ($2\arcsec \leqslant r_{e} \leqslant 21\arcsec$) and surface brightness ($23 \leqslant \overline{\mu}_{\rm eff}(g) \leqslant 28.5\ \mathrm{mag\ arcsec^{-2}}$). Detection completeness is defined as the number of detected objects divided by the number of injected objects, whereas the deblending completeness is defined as the fraction of objects remaining after the deblending cuts. 

We find that completeness mainly depends on size and surface brightness. As shown in Figure \ref{fig:completeness}, the detection completeness remains high across different sizes. It drops below 50\% when the surface brightness gets fainter than $\sbeff = 27.5\ \sbunit$. The deblending completeness is very high at the bright end but starts to decline with increasing size and decreasing surface brightness. Mock galaxies fainter than $\sbeff > 27.5\ \sbunit$ are mostly removed by the deblending step, likely due to the blending between compact sources and the mock galaxy. Interestingly, the detection completeness and deblending completeness both drop below 40\% around $\sbeff=27.5\ \sbunit$. In this sense, the detection and deblending cooperate well in terms of completeness. 

The combined completeness is shown in the right panel of Figure \ref{fig:completeness}. The dashed lines highlight the 70\%, 50\%, and 20\% completeness contours. The completeness drops as we go to fainter surface brightness and smaller size. For the size range of $3\arcsec < r_e < 14\arcsec$, we are $>70\%$ complete to $\sbeff < 26.5\ \sbunit$ and $>50\%$ complete to $\sbeff \leqslant 27.0\ \sbunit$. Although \sersic{} galaxies do not necessarily resemble the morphology of real LSBGs, our completeness based on the \sersic{} model tests still sets a baseline for real completeness. In the future, we might be more capable of generating realistic LSBGs using deep learning techniques.

We compare our completeness with that of other LSBG searches. Our completeness is most similar to \citetalias{Greco2018}, where they also reached $\sim 50\%$ at $\sbeff \approx 27.0\ \sbunit$ and $3\arcsec < r_e < 12\arcsec$ \citep{Kado-Fong2021,Greene2022}, but their completeness decreased more quickly with increasing size than ours does. This is because HSC PDR2 has a better sky subtraction for LSB studies than S16A in \citetalias{Greco2018}, and our initial selection is more inclusive for large objects. \citet{CarlstenELVES2022} searched for LSBGs in Dark Energy Camera Legacy Survey (DECaLS; \citealt{Dey2019}) and Canada–-France-–Hawaii Telescope (CFHT) data using a similar search algorithm to \citetalias{Greco2018} and ours. They reached a completeness of $\sim 50\%$ at $\mu_0(g)\approx 26.5\ \sbunit$ or equivalently $\sbeff = 27.5\ \sbunit$ assuming a Sersic index $n \approx 1$. \citet{vdBurg2016} surveyed eight galaxy clusters at $0.044 < z < 0.063$ using CFHT data and achieved $\sim 50\%$ completeness at $\sbeffr\approx 26.0\ \sbunit$ or equivalently $\sbeff \approx 26.6\ \sbunit$. \citet{vdBurg2017} searched for LSBGs from the ESO Kilo-Degree Survey data and achieve similar completeness as in \citet{vdBurg2016}. \citet{ManceraPina2019a} also searched for UDGs in nearby clusters using the 2.5 m Issac Newton Telescope (INT) and achieve $\sim 50\%$ completeness at $\sbeffr\approx 26.5\ \sbunit$ or equivalently $\sbeff \approx 27.1\ \sbunit$. \citet{Zaritsky2021} focused on searching for LSBGs in the SDSS Stripe82 region using DECaLS images and reach $\sim 25\%$ completeness at $\mu_{0}(g) \approx 25.5\ \sbunit$ or equivalently $\sbeff \approx 26.5\ \sbunit$ for $n=1$. Their lower completeness might be explained by the fact that they remove a large fraction of the sky contaminated by Galactic dust. \citet{Tanoglidis2021} conducted an LSBG search using the Dark Energy Survey (DES; \citealt{Abbot2018}) data and concluded an overall completeness of $\sim 40\%$ by comparing their catalog with the deeper observations in the Fornax cluster. \citet{Kado-Fong2021} estimated the completeness of \citet{Tanoglidis2021} by comparing with \citetalias{Greco2018} and find a completeness similar to \citetalias{Greco2018} at $\sbeff = 25.5\ \sbunit$ but a much lower completeness for $\sbeff > 26.5\ \sbunit$. The Dragonfly Wide Field Survey \citep{Danieli2020} reached $\sim 29\ \sbunit$ for $>3\sigma$ detection on scales $\sim 1\arcmin$, which is deeper than all other surveys by sacrificing spatial resolution. In summary, our search achieves overall high completeness compared with other works, and we demonstrate the great power of HSC data on LSBG studies \citep[e.g.,][]{Huang2018,Kado-Fong2018}. 

\begin{figure*}
	\vbox{ 
		\centering
		\includegraphics[width=1\linewidth]{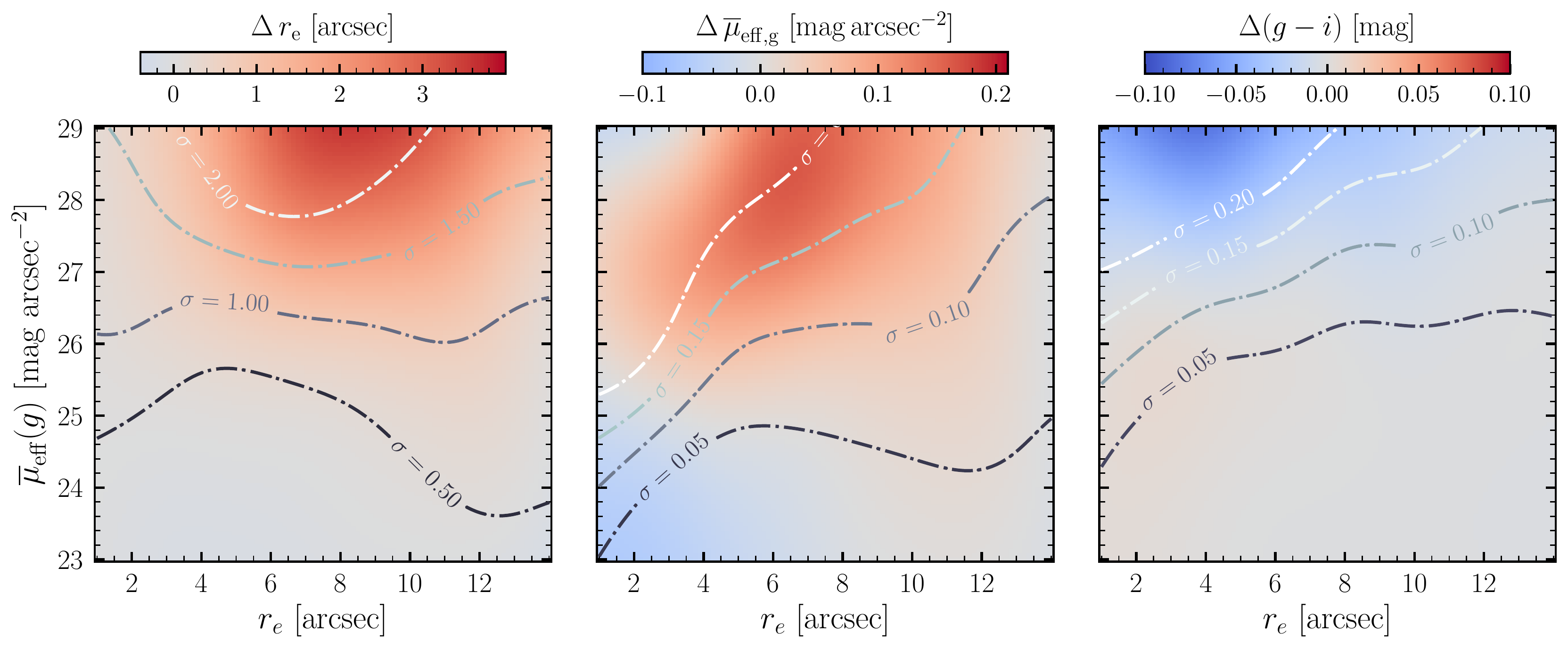}
	}
    \caption{Measurement bias (defined as ``truth $-$ measurement'') and uncertainty as a function of the measured angular size $r_e$ and surface brightness $\sbeff$. The colors show the bias term while the dashed contours show the constant uncertainty lines. We derive the bias in the total magnitudes from the combined bias in $r_e$, $\sbeff$, and color. We find that the bias and uncertainty in our measurement increase toward the fainter end. We apply the bias correction to the real LSBGs before studying their physical properties.}
    \label{fig:meas_err}
\end{figure*}

\vspace{1em}

The size, magnitude, surface brightness, and shape of LSBGs are hard to characterize because of their low-surface-brightness nature. We characterize the quality of the structural measurements by comparing the Spergel modeling results with the ground truth for mock \sersic{} galaxies. We find that the measured effective radius $r_e$ tends to be smaller than the truth, and the bias depends on the surface brightness and the angular size. For surface brightness fainter than $>27\ \sbunit$, the measured size can be much smaller than the truth. As a result, the measured total magnitude $m$ is also too faint. For the measured average surface brightness $\overline{\mu}_{\rm eff}$, the trade-off between the total magnitude and size makes it less biased than the total magnitude. Similar to \citet{Zaritsky2021}, we derive the measurement bias and uncertainties in size, surface brightness, total magnitude, and color as a function of size and surface brightness. We refer interested readers to Appendix \ref{ap:comp_meas_unc} for details. The main results are summarized in Figure \ref{fig:meas_err}, where the colors show the bias terms as a function of the measured size and surface brightness. The size and surface brightness bias grow with increasing surface brightness. The bias term is not a monotonic function of the size because the size here in Figure \ref{fig:meas_err} is not intrinsic. It is the bias in size measurement that makes galaxies with large intrinsic size pile up around $r_e\sim 6\arcsec$. We find the bias in $g-i$ color to be quite small. The $1\sigma$ measurement errors $\sigma(X)$ are shown in Figure \ref{fig:meas_err} as contours, and they have the same units as the bias. We set the minimum uncertainties to be $\sigma(r_e) \geqslant 0\farcs3,\ \sigma(\overline{\mu}_{\rm eff}) \geqslant 0.05\ \sbunit,\ \sigma(g-i) \geqslant 0.05$ to avoid meaninglessly small uncertainties. 

We apply the bias corrections to the real LSBGs, estimate the measurement uncertainties, and evaluate the completeness based on the bias-corrected properties. 
The values presented in our catalogs (Table \ref{tab:catalog}) are already bias-corrected, but we do provide the bias values for reference. 

\begin{deluxetable}{lll}
\tabletypesize{\footnotesize}
\label{tab:catalog}
\tablecaption{UDG and UPG Catalog Description} 
\tablehead{\colhead{Column Name} & \colhead{Unit} & \colhead{Description}} 
\startdata
Type  &  &  Galaxy type\\
ID                       &         & Unique LSBG ID \\
R.A.                       & deg     & Source R.A. in decimal degrees \\
 & & (J2000) \\
Decl. & deg     & Source Decl. in decimal degrees\\
& & (J2000) \\
$r_e$         & arcsec  & Circularized effective radius  \\
$\sigma(r_e)$ & arcsec  & Uncertainty in $r_e$ \\
$\overline{\mu}_{\mathrm{eff}}(g)$               & $\sbunit$ & $g$-band average surface brightness \\
& & within $r_e$ \\
$\sigma(\overline{\mu}_{\mathrm{eff}}(g))$       & $\sbunit$ & Uncertainty in $\overline{\mu}_{\mathrm{eff}}(g)$           \\
$m_g$                    & mag     & $g$-band apparent magnitude     \\
$\sigma(m_g)$            & mag     & Uncertainty in $m_g$            \\
$g-i$                    & mag     & $g-i$ color                     \\
$\sigma(g-i)$            & mag     & Uncertainty of $g-i$ color      \\
$g-r$                    & mag     & $g-r$ color                     \\
$\sigma(g-r)$            & mag     & Uncertainty of $g-r$ color      \\
$\nu$                    &         & Spergel index              \\
$\varepsilon$            &         & Ellipticity                     \\
$A_g$                    & mag     & $g$-band Galactic extinction \\
$A_r$                    & mag     & $r$-band Galactic extinction \\
$A_i$                    & mag     & $i$-band Galactic extinction \\
comp & & Completeness \\
weight & & Importance weight \\
$r_{e,\rm phys}$ & kpc  & Circularized physical effective radius  \\
$\log\, M_\star$ & $M_\odot$ & log stellar mass \\
$\sigma(\log\, M_\star)$ & $M_\odot$ & Uncertainty in $\log\, M_\star$ \\
host\_name & & Host galaxy name \\
host\_ra & deg & Host galaxy R.A. in decimal degrees \\
& & (J2000) \\
host\_dec & deg & Host galaxy Decl. in decimal degrees \\
& & (J2000) \\
host\_z &  & Host galaxy redshift \\
host\_log\_m\_star & $M_\odot$ & Host galaxy log10 stellar mass\\
host\_r\_vir & kpc & Host galaxy virial radius \\
host\_g\_i & mag & Host galaxy $g-i$ color \\
sep\_to\_host & arcmin & Angular separation between \\
& & host and UDG/UPG \\
\enddata
\tablecomments{
The table is published in its entirety in machine-readable format.
Magnitudes are on the AB system and have been corrected for measurement biases and Galactic
extinction. The information on the host galaxies is from NASA-Sloan Atlas. We provide Galactic extinction corrections, which are derived from
the \citet{Schlafly2011} recalibration of the \citet{SFD1998} dust maps. 
}
\end{deluxetable}

\section{Mass--Size Outliers Around Milky Way Analogs}\label{sec:sample_construction}

The goal of this paper is to study the mass--size outliers in the satellite systems of MW-like galaxies. However, distance information is needed to convert the observed size and magnitude to the physical size and stellar mass. It is well-known that getting the distances to LSBGs is one of the major obstacles to studying their properties. Besides direct distance measurements (e.g., SAGA and ELVES; \citealt{Kadowaki2021}), it is common to assume a distance to an LSBG by associating it with a host galaxy based on the projected angular distance \citep[e.g.,][]{vanDokkum2015,vdBurg2016,Wang2021,Zaritsky2022,Nashimoto2022}. Statistical background subtraction is then needed to account for the contribution from background and foreground sources. Furthermore, the cross-correlation between an LSBG sample and a host galaxy sample can also reveal the distance distribution of LSBGs \citep{Greene2022}. These methods are also complemented by recent machine-learning techniques \citep{Baxter2021,xSAGA-I}. 

In this section, we cross-match our LSBG catalog with MW-like galaxies in the NSA catalog. For the LSBGs matched with MW analogs, we run the deblending step to exclude false positives (\S\ref{sec:deblending}) and then run Spergel modeling to measure the properties of LSBGs (\S\ref{sec:modeling}). In the end, we construct the samples of mass--size outliers among the satellites of MW analogs. Table \ref{tab:steps_flow} summarizes the number of objects after each step. 

\subsection{Matching with Milky Way Analogs}\label{sec:match}
The properties of the Milky Way itself vary in the literature \citep{Licquia2015,Bland-Hawthorn2016}, and the definitions of MW analogs vary between studies. In the SAGA survey \citep{SAGA-I,SAGA-II}, MW analogs are selected based on the absolute $K$-band magnitude $-23 > M_K > -24.6$~mag (corresponding to a stellar mass range of $10.2 < \log\, M_\star/M_\odot < 11.0$) and redshift $0.005 < z < 0.01$ (20--40 Mpc). SAGA generally requires that the MW analogs be in isolation (without nearby bright galaxies). In the ELVES survey \citep{ELVES-I,ELVES-II,CarlstenELVES2022}, the requirements for a MW-like host are loosened to be $M_K < -22.1$~mag ($M_\star > 10^{9.9}\ M_\odot$) because the probed volume by ELVES ($D<12$ Mpc) is smaller than that of SAGA. In this work, we choose the stellar mass range of our MW analog sample to be $10.2 < \log\, M_\star/M_\odot < 11.2$ based on NSA stellar mass, which is simply a 1 dex bin centered at the measured stellar mass of the Milky Way ($M_{\star, \mathrm{MW}}\approx 10^{10.7}\ M_\odot$, \citealt{Licquia2015}). MW analogs selected using this definition are very close to those in SAGA but are slightly less massive than the ELVES hosts as ELVES has several groups more massive than SAGA hosts. We do not discriminate between isolated or paired hosts. We acknowledge that defining MW analogs is tricky because of the steep mass function of galaxies and limited survey volumes. A slight change in the mass range could alter the number of MW analogs, and how the galaxies populate the brighter end of mass function also matters (see \citealt{CarlstenELVES2022}). For example, it is also possible to select MW analogs in linear stellar mass such as $1.5 \times 10^{10} < M_\star < 8.5\times 10^{10}\,M_\odot$. Compared to our logarithmic selection, such a linear selection excludes 10\% of high-mass MW analogs at $10.9 < \log M_\star/M_\odot < 11.2$, and will result in lower abundances of UDGs and UPGs. Better ways to define MW analogs are needed and will certainly benefit the community.

Mass--size outliers, by definition, are relatively scarce compared with ``normal'' satellites around MW analogs \citep{SAGA-II,CarlstenELVES2022}. It is therefore helpful to probe a larger volume to obtain better statistics. We choose our redshift range to be $0.01 < z < 0.04$, which ensures that we can detect a significant number of LSB satellite candidates around MW-like hosts. The depth of HSC limits our detection to below $z \approx 0.04$ because dwarf galaxies will be too small and too faint to be detected beyond that distance. Our search only includes objects larger than $r_e\approx 2\arcsec$, which corresponds to $M_\star \sim 10^{8.5}\ M_\odot$ at $z=0.04$ assuming the \textit{average} mass--size relation from \citet{ELVES-I}. We also exclude galaxies at $z<0.01$ because (1) the number of mass--size outliers will be very small due to the small volume; (2) their large angular size makes them more vulnerable to being shredded during detection, so including them will introduce a number of spurious LSB objects. 

After applying the stellar mass and redshift cuts to the NSA catalog (\S\ref{sec:data}), there are 23,218 MW-like galaxies. We then select those within the HSC PDR2 footprint and match them to the initial LSBG sample (described in \S \ref{sec:detection}, before the deblending step) as follows. 

For a given MW-like galaxy, we calculate its virial mass based on the stellar-to-halo mass relation (SHMR) in \citet{Behroozi2010} using \href{https://halotools.readthedocs.io/en/latest/index.html}{\code{halotools}} \citep{Hearin2017}. The viral mass is defined to be the enclosed mass with $\rho > \Delta_{\rm vir}(z) \rho_c(z)$, where the virial overdensity only depends on cosmological parameters and redshift \citep{Bryan1998}. Then we convert virial mass to virial radius $R_{\rm vir}$ using the virial overdensity. It turns out that 40\% of our hosts have virial radii larger than 300 kpc, which is commonly used for the virial radius of the Milky Way. We note that the SHMR from abundance matching might overestimate the halo mass for massive spirals \citep[e.g.,][]{Posti2019,Mancera2022b}, implying that our sample might include LSBGs beyond the virial radius. We also compare different SHMRs \citep[e.g.,][]{Moster2013} and do not find significant differences in the distribution of $R_{\rm vir}$.

Then we identify any LSBG that falls inside the projected virial radius of a host. About 40\% of LSBGs are matched with 2 hosts and about 15\% of them are matched with 3 or more hosts. If an LSBG is matched to multiple hosts, we assign it to the nearest host based on the separation normalized by the host virial radius. There are 922 MW-like hosts and 10,579 LSBG candidates associated with them. These MW analogs occupy a sky area of 89.19 deg$^{2}$ (out to $1\ R_{\rm vir}$).

As described in \S\ref{sec:deblending}, we perform a deblending step to effectively remove spurious objects. There are 2673 objects left after the deblending cuts. Next, we model these remaining objects using the Spergel profile as described in \S\ref{sec:modeling} and obtain their photometry and structural parameters from the best-fit models. We also remove duplicate objects. In the final catalog, we are left with 2510 LSBG candidates associated with 689 MW analogs (out of 922) as our final LSBG sample. The host number drops from 922 to 689 because many hosts in the prior sample have no true positives as their satellites. We note that these LSBGs are matched to MW hosts in projection but do not have direct distance measurements, and we have not applied any statistical background correction at this stage. 

We apply the measurement bias correction (\S \ref{sec:comp_meas}) and assign completeness after bias correction. Lastly, we correct for the effect of Galactic extinction on colors based on \citet{SFD1998} and \citet{Schlafly2011}. The stellar masses of cross-matched LSBG are derived from the Spergel model fitting and the relation between color and mass-to-light ratio $M_{\star}/L$ from \citet{Into2013}:

\begin{align*}
&\log \left(M_{\star} / L_{g}\right)=1.774\,(g-r)-0.783, \\
&\log \left(M_{\star} / L_{g}\right)=1.297\,(g-i)-0.855.
\end{align*}
\citet{Kado-Fong2022} show that the UDG population can be well described by the color--$M_\star/L$ relation from \citet{Into2013}, which is also used in ELVES \citep{CarlstenELVES2022}. We note that the stellar mass derived using the color--$M_\star/L$ relation from \citet{Herrmann2016} and \citet{Du2020} is $\sim 0.2$ dex lower than the one derived using the \citet{Into2013} relation. In our case, because we have both the $g-r$ and $g-i$ colors available from the model fitting, we use the average $M_{\star}/L$ derived from the two colors to calculate stellar mass. We assume the solar absolute magnitude in the $g$ band to be 5.03 \citep{Willmer2018}.

\subsection{Mass--Size Outlier Sample}\label{sec:sample}
The concept of UDG was proposed to describe galaxies with unusually large sizes ($r_e>1.5$ kpc) and low central surface brightnesses ($\sbcen > 24.0\ \sbunit$; \citealt{vanDokkum2015}). Although UDGs are large in size, the constant size cut does not account for the fact that the galaxy size is strongly correlated with its mass, known as the mass--size relation \citep[e.g.,][]{Graham2003,Trujillo2007,vanDokkum2013,Cappellari2013,Lange2015}. For example, a galaxy with $r_e = 1.5$ kpc is very extended if its stellar mass is $10^7\ M_\odot$, but is normal-sized if its stellar mass is $10^{8.5}\ M_\odot$.
Thus, a more physically useful definition of puffy dwarf galaxies (i.e., mass--size outliers) would be based on a mass--size relation and its scatter. Following this thread, we propose a new definition of mass--size outliers, dubbed \textit{ultra-puffy galaxies} (UPGs). 

It is worth noting that in several works \citep[e.g.,][]{Lim2020,Venhola2022}, UDGs in galaxy clusters are defined as outliers in the scaling relations between size, surface brightness, and total luminosity. A UDG definition based on the mass--size relation is also used in simulation studies where the absolute sizes were uniformly too large such that size outliers could only be selected relative to the simulated mass--size relation \citep[e.g.,][]{Benavides2021,Benavides2022}. In this work, we are proposing such a definition of mass--size outliers to be used universally, not limited to cluster environments or simulations.

In this section, we describe the definition of UPG and contrast it with UDG. Then we select UDGs and UPGs from our LSBGs matched with MW hosts. The catalogs and mosaic images of UDGs and UPGs are available online\footnote{\url{https://astrojacobli.github.io/research/BeyondUDG/}}.

\subsubsection{UPGs}
To select mass--size outliers, a mass--size relation and its scatter are needed. In general, the slope of the mass--size relation in the nearby Universe has been shown to be color- and morphology-dependent for galaxies above $M_\star > 10^{9}\ M_\odot$: blue star-forming galaxies have a shallower mass--size relation than do red quenched galaxies \citep[e.g.,][]{Lange2015}. In the dwarf galaxy regime ($10^{5.5} < M_\star/M_\odot < 10^{8.5}$), \citet{ELVES-I} derive the mass--size relation from the satellites of MW analogs in the Local Volume and show that the average mass--size relation is quite universal: the slope and intercept are not sensitive to the color or morphology of dwarf galaxies. Residuals from the mass--size relation follow a log-normal distribution reasonably well. This fact allows us to define a subset of dwarf galaxies that are outliers with respect to the average mass--size relation.


Taking the measured average mass--size relation from \citet{ELVES-I} $\log\, (r_e/\mathrm{pc}) = 0.25\, \log\, (M_\star/M_\odot) + 1.07$ and a scatter of $\sigma=0.18$ dex, we define a population of UPGs to be galaxies that are $>1.5\sigma$ above the average mass--size relation. We note that the mass--size relation in \citet{ELVES-I} is derived for a mass range of $10^{5.5}\ M_\odot < M_\star < 10^{8.5}\ M_\odot$, so the mass--size relation is extrapolated to $M_\star \sim 10^9\ M_\odot$ to define our UPG sample. We choose $1.5\sigma$ because it gives us a statistically significant sample. One can certainly select $>2\sigma$ or $>3\sigma$ UPGs if a larger LSBG sample is available. The UPGs are not absolutely large in size or absolutely faint in surface brightness, but they are outliers in size for their stellar masses. For example, a UPG with $M_\star = 10^7\ M_\odot$ has a size larger than 1.2 kpc and surface brightness fainter than $\sbeff > 25.5\ \sbunit$ for $g-i=0.4$ and $\sbeff > 26.8\ \sbunit$ for $g-i=0.8$. A UPG with $M_\star = 10^{8.5}\ M_\odot$ has a size larger than 2.7 kpc and surface brightness fainter than $\sbeff > 23.7\ \sbunit$ for $g-i=0.4$ and $\sbeff > 25.0\ \sbunit$ for $g-i=0.8$. The definition of UPG based on the observed mass--size relation is physically motivated, and better allows us to study size outliers at a given stellar mass. We further discuss the advantages and disadvantages of UDG and UPG in \S\ref{sec:mass-size} and \S\ref{sec:discussion}.

There are 362 LSBG candidates that fall $1.5\sigma$ above the ridge line of the mass--size relation. After removing spurious objects (e.g., shredded galaxy outskirts, cirrus, tidal feature, blends) by another round of visual inspection and removing objects with completeness less than 0.1, we have 337 galaxies in our UPG sample.\footnote{For context, there are 155 (39) galaxies being $2\sigma$ ($3\sigma$) above the average mass--size relation.} These UPGs are associated with 239 hosts. The total sky area occupied by UPG hosts (out to 1 $R_{\rm{vir}}$) is 32.37 deg$^{2}$. The catalogs are available online in machine-readable format\footnote{\url{https://astrojacobli.github.io/research/BeyondUDG/}}, and we demonstrate the catalog format in Table \ref{tab:catalog}. 

\subsubsection{UDGs}
We also select a UDG sample from our LSBGs in order to compare it with existing observations and simulations in different environments. The UDG sample is usually defined based on the surface brightness and physical size of the galaxy. However, there are many different criteria in the literature: \citet{vanDokkum2015} define UDGs to have effective radii $r_e$ larger than 1.5 kpc and central surface brightness $\mu_0(g)$ fainter than $24.0\ \sbunit$; other groups also use the surface brightness at $r_e$ \citep[e.g.,][]{DiCintio2017,Cardona-Barrero2020} or the average surface brightness within $r_e$ to define UDGs \citep[e.g.,][]{Koda2015,Yagi2016,vdBurg2016,Leisman2017,ManceraPina2018,ManceraPina2019a,Martin2019,Karunakaran2022b}. The size criterion also varies from $r_e > 1$ kpc to $r_e > 1.5$ kpc. \citet{vanNest2022} explore these definitions in simulations and find that different definitions of ``UDG'' can drastically change the selected subset of dwarfs, therefore affecting our understanding of the UDG population.

In practice, the measured central surface brightness might be biased by nuclear star clusters \citep{Neumayer2020,ELVES-II,Somalwar2020} or other contaminants. Therefore in this work, we use the average surface brightness within the effective radius $\sbeff$ to define UDGs. The difference between the average surface brightness and the central surface brightness can be analytically calculated for a \sersic{} profile: $\overline{\mu}_{\mathrm{eff}} - \mu_0 = 1.124$ for $n=1$ and 0.796 for $n=0.8$ \citep{Graham2005,Yagi2016}. As dwarf galaxies and UDGs typically have \sersic{} indices of $0.8 < n < 1.2$ \citep[e.g.,][]{vanDokkum2015,ELVES-I}, we take $\overline{\mu}_{\mathrm{eff}} - \mu_0 = 1.0$ as an average value to convert $\mu_0$ to $\mu_{\mathrm{eff}}$. In this work, we define UDGs to be galaxies with $r_e+\sigma(r_e) > 1.5$ kpc and $\sbeff + \sigma(\sbeff) > 25\ \sbunit$ to take the $1\sigma$ measurement errors into account. As a result, there are a few objects with nominally smaller sizes but large uncertainties that scatter into our UDG sample. This definition maximizes the consistency with the definition in \citet{vanDokkum2015} while not losing UDGs harboring nuclear star clusters.

Among the 2510 LSBG candidates around MW analogs, there are 432 objects satisfying the UDG definition. We did a final visual inspection for these objects and excluded 16 objects that are false positives including blends and Galactic cirrus. We also removed another 4 objects having completeness less than 0.1. In the end, we obtained our UDG sample with 412 objects (associated with 258 hosts) after searching around 689 MW analogs. The total sky area occupied by UDG hosts (out to 1 $R_{\rm{vir}}$) is 32.71 deg$^{2}$. We describe the properties of the mass--size outliers (including UPGs and UDGs) in \S\ref{sec:mass-size}.

\subsection{Background Contamination Fraction}\label{sec:bkg}
Although we have matched LSBGs to MW analogs by proximity in projection, the physical affiliation of LSBGs to the host is not guaranteed. As a result, a certain fraction of galaxies in our mass--size outlier samples are foreground or background galaxies that fall within the virial radius of the host by chance. Given the volume and the surface brightness range of our search, contamination is most likely to be dominated by background galaxies. In this section, we try to estimate the contamination fraction empirically by randomly matching LSBGs to our MW hosts and measuring the number density of apparent UPGs and UDGs.

We randomly selected a continuous patch of sky of 24 deg$^{2}$ in HSC PDR2 regardless of whether it contains MW analogs\footnote{$345\ \deg < \rm{R.A.} < 351\ \deg$, $-1.5\ \deg < \rm{Dec} < 2.5\ \deg$, which is not dominated by Galactic cirrus and stars. Such a patch of sky with 24 deg$^2$ is already enough for us to derive a background contamination fraction with small Poisson noise.}. Then we performed the same deblending and modeling steps for the 2707 LSBG candidates detected in this region. We also removed objects that are already in our mass--size outlier samples and objects with completeness less than 0.1. We also did the same visual inspection as for UDGs and UPGs to remove false positives. In the end, we obtained 480 LSBGs (excluding false positives) representing a population of possible contaminants for the UDG and UPG samples. Next, because both UDG and UPG are defined based on the physical size, we randomly associated the 480 LSBGs with the 922 MW analogs that are surveyed in this paper. Then we calculated a physical size for each LSBG and classify an LSBG as an ``artificial`` UDG or UPG if it satisfied the corresponding definition. We repeated such random matching 200 times. In the end, we obtained 7,625 artificial UDGs and 8267 artificial UPGs. These UDGs and UPGs are ``artificial'' only in the sense of being artificially associated with random MW hosts. 

In this way, the number density of artificial UDGs is estimated to be $S_{\rm UDG} = 1.60\pm0.25\ \mathrm{deg}^{-2}$, and the number density of artificial UPGs is $S_{\rm UPG} = 1.72\pm0.23\ \mathrm{deg}^{-2}$. The catalogs of artificial UDG and UPGs are also available online\footnote{\url{https://astrojacobli.github.io/research/BeyondUDG/}}. Using the number densities and the probed area of our survey (89.19 deg$^{2}$ occupied by 922 hosts), we calculate the contamination fraction for both the UDG and UPG samples. For the UDG sample, we find $f_{\rm contam}^{\rm UDG} \approx 35\%\pm5\%$; for the UPG sample, the contamination fraction is $f_{\rm contam}^{\rm UPG} \approx 45\%\pm6\%$. We take the contamination fraction into account when calculating the abundances in \S\ref{sec:results}.

\section{Results}\label{sec:results}
In this section, we compare the distributions of UDGs and UPGs on the size--surface brightness and mass--size planes to gain intuition on the differences between UDG and UPG definitions. Then we present the abundances of UDGs and UPGs and compare them with the literature. 

\subsection{Properties of Mass--Size Outliers}\label{sec:mass-size}

\begin{figure*}
	\vbox{ 
		\centering
		\includegraphics[width=1\linewidth]{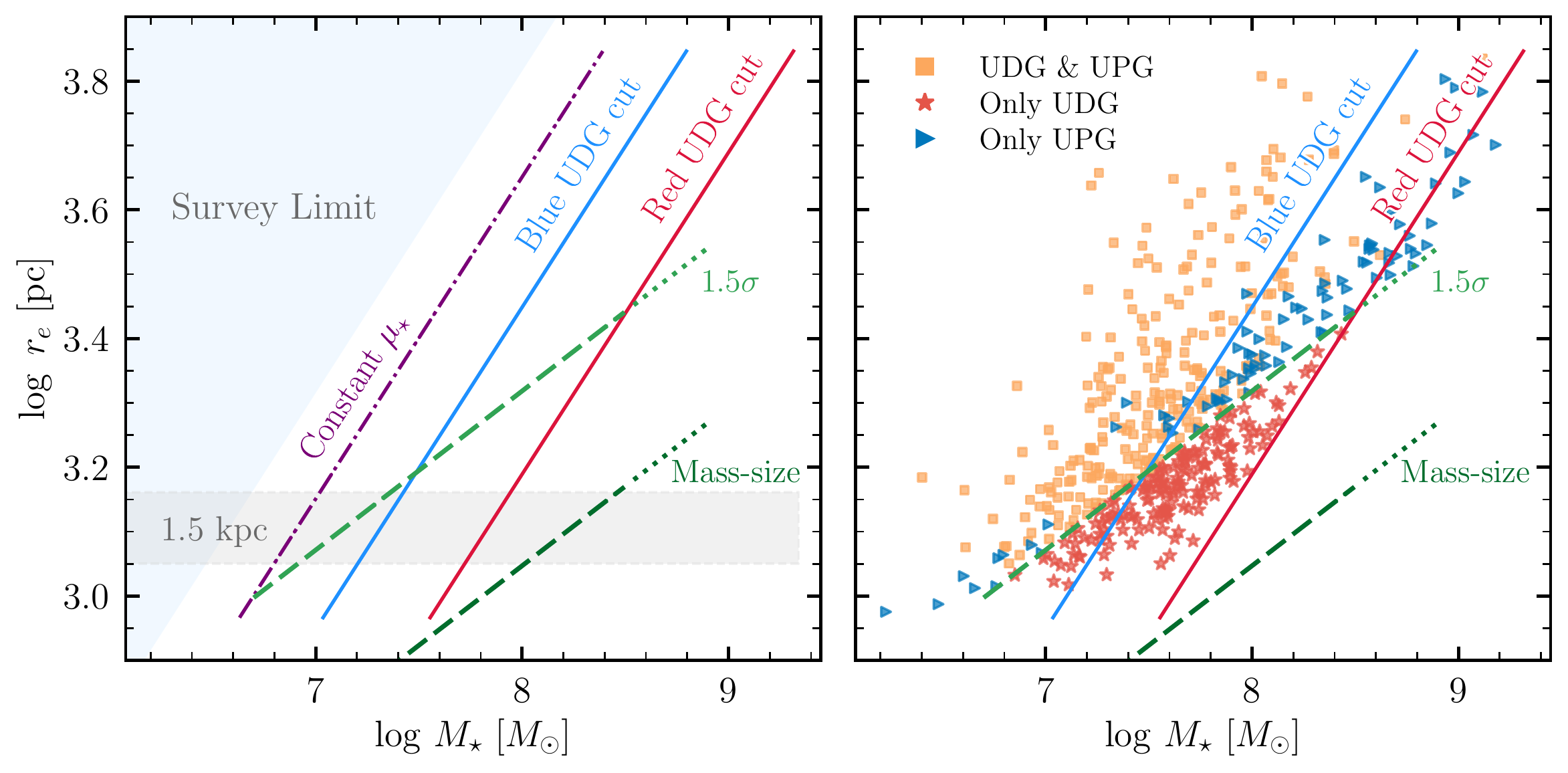}
	}
    \caption{\textit{Left panel}: schematic diagram showing the UDG size cut (gray band), survey limit (blue shade), average mass--size relation (dark green dashed line), and $1.5\sigma$ above it (light green line). The solid lines are constant surface brightness cuts at $\sbeff=25\ \sbunit$ for two different colors $g-i=0.4$ (blue) and $g-i=0.8$ (red). The constant surface stellar mass density line is in purple, which is parallel to the solid lines. The dotted part of the green lines indicates where we linearly extrapolate the mass--size relation to define UPGs. \textit{Right panel}: distributions of UDGs and UPGs on the mass--size plane. UDGs that are also classified as UPGs are shown as orange squares, while UDGs that do not satisfy the UPG definition are marked as red stars. Conversely, blue triangles are UPGs that are not classified as UDGs. The UDG sample includes a significant number of galaxies that are not mass--size outliers (falling below the $1.5\sigma$ line), which are red in color because the surface brightness cuts are different on the mass--size plane for blue and red galaxies.
    }
    \label{fig:mass_size}
\end{figure*}

\begin{figure*}
	\vbox{ 
		\centering
		\includegraphics[width=1\linewidth]{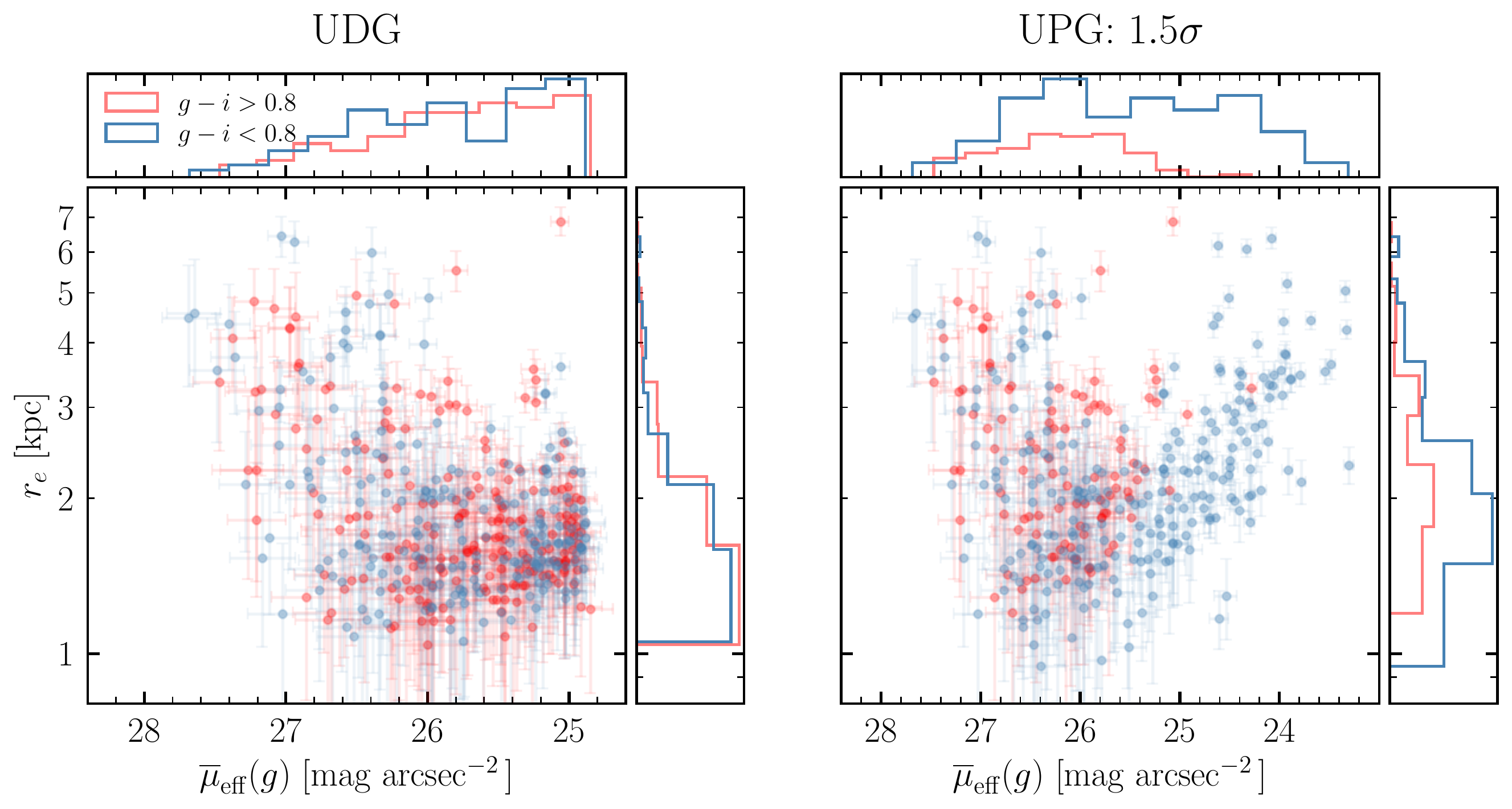}
	}
    \caption{Distribution of UDGs (\textit{left}) and UPGs (\textit{right}) on the size--surface brightness plane. The UPGs are defined to be galaxies that are $1.5\sigma$ above the average mass--size relation in \citet{ELVES-I}. The samples are split into two color bins and shown in red ($g-i>0.8$) and blue ($g-i<0.8$). The marginal histograms are not normalized to highlight the relative number of red and blue galaxies. Compared with the UDG sample, the UPG sample includes blue galaxies with surface brightness higher than the UDG cut ($\sbeff < 25\ \sbunit$) and excludes red galaxies at $25 < \sbeff < 26\ \sbunit$, due to the color--$M_\star/L$ dependence.
    }
    \label{fig:udg_upg_re_mu}
\end{figure*}

We show UDGs and UPGs on the mass--size plane in Figure \ref{fig:mass_size}. To orient the reader to the mass--size plane, we first show a schematic diagram in the left panel where regions relevant to the definitions of UDG and UPG are highlighted. The light blue shade shows the survey limit; we cannot effectively probe LSBGs fainter than $\sbeff \gtrsim 27.5\ \sbunit$ (see \S\ref{sec:comp_meas}). 
For defining UDGs, a physical size cut and a surface brightness cut are needed. We show the $r_e = 1.5$ kpc cut as a horizontal gray band (spanning from 1.1 to 1.5 kpc) to acknowledge the fact that smaller galaxies with large size uncertainties are included in the UDG sample. The slanted solid lines show constant surface brightness lines with $\sbeff = 25.0\ \sbunit$ for two different colors ($g-i=0.4$ and $g-i=0.8$). For our redshift range, the cosmological dimming effect is negligible and the surface brightness is constant with distance. The constant surface brightness lines depend on color because of the color--$M_\star/L$ relation. We also show the constant surface stellar mass density ($\mu_\star$) line, which is parallel to the constant surface brightness lines because the surface mass density is proportional to surface brightness multiplied by the mass-to-light ratio. Therefore, at a given size and surface brightness, blue UDGs are less massive than red UDGs; at a given stellar mass and surface brightness, blue UDGs are larger than red UDGs in size. Such a constant surface brightness cut leads to a UDG population that is inhomogeneous in surface mass density. 

In the left panel of Figure \ref{fig:mass_size}, the green dashed lines show the average mass--size relation and the $1.5\sigma$ line above the average relation. UPGs would lie above the $1.5\sigma$ line by construction, and there will be an overlap between UDGs and UPGs. The mass--size relation is shallower than the constant surface brightness and surface mass density lines. 

We highlight the similarities and differences between the UDGs and UPGs in the right panel. UDGs that are also classified as UPGs are shown as orange squares, whereas UDGs that do not satisfy the UPG definition are marked as red stars. Blue triangles correspond to UPGs that are not classified as UDGs. Limited by the detection limit and the fact that our hosts lie beyond the Local Volume, our mass limit is $M_\star \approx 10^{6.5}\ M_\odot$, which is $\sim 1$ dex higher than ELVES. 
Compared with the UPG sample, we find that the UDG sample (orange squares + red stars) comprises a number of galaxies below the $1.5\sigma$ line, but also loses several low-mass galaxies that are above the $1.5\sigma$ line but have sizes smaller than 1.5 kpc. For the UDG sample, the stellar mass range only reaches $M_\star\approx 10^{8.3}\ M_\odot$ because galaxies with higher stellar mass are too bright to satisfy the UDG surface brightness criterion. On the contrary, as the UPG definition does not have a hard surface brightness cut, it includes many galaxies that are more massive than UDGs provided they have an extraordinarily large size. We also note that the mass--size relation from \citet{ELVES-I} is extrapolated to define UPGs in the mass range $10^{8.5}\ M_\odot < M_\star < 10^9\ M_\odot$, where about 10\% of our UPGs are located (dotted green lines in Figure \ref{fig:mass_size}). However, the mass--size relation might show color and morphology dependence for $M_\odot > 10^{8.5}\ M_\odot$. As a consequence, disk galaxies might contribute to the UPG sample at the high-mass end. We visually checked the UPG sample and did find some blue UPGs resembling disk galaxies at the high-mass end. A spectroscopic follow-up would be required to confirm whether they are just blue background galaxies. We also note that 7\% of our UDGs and 3\% of our UPGs have \sersic{} indices $n>1.5$. We believe these galaxies with large \sersic{} indices are mostly interlopers given the fact that $\sim$40\% of UDGs and UPGs are interlopers (see \S\ref{sec:bkg}). We do not consider these galaxies to derive our conclusions given their small contribution to the whole sample.

We also show the UDG and UPG samples on the size--surface brightness plane in Figure \ref{fig:udg_upg_re_mu}. The galaxies are split into two color bins and are shown in blue ($g-i < 0.8$) and red ($g-i > 0.8$), respectively. We do not apply any background contamination correction in Figure \ref{fig:udg_upg_re_mu} because such corrections can only be done in a statistical sense. The error bars correspond to $1\sigma$ measurement uncertainties (\S\ref{sec:comp_meas}). The two marginal plots show the unnormalized histograms of galaxies in the two color bins. The numbers of red and blue galaxies are similar in the UDG sample, but there are more blue galaxies than red ones in the UPG sample. As there is no hard surface brightness cut for UPGs, the UPG sample includes many blue galaxies with brighter surface brightness ($23 \lesssim \sbeff < 25\ \sbunit$). For blue UPGs (shown as blue dots), the apparent correlation between $\sbeff$ and $r_e$ is merely due to the fact that blue UPGs pile up around the $1.5\sigma$ line above the mass--size relation. Due to the surface brightness cut ($\sbeff > 23.0\ \sbunit$) in the deblending step, there is little chance for an LSB disk galaxy with a bright bulge to scatter into the UPG sample. The UPG sample also loses a significant fraction of red galaxies at $25 < \sbeff < 26\ \sbunit$ because their stellar masses are high but their sizes are too small to be mass--size outliers (i.e., red stars in Figure \ref{fig:mass_size}).

\subsection{Abundance of Mass--Size Outliers}\label{sec:n_udg}

\begin{figure*}
	\vbox{ 
		\centering
		\includegraphics[width=1\linewidth]{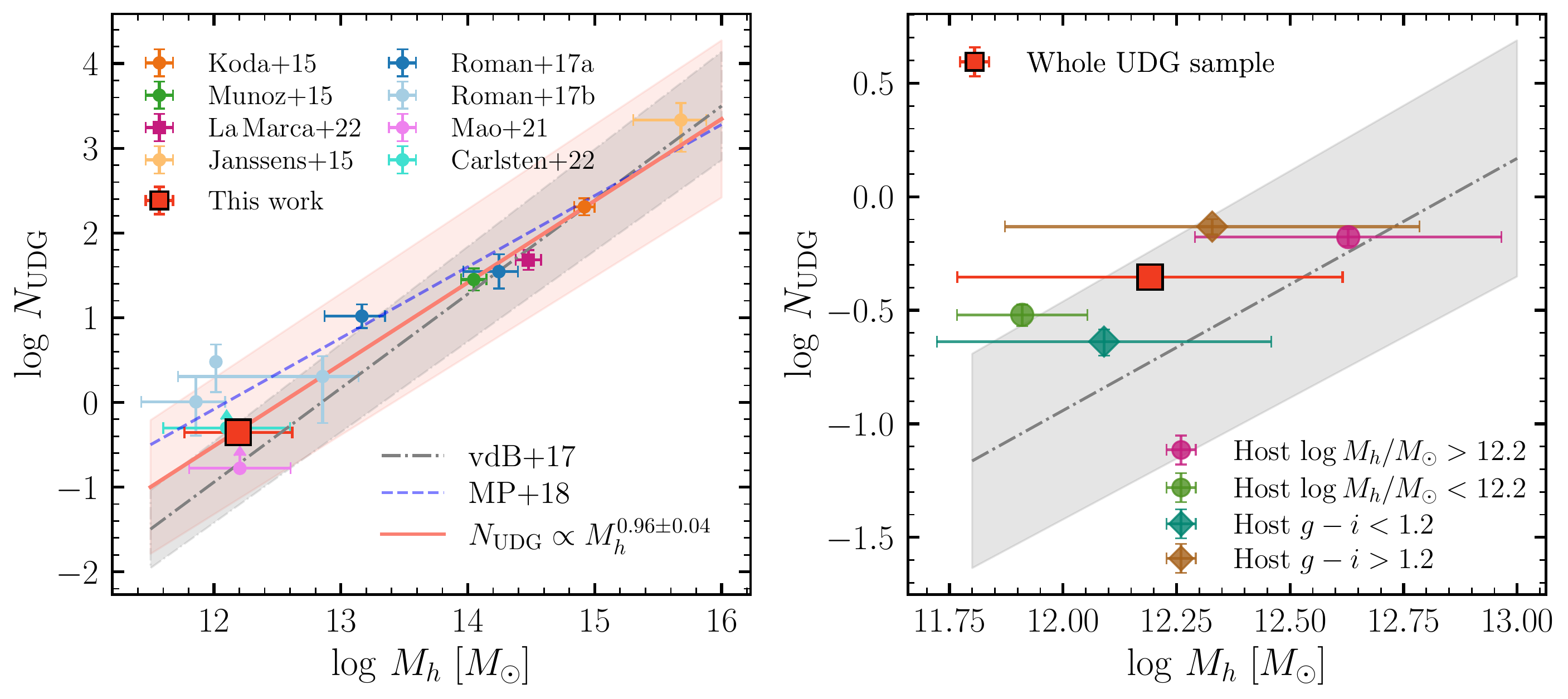}
	}
    \caption{Abundance of UDGs as a function of the host halo mass. \textit{Left}: We compile the UDG abundance measurements from the literature covering a wide range of host mass. The UDG abundance from our work $N_{\rm UDG} = 0.44\pm 0.05$ is shown as a red square, and the power-law relations from \citet{vdBurg2017} and \citet{ManceraPina2018} are shown in gray and blue respectively. Our fitting result is shown in pink. Our UDG abundance is consistent with ELVES but significantly higher than that in SAGA. After including more measurements at the lower halo-mass end, the power law is shallower than the relation reported in \citet{vdBurg2017}. \textit{Right}: We split the UDG sample into bins based on the host halo mass (circles) and $g-i$ color (diamonds). The UDG abundance is higher for more massive hosts and redder hosts. }
    \label{fig:n_udg}
\end{figure*}

As demonstrated by many previous studies \citep[e.g.,][]{vdBurg2016,vdBurg2017,Roman2017a,Karunakaran2022b}, the average number of UDGs per host scales with host halo mass. While much literature focuses on finding UDGs in clusters and large groups, there are fewer constraints on the UDG abundance at the lower halo-mass end. In this section, we calculate the UDG and UPG abundances of MW analogs and compare them with other surveys.

We define the UDG (UPG) abundance as the average number of UDGs (UPGs) per host galaxy. We searched 922 MW analogs. In our UDG sample, there are 412 UDGs associated with 258 hosts. After correcting for background contamination (see \S \ref{sec:bkg}) and completeness (see \S \ref{sec:comp_meas}), the UDG abundance of MW analogs is $N_{\rm UDG} = 0.44\pm 0.05$ per host. The UPG abundance in our sample is $N_{\rm UPG} = 0.31\pm 0.04$ per host. Here we neglect the fact that the number of satellites contained within the virial sphere is different from the number of satellites within a projected cylinder with virial radius (the so-called deprojection factor; \citealt{vdBurg2017}).

We compare our UDG abundance with other surveys focused on MW analogs. In the SAGA survey \citep{SAGA-II}, 6 satellite galaxies (out of 127 spectroscopy-confirmed satellites around 36 MW analogs) satisfy the definition of UDG. Therefore the UDG abundance in \citet{SAGA-II} is about $N_{\rm UDG}^{\rm SAGA}\approx 0.17\pm0.07$. In the ELVES survey \citep{CarlstenELVES2022}, there are 13 UDGs out of 351 satellites with secure distances ($P_{\rm sat} > 0.8$) in 30 MW analogs, leading to a UDG abundance of $N_{\rm UDG}^{\rm ELVES} \approx 0.43\pm0.12$. If we only take UDGs more massive than our detection limit $M_\star > 10^{6.5}\ M_\odot$, we obtain slightly lower UDG abundances of $N_{\rm UDG, 6.5}^{\rm ELVES} \approx 0.36\pm0.11$. Because both SAGA and ELVES select satellites based on distance measurements and do not correct for completeness, we interpret their UDG abundances as lower limits. \citet{Roman2017b} identified 11 UDGs around three galaxy groups in the IAC Stripe 82 Legacy Survey \citep{Fliri2016}, but it is hard to calculate a UDG abundance due to small-number statistics. Our UDG abundance is consistent with the value in the ELVES survey, but higher than the value from SAGA. This might be explained by the different photometric depths of data used. As pointed out by \citet{CarlstenELVES2022} and \citet{Font2022}, the SAGA survey reaches a surface brightness depth of $\sbeffr\approx 25\ \sbunit$, which is $\sim 2.5\ \sbunit$ shallower than the ELVES survey and this work. It is probable that SAGA missed some fraction of red and lower surface brightness galaxies. On the other hand, ELVES includes several more massive hosts than our search and SAGA do, and consequently bias the UDG abundance.

We plot the UDG abundances from these surveys in the left panel of Figure \ref{fig:n_udg}, together with the results for larger galaxy groups and clusters \citep{Koda2015,Munoz2015,Roman2017a,Roman2017b,Janssens2017,vdBurg2017,LaMarca2022}. 
The power-law regression result from \citet{vdBurg2017} $N_{\rm UDG} \propto M_h^{1.11\pm 0.07}$ is shown in gray. We also show the result from \citet{ManceraPina2018} $N_{\rm UDG} \propto M_h^{0.84\pm 0.07}$ as a blue-dashed line where they homogeneously combined the samples in \citet{vdBurg2016} and \citet{Roman2017b}. The UDG abundance of this work is highlighted as the red square. As shown in Figure \ref{fig:n_udg}, the scatter in the $N_{\rm UDG}-M_h$ relation gets larger at the lower halo-mass end, but this might be due to the small statistics in prior studies and the differences in completeness. Our UDG abundance is marginally higher than the prediction from \citet{vdBurg2017} but is still consistent considering the large scatter of the power law. 

Taking all data points from the literature (as shown in Figure \ref{fig:n_udg}), we use the orthogonal distance regression (ODR) to fit a power-law between $N_{\rm UDG}$ and host halo mass $M_h$ and find $N_{\rm UDG} \propto M_h^{0.96\pm 0.04}$, shown as the red solid line in Figure \ref{fig:n_udg}. The slope of the power law is shallower than \citet{vdBurg2017} ($\beta=1.11\pm0.07$) but steeper than \citet{Roman2017b} ($\beta=0.85\pm0.05$). \citet{ManceraPina2018} surveyed eight clusters and find a sublinear power law. Recently \citet{Karunakaran2022b} performed a similar analysis including SAGA, ELVES, and \citet{Nashimoto2022} and found a slightly shallower power law of $\beta=0.87\pm0.07$. We note that their surface brightness cut for UDGs is brighter than ours and thus they include more ``UDGs'' for MW-like hosts and have a shallower power law. One should be aware of the fact that literature results are of different depths, often not completeness corrected and background subtracted, and UDG definitions also vary across different studies. These effects could significantly bias the results when combining them. Therefore we caution that the UDG abundance and the power-law index here should not be over-interpreted.

In the right panel of Figure \ref{fig:n_udg}, we split the UDG sample into different bins based on their host halo mass (shown as circles) and host color (shown as diamonds). We find that the UDG abundance is slightly higher for hosts with higher halo mass and redder color, but the trend is stronger with host color. 

Because there is little literature on UPGs, we only compare our UPG abundance with ELVES. In ELVES, we take the 351 satellites with secure distances in 30 MW analogs and identify 36 UPGs associated with 20 hosts. Thus the UPG abundance fraction in ELVES is $N_{\rm UPG}^{\rm ELVES} = 1.20 \pm 0.20$. When only selecting UPGs with $M_\star > 10^{6.5}\ M_\odot$, we obtain 23 UPGs in 30 MW analogs, thus $N_{\rm UPG, 6.5}^{\rm ELVES} = 0.77 \pm 0.16$, which is slightly higher than our UPG abundance.

\vspace{1em}

Another interesting quantity is the fraction of satellites that are mass--size outliers in a group or cluster. This fraction represents how efficient an environment is at producing puffy satellites. We denote this quantity as the UDG (UPG) fraction hereafter, which characterizes the tail of the size distribution of satellites. In order to calculate this fraction, we assume that MW analogs have $\sim 6 \pm 1$ satellites with $M_\star > 10^{6.5}\ M_\odot$ \citep{CarlstenELVES2022}. This lower limit in stellar mass is roughly our detection limit (see Fig. \ref{fig:mass_size}). Then for our samples, the UDG (UPG) fraction is $f_{\rm UDG} \approx 0.07 \pm 0.02$ and $f_{\rm UPG} \approx 0.05 \pm 0.02$. In ELVES, among 351 satellites with secure distances in 30 MW analogs, we identify 13 UDGs associated with 10 hosts and 36 UPGs associated with 20 hosts. Thus the UDG fraction in ELVES is $f_{\rm UDG}^{\rm ELVES} \approx 0.04 \pm 0.01$, the UPG fraction in ELVES is $f_{\rm UPG}^{\rm ELVES} \approx 0.10 \pm 0.02$. We further only select ELVES galaxies with $M_\star > 10^{6.5}\ M_\odot$ and find the same UDG and UPG fractions. We also compare our results with results in the Virgo cluster. Taking the data from the Next Generation Virgo Cluster Survey (NGVS; \citealt{Ferrarese2020}) and only considering satellites with $M_\star > 10^{6.5}\ M_\odot$, there are 14 UDGs and 17 UPGs out of 218 identified satellites (with no completeness correction). Thus their UDG (UPG) fraction is $f_{\rm UDG}^{\rm Virgo} \approx 0.06\pm 0.02$ and $f_{\rm UPG}^{\rm Virgo} \approx 0.08\pm 0.02$. However, we note that \citet{Ferrarese2020} only survey the central 4 deg$^2$ of the Virgo cluster, which probably biases the UDG (UPG) fraction.
We interpret our UDG and UPG fractions to be consistent with ELVES and NGVS. However, we emphasize that it is challenging to compare across various searches due to the differences in the host sample selection, depth, completeness, and contamination estimation. In the future, a more systematic and careful comparison of UDG (UPG) fractions in different environments is needed.

\section{Robust Identification of Mass--Size Outliers}\label{sec:discussion}


In Figure \ref{fig:mass_size}, we find that combining a hard physical size cut with a surface brightness cut carves out an interesting region of the mass--size plane that is complicated in two ways. On the one hand, there are many galaxies classified as UDGs that are not size outliers at $M_\star \sim 10^{7.5}\ M_\odot$ (highlighted as red stars in Fig. \ref{fig:mass_size}). On the other hand, the hard surface brightness cut corresponds to very different surface mass densities between red and blue UDGs, making it hard to directly compare them, as they occupy different regions on the mass--size plane. For a given stellar mass and size (e.g., at $M_\star \sim 10^{7.5}\ M_\odot$ and $r_e = 1.5$ kpc), the hard surface brightness cut in the UDG definition preferentially removes blue galaxies because their surface brightnesses are higher and their $M_\star/L$ ratios are lower. Thus, red UDGs are over-represented. If one calculates the quenched fraction of UDGs as a function of the stellar mass, it will be biased high because of this constant surface brightness cut in the UDG definition (see \citealt{Li2023}). These limitations of the original UDG definition are also discussed in \citet{Trujillo2017} and \citet{ManceraPina2018}.

It is worth emphasizing that the concept of UDG was originally proposed in a cluster context \citep[e.g.,][]{vanDokkum2015} where UDGs are far less diverse in stellar populations compared with field UDGs, hence the sample selection suffers less from the effect of color--$M_\star/L$ relation. However, UDGs are not all red and quiescent. Blue UDGs are found in less dense environments and fields. Therefore, the concept of UDG might not be optimal for systematic studies outside of a cluster environment; more specifically, to study diffuse dwarf galaxies with a range of colors in different environments, a more stellar-population agnostic criterion must be adopted to select samples.
The UPG is defined based on the average mass--size relation and thus alleviates some of the issues in the UDG definition. UPGs are defined to lie $1.5\sigma$ above the average mass--size relation of the satellites in MW analogs in the Local Volume. The specific value above the average mass--size relation can be varied to probe different parts of the size distribution. We advocate studying the UPG population in both observations and simulations to explore samples with different ``puffiness'' (e.g., comparing UPG at $1.5\sigma$ and $2\sigma$ from the mass--size relation).

Because the UPG is a physically motivated selection, to use the UPG criteria, one needs to know the distance, color, a color--$M_\star/L$ relation to convert observed magnitudes to stellar mass, and a mass--size relation and the scatter therein. None of these are easy to obtain or free from systematic errors. Both size and stellar mass could have large uncertainties in the LSB regime. One also needs to apply a surface brightness cut (such as $\sbeff > 23.0\ \sbunit$ in this work) to make sure that the UPG sample is not dominated by disk galaxies with high surface brightness due to the bulges. In \S\ref{sec:bkg}, we find that the UPG sample has a higher contamination fraction than the UDG sample. As the UPG definition does not preferentially exclude blue LSBGs, many background disk galaxies can be misidentified as UPGs. Based on our tests, a UPG sample with a more aggressive cut (e.g., $>2\sigma$ above the average mass--size relation) would have a higher purity, but building a significant sample of these is only possible when one has a much larger and deeper LSBG sample. 

Our definition of UPG greatly benefits from the well-measured mass--size relation in \citet{ELVES-I} and the fact that this mass--size relation is independent of galaxy color and morphology. However, such mass--size relations are not always available for all galaxy mass ranges and environments (in the field or in groups and clusters). \citet{ELVES-I} derive the mass--size relation of satellite galaxies of MW analogs below $M_\star \sim 10^{8.5}\ M_\odot$, and find it similar to the mass--size relation of field late-type dwarfs in \citet{Karachentsev2013}. \citet{Lange2015} have relatively good constraints on the mass--size relation above $M_\star \sim 10^{9}\ M_\odot$. However, the mass--size relation at $10^{8.5}\ M_\odot < M_\star < 10^{9.5}\ M_\odot$ has been shown to be very shallow for red galaxies \citep{SmithCastelli2008,Misgeld2008,Misgeld2011,Eigenthaler2018}, and its dependence on color, morphology, or environment is still unclear. In this work, we simply extrapolate the \citet{ELVES-I} mass--size relation to $10^9\ M_\odot$. However, it is not clear whether the mass--size relation at $10^{8.5}\ M_\odot < M_\star < 10^{9}\ M_\odot$ is sensitive to color or morphology. Future work should revisit the mass--size relation in this regime. 

At the same time, whether the size distribution for a given mass can be well-described by a Gaussian distribution is still a question. Fig. 9 in \citet{ELVES-I} shows the residual in $\log(r_e)$ after fitting the mass--size relation, where the distribution of residuals is roughly Gaussian but skewed toward the large-size end. With enough statistics in the future, one might be able to characterize the shape of the large-size tail and define UPGs based on the percentiles of the size distribution \citep{Greene2022}. However, there are also other size definitions that may help reduce the scatter in the mass--size relation \citep[e.g.,][]{Miller2019,Mowla2019,Trujillo2020,Chamba2022} and can further refine the UPG definition proposed in this work.

In the right panel of Figure \ref{fig:mass_size}, we also plot the lines of constant average surface mass density, which are steeper than the mass--size relation but have the same slope as constant surface brightness lines. Hence, a possible improvement to the selection of diffuse galaxies could be to select galaxies with low-surface mass density. Similar to the method proposed in this work, this method also alleviates the artifacts introduced by the color--$M_\star/L$ relation on the mass--size plane. We defer the exploration of this definition to future work.

\section{Summary}\label{sec:summary}
In this work, we perform a search for low surface brightness galaxies in HSC PDR2 data and construct samples for mass--size outliers around MW analogs at $0.01 < z < 0.04$. Besides the UDGs, we define ``ultra-puffy galaxies'' to be $1.5\sigma$ above the average mass--size relation. We calculate the abundances of mass--size outliers and compare them with literature. We also argue that UPGs better represent the large-size tail of the dwarf galaxy population. Our main findings and prospects are summarized below. 

\begin{enumerate}
    \item Using the HSC PDR2 data ($\sim 300\ \mathrm{deg}^{2}$), we conduct a systematic search for LSBGs using the method in \citet{Greco2018} complemented by our new deblending and modeling methods. Utilizing the nonparametric models generated by \code{scarlet}, we design a metric based on morphological parameters to effectively remove false positives in the initial LSBG sample. We measure the structural properties of LSBGs using parametric modeling and carefully characterize the measurement biases and uncertainties. The completeness of the search is also derived by injecting mock galaxies. Our LSBG search achieves high completeness compared with other searches and demonstrates the power of HSC (and future LSST) data for LSB science. 
    
    \item By matching LSBGs with MW analogs in the NSA catalog, we construct samples of mass--size outliers including UDGs and UPGs. We display their distributions on the mass--size plane (Figure \ref{fig:mass_size}) and the size--surface brightness plane (Figure \ref{fig:udg_upg_re_mu}). The UPG sample contains more blue and higher surface brightness galaxies than the UDG sample. In contrast, the UDG sample includes many red galaxies below $1.5\sigma$ from the average mass--size relation.
    
    \item After correcting for background contamination and completeness, the UDG abundance in MW analogs is $N_{\rm UDG} = 0.44\pm 0.05$ per host and the UPG abundance is $N_{\rm UPG} = 0.31 \pm 0.04$ per host. Our UDG abundance agrees with ELVES quite well but is much higher than that in SAGA. We obtain a UPG abundance lower than that in ELVES.
    
    \item Combining data from studies in denser environments, we find that the UDG abundance follows a sublinear power law with the host halo mass. The power-law slope agrees with other studies considering the errors. We caution that each study adopts different UDG definitions and may or may not correct for completeness, which could bias the results.
    
    \item For an MW analog, on average, about $7\%\pm2\%$ of its satellites are UDGs and about $5\%\pm2\%$ of its satellites are UPGs. Our UDG and UPG fractions are consistent with ELVES and NGVS.
    
    \item We advocate the concept of UPG, which is physically motivated, does not introduce artifacts in the mass--size distribution, and better represents the large-size tail of dwarf galaxy population. 
\end{enumerate}

This paper presents the UDG and UPG sample and explores their abundances. In our Paper II \citep{Li2023}, we study their size and spatial distributions and their star formation status. This study is focused on a small subset of LSBG candidates matched with MW analogs. In future work, we would like to exploit the full sample to study interesting topics on LSBGs including their redshift distributions, nucleation fractions, and intrinsic shapes. We will also explore machine-learning techniques to help us classify LSBGs and estimate structural parameters. With the upcoming LSST \citep{lsst2009,LSST2019}, we will be able to study the LSBG population with much greater statistical power and gain new insights into the formation and evolution of mass--size outliers. 

\section*{Acknowledgment}
We thank the anonymous reviewer for useful comments, which make the manuscript clearer.
J.L. is grateful for the discussions with Meng Gu, ChangHoon Hahn, and Sihao Cheng. The authors thank Yao-Yuan Mao for his \href{https://github.com/yymao/decals-image-list-tool}{tool} for visualizing the image cutouts. J.E.G. gratefully acknowledges support from NSF grant AST-1007052. J.H.K. acknowledges the support from the National Research Foundation of Korea (NRF) grant, No. 2021M3F7A1084525, funded by the Korean government (MSIT). S.D. is supported by NASA through Hubble Fellowship grant HST-HF2-51454.001-A awarded by the Space Telescope Science Institute, which is operated by the Association of Universities for Research in Astronomy, Incorporated, under NASA contract NAS5-26555.

The Hyper Suprime-Cam (HSC) collaboration includes the astronomical communities of Japan and Taiwan and Princeton University. The HSC instrumentation and software were developed by the National Astronomical Observatory of Japan (NAOJ), Kavli Institute for the Physics and Mathematics of the Universe (Kavli IPMU), University of Tokyo, High Energy Accelerator Research Organization (KEK), Academia Sinica Institute for Astronomy and Astrophysics in Taiwan (ASIAA), and Princeton University.  
Funding was contributed by the FIRST program from the Japanese Cabinet Office, Ministry of Education, Culture, Sports, Science and Technology (MEXT), Japan Society for the Promotion of Science (JSPS), Japan Science and Technology Agency (JST), Toray Science Foundation, NAOJ, Kavli IPMU, KEK, ASIAA, and Princeton University. The authors are pleased to acknowledge that the work reported on in this paper was substantially performed using the Princeton Research Computing resources at Princeton University, which is a consortium of groups led by the Princeton Institute for Computational Science and Engineering (PICSciE) and the Office of Information Technology's Research Computing.

\vspace{1em}
\software{\href{http://www.numpy.org}{\code{NumPy}} \citep{Numpy},
          \href{https://www.astropy.org/}{\code{Astropy}} \citep{astropy}, \href{https://www.scipy.org}{\code{SciPy}} \citep{scipy}, \href{https://matplotlib.org}{\code{Matplotlib}} \citep{matplotlib},
          \href{https://statmorph.readthedocs.io/en/latest/}{\code{statmorph}} \citep{statmorph},
          \href{https://bdiemer.bitbucket.io/colossus/index.html}{\code{Colossus}} \citep{Colossus},
          \href{https://pmelchior.github.io/scarlet/}{\code{scarlet}} \citep{Melchior2018}, \href{https://github.com/dr-guangtou/unagi}{\code{unagi}}.
          }

\bibliography{citation}{}

\begin{thebibliography}{}
\expandafter\ifx\csname natexlab\endcsname\relax\def\natexlab#1{#1}\fi
\providecommand{\url}[1]{\href{#1}{#1}}
\providecommand{\dodoi}[1]{doi:~\href{http://doi.org/#1}{\nolinkurl{#1}}}
\providecommand{\doeprint}[1]{\href{http://ascl.net/#1}{\nolinkurl{http://ascl.net/#1}}}
\providecommand{\doarXiv}[1]{\href{https://arxiv.org/abs/#1}{\nolinkurl{https://arxiv.org/abs/#1}}}

\bibitem[{{Abbott} {et~al.}(2018){Abbott}, {Abdalla}, {Allam}, {Amara},
  {Annis}, {Asorey}, {Avila}, {Ballester}, {Banerji}, {Barkhouse}, {Baruah},
  {Baumer}, {Bechtol}, {Becker}, {Benoit-L{\'e}vy}, {Bernstein}, {Bertin},
  {Blazek}, {Bocquet}, {Brooks}, {Brout}, {Buckley-Geer}, {Burke}, {Busti},
  {Campisano}, {Cardiel-Sas}, {Carnero Rosell}, {Carrasco Kind}, {Carretero},
  {Castander}, {Cawthon}, {Chang}, {Chen}, {Conselice}, {Costa}, {Crocce},
  {Cunha}, {D'Andrea}, {da Costa}, {Das}, {Daues}, {Davis}, {Davis}, {De
  Vicente}, {DePoy}, {DeRose}, {Desai}, {Diehl}, {Dietrich}, {Dodelson},
  {Doel}, {Drlica-Wagner}, {Eifler}, {Elliott}, {Evrard}, {Farahi}, {Fausti
  Neto}, {Fernandez}, {Finley}, {Flaugher}, {Foley}, {Fosalba}, {Friedel},
  {Frieman}, {Garc{\'\i}a-Bellido}, {Gaztanaga}, {Gerdes}, {Giannantonio},
  {Gill}, {Glazebrook}, {Goldstein}, {Gower}, {Gruen}, {Gruendl}, {Gschwend},
  {Gupta}, {Gutierrez}, {Hamilton}, {Hartley}, {Hinton}, {Hislop}, {Hollowood},
  {Honscheid}, {Hoyle}, {Huterer}, {Jain}, {James}, {Jeltema}, {Johnson},
  {Johnson}, {Kacprzak}, {Kent}, {Khullar}, {Klein}, {Kovacs}, {Koziol},
  {Krause}, {Kremin}, {Kron}, {Kuehn}, {Kuhlmann}, {Kuropatkin}, {Lahav},
  {Lasker}, {Li}, {Li}, {Liddle}, {Lima}, {Lin}, {L{\'o}pez-Reyes}, {MacCrann},
  {Maia}, {Maloney}, {Manera}, {March}, {Marriner}, {Marshall}, {Martini},
  {McClintock}, {McKay}, {McMahon}, {Melchior}, {Menanteau}, {Miller},
  {Miquel}, {Mohr}, {Morganson}, {Mould}, {Neilsen}, {Nichol}, {Nogueira},
  {Nord}, {Nugent}, {Nunes}, {Ogando}, {Old}, {Pace}, {Palmese},
  {Paz-Chinch{\'o}n}, {Peiris}, {Percival}, {Petravick}, {Plazas}, {Poh},
  {Pond}, {Porredon}, {Pujol}, {Refregier}, {Reil}, {Ricker}, {Rollins},
  {Romer}, {Roodman}, {Rooney}, {Ross}, {Rykoff}, {Sako}, {Sanchez}, {Sanchez},
  {Santiago}, {Saro}, {Scarpine}, {Scolnic}, {Serrano}, {Sevilla-Noarbe},
  {Sheldon}, {Shipp}, {Silveira}, {Smith}, {Smith}, {Smith}, {Soares-Santos},
  {Sobreira}, {Song}, {Stebbins}, {Suchyta}, {Sullivan}, {Swanson}, {Tarle},
  {Thaler}, {Thomas}, {Thomas}, {Troxel}, {Tucker}, {Vikram}, {Vivas},
  {Walker}, {Wechsler}, {Weller}, {Wester}, {Wolf}, {Wu}, {Yanny}, {Zenteno},
  {Zhang}, {Zuntz}, {DES Collaboration}, {Juneau}, {Fitzpatrick}, {Nikutta},
  {Nidever}, {Olsen}, {Scott}, \& {NOAO Data Lab}}]{Abbot2018}
{Abbott}, T.~M.~C., {Abdalla}, F.~B., {Allam}, S., {et~al.} 2018, \apjs, 239,
  18, \dodoi{10.3847/1538-4365/aae9f0}

\bibitem[{{Abraham} {et~al.}(2003){Abraham}, {van den Bergh}, \&
  {Nair}}]{Abraham2003}
{Abraham}, R.~G., {van den Bergh}, S., \& {Nair}, P. 2003, \apj, 588, 218,
  \dodoi{10.1086/373919}

\bibitem[{{Aihara} {et~al.}(2018){Aihara}, {Arimoto}, {Armstrong}, {Arnouts},
  {Bahcall}, {Bickerton}, {Bosch}, {Bundy}, {Capak}, {Chan}, {Chiba}, {Coupon},
  {Egami}, {Enoki}, {Finet}, {Fujimori}, {Fujimoto}, {Furusawa}, {Furusawa},
  {Goto}, {Goulding}, {Greco}, {Greene}, {Gunn}, {Hamana}, {Harikane},
  {Hashimoto}, {Hattori}, {Hayashi}, {Hayashi}, {He{\l}miniak}, {Higuchi},
  {Hikage}, {Ho}, {Hsieh}, {Huang}, {Huang}, {Ikeda}, {Imanishi}, {Inoue},
  {Iwasawa}, {Iwata}, {Jaelani}, {Jian}, {Kamata}, {Karoji}, {Kashikawa},
  {Katayama}, {Kawanomoto}, {Kayo}, {Koda}, {Koike}, {Kojima}, {Komiyama},
  {Konno}, {Koshida}, {Koyama}, {Kusakabe}, {Leauthaud}, {Lee}, {Lin}, {Lin},
  {Lupton}, {Mandelbaum}, {Matsuoka}, {Medezinski}, {Mineo}, {Miyama},
  {Miyatake}, {Miyazaki}, {Momose}, {More}, {More}, {Moritani}, {Moriya},
  {Morokuma}, {Mukae}, {Murata}, {Murayama}, {Nagao}, {Nakata}, {Niida},
  {Niikura}, {Nishizawa}, {Obuchi}, {Oguri}, {Oishi}, {Okabe}, {Okamoto},
  {Okura}, {Ono}, {Onodera}, {Onoue}, {Osato}, {Ouchi}, {Price}, {Pyo}, {Sako},
  {Sawicki}, {Shibuya}, {Shimasaku}, {Shimono}, {Shirasaki}, {Silverman},
  {Simet}, {Speagle}, {Spergel}, {Strauss}, {Sugahara}, {Sugiyama}, {Suto},
  {Suyu}, {Suzuki}, {Tait}, {Takada}, {Takata}, {Tamura}, {Tanaka}, {Tanaka},
  {Tanaka}, {Tanaka}, {Terai}, {Terashima}, {Toba}, {Tominaga}, {Toshikawa},
  {Turner}, {Uchida}, {Uchiyama}, {Umetsu}, {Uraguchi}, {Urata}, {Usuda},
  {Utsumi}, {Wang}, {Wang}, {Wong}, {Yabe}, {Yamada}, {Yamanoi}, {Yasuda},
  {Yeh}, {Yonehara}, \& {Yuma}}]{Aihara2018}
{Aihara}, H., {Arimoto}, N., {Armstrong}, R., {et~al.} 2018, Publications of
  the Astronomical Society of Japan, 70, S4, \dodoi{10.1093/pasj/psx066}

\bibitem[{{Akhlaghi} \& {Ichikawa}(2015)}]{Akhlaghi2015}
{Akhlaghi}, M., \& {Ichikawa}, T. 2015, \apjs, 220, 1,
  \dodoi{10.1088/0067-0049/220/1/1}

\bibitem[{{Amorisco} \& {Loeb}(2016)}]{Amorisco2016}
{Amorisco}, N.~C., \& {Loeb}, A. 2016, \mnras, 459, L51,
  \dodoi{10.1093/mnrasl/slw055}

\bibitem[{{Astropy Collaboration} {et~al.}(2013){Astropy Collaboration},
  {Robitaille}, {Tollerud}, {Greenfield}, {Droettboom}, {Bray}, {Aldcroft},
  {Davis}, {Ginsburg}, {Price-Whelan}, {Kerzendorf}, {Conley}, {Crighton},
  {Barbary}, {Muna}, {Ferguson}, {Grollier}, {Parikh}, {Nair}, {Unther},
  {Deil}, {Woillez}, {Conseil}, {Kramer}, {Turner}, {Singer}, {Fox}, {Weaver},
  {Zabalza}, {Edwards}, {Azalee Bostroem}, {Burke}, {Casey}, {Crawford},
  {Dencheva}, {Ely}, {Jenness}, {Labrie}, {Lim}, {Pierfederici}, {Pontzen},
  {Ptak}, {Refsdal}, {Servillat}, \& {Streicher}}]{astropy}
{Astropy Collaboration}, {Robitaille}, T.~P., {Tollerud}, E.~J., {et~al.} 2013,
  \aap, 558, A33, \dodoi{10.1051/0004-6361/201322068}

\bibitem[{Barbary(2016)}]{Barbary2016}
Barbary, K. 2016, Journal of Open Source Software, 1(6), 58,
  \dodoi{10.21105/joss.0005}

\bibitem[{{Baxter} {et~al.}(2021){Baxter}, {Cooper}, \&
  {Fillingham}}]{Baxter2021}
{Baxter}, D.~C., {Cooper}, M.~C., \& {Fillingham}, S.~P. 2021, \mnras, 503,
  1636, \dodoi{10.1093/mnras/stab523}

\bibitem[{{Behroozi} {et~al.}(2010){Behroozi}, {Conroy}, \&
  {Wechsler}}]{Behroozi2010}
{Behroozi}, P.~S., {Conroy}, C., \& {Wechsler}, R.~H. 2010, \apj, 717, 379,
  \dodoi{10.1088/0004-637X/717/1/379}

\bibitem[{{Benavides} {et~al.}(2022){Benavides}, {Sales}, {Abadi}, {Marinacci},
  {Vogelsberger}, \& {Hernquist}}]{Benavides2022}
{Benavides}, J.~A., {Sales}, L.~V., {Abadi}, M.~G., {et~al.} 2022, arXiv
  e-prints, arXiv:2209.07539.
\newblock \doarXiv{2209.07539}

\bibitem[{{Benavides} {et~al.}(2021){Benavides}, {Sales}, {Abadi}, {Pillepich},
  {Nelson}, {Marinacci}, {Cooper}, {Pakmor}, {Torrey}, {Vogelsberger}, \&
  {Hernquist}}]{Benavides2021}
---. 2021, Nature Astronomy, 5, 1255, \dodoi{10.1038/s41550-021-01458-1}

\bibitem[{{Bertin}(2011)}]{Bertin2011}
{Bertin}, E. 2011, in Astronomical Society of the Pacific Conference Series,
  Vol. 442, Astronomical Data Analysis Software and Systems XX, ed. I.~N.
  {Evans}, A.~{Accomazzi}, D.~J. {Mink}, \& A.~H. {Rots}, 435

\bibitem[{{Bertin} \& {Arnouts}(1996)}]{Bertin1996}
{Bertin}, E., \& {Arnouts}, S. 1996, \aaps, 117, 393,
  \dodoi{10.1051/aas:1996164}

\bibitem[{{Bland-Hawthorn} \& {Gerhard}(2016)}]{Bland-Hawthorn2016}
{Bland-Hawthorn}, J., \& {Gerhard}, O. 2016, \araa, 54, 529,
  \dodoi{10.1146/annurev-astro-081915-023441}

\bibitem[{{Blanton} {et~al.}(2011){Blanton}, {Kazin}, {Muna}, {Weaver}, \&
  {Price-Whelan}}]{Blanton2011}
{Blanton}, M.~R., {Kazin}, E., {Muna}, D., {Weaver}, B.~A., \& {Price-Whelan},
  A. 2011, \aj, 142, 31, \dodoi{10.1088/0004-6256/142/1/31}

\bibitem[{{Blanton} {et~al.}(2005){Blanton}, {Lupton}, {Schlegel}, {Strauss},
  {Brinkmann}, {Fukugita}, \& {Loveday}}]{Blanton2005}
{Blanton}, M.~R., {Lupton}, R.~H., {Schlegel}, D.~J., {et~al.} 2005, \apj, 631,
  208, \dodoi{10.1086/431416}

\bibitem[{{Blanton} \& {Roweis}(2007)}]{Blanton2007}
{Blanton}, M.~R., \& {Roweis}, S. 2007, \aj, 133, 734, \dodoi{10.1086/510127}

\bibitem[{{Bosch} {et~al.}(2018){Bosch}, {Armstrong}, {Bickerton}, {Furusawa},
  {Ikeda}, {Koike}, {Lupton}, {Mineo}, {Price}, {Takata}, {Tanaka}, {Yasuda},
  {AlSayyad}, {Becker}, {Coulton}, {Coupon}, {Garmilla}, {Huang}, {Krughoff},
  {Lang}, {Leauthaud}, {Lim}, {Lust}, {MacArthur}, {Mand elbaum}, {Miyatake},
  {Miyazaki}, {Murata}, {More}, {Okura}, {Owen}, {Swinbank}, {Strauss},
  {Yamada}, \& {Yamanoi}}]{Bosch2018}
{Bosch}, J., {Armstrong}, R., {Bickerton}, S., {et~al.} 2018, \pasj, 70, S5,
  \dodoi{10.1093/pasj/psx080}

\bibitem[{{Bryan} \& {Norman}(1998)}]{Bryan1998}
{Bryan}, G.~L., \& {Norman}, M.~L. 1998, \apj, 495, 80, \dodoi{10.1086/305262}

\bibitem[{{Caldwell} \& {Bothun}(1987)}]{Caldwell1987}
{Caldwell}, N., \& {Bothun}, G.~D. 1987, \aj, 94, 1126, \dodoi{10.1086/114550}

\bibitem[{{Cappellari}(2013)}]{Cappellari2013}
{Cappellari}, M. 2013, \apjl, 778, L2, \dodoi{10.1088/2041-8205/778/1/L2}

\bibitem[{{Cardona-Barrero} {et~al.}(2020){Cardona-Barrero}, {Di Cintio},
  {Brook}, {Ruiz-Lara}, {Beasley}, {Falc{\'o}n-Barroso}, \&
  {Macci{\`o}}}]{Cardona-Barrero2020}
{Cardona-Barrero}, S., {Di Cintio}, A., {Brook}, C. B.~A., {et~al.} 2020,
  \mnras, 497, 4282, \dodoi{10.1093/mnras/staa2094}

\bibitem[{{Carlsten} {et~al.}(2022{\natexlab{a}}){Carlsten}, {Greene},
  {Beaton}, {Danieli}, \& {Greco}}]{CarlstenELVES2022}
{Carlsten}, S.~G., {Greene}, J.~E., {Beaton}, R.~L., {Danieli}, S., \& {Greco},
  J.~P. 2022{\natexlab{a}}, \apj, 933, 47, \dodoi{10.3847/1538-4357/ac6fd7}

\bibitem[{{Carlsten} {et~al.}(2022{\natexlab{b}}){Carlsten}, {Greene},
  {Beaton}, \& {Greco}}]{ELVES-II}
{Carlsten}, S.~G., {Greene}, J.~E., {Beaton}, R.~L., \& {Greco}, J.~P.
  2022{\natexlab{b}}, \apj, 927, 44, \dodoi{10.3847/1538-4357/ac457e}

\bibitem[{{Carlsten} {et~al.}(2021){Carlsten}, {Greene}, {Greco}, {Beaton}, \&
  {Kado-Fong}}]{ELVES-I}
{Carlsten}, S.~G., {Greene}, J.~E., {Greco}, J.~P., {Beaton}, R.~L., \&
  {Kado-Fong}, E. 2021, \apj, 922, 267, \dodoi{10.3847/1538-4357/ac2581}

\bibitem[{{Chabrier}(2003)}]{Chabrier2003}
{Chabrier}, G. 2003, \pasp, 115, 763, \dodoi{10.1086/376392}

\bibitem[{{Chamba} {et~al.}(2022){Chamba}, {Trujillo}, \&
  {Knapen}}]{Chamba2022}
{Chamba}, N., {Trujillo}, I., \& {Knapen}, J.~H. 2022, arXiv e-prints,
  arXiv:2209.05497.
\newblock \doarXiv{2209.05497}

\bibitem[{{Chan} {et~al.}(2018){Chan}, {Kere{\v{s}}}, {Wetzel}, {Hopkins},
  {Faucher-Gigu{\`e}re}, {El-Badry}, {Garrison-Kimmel}, \&
  {Boylan-Kolchin}}]{Chan2018}
{Chan}, T.~K., {Kere{\v{s}}}, D., {Wetzel}, A., {et~al.} 2018, \mnras, 478,
  906, \dodoi{10.1093/mnras/sty1153}

\bibitem[{{Cohen} {et~al.}(2018){Cohen}, {van Dokkum}, {Danieli}, {Romanowsky},
  {Abraham}, {Merritt}, {Zhang}, {Mowla}, {Kruijssen}, {Conroy}, \&
  {Wasserman}}]{Cohen2018}
{Cohen}, Y., {van Dokkum}, P., {Danieli}, S., {et~al.} 2018, \apj, 868, 96,
  \dodoi{10.3847/1538-4357/aae7c8}

\bibitem[{{Conselice}(2003)}]{Conselice2003}
{Conselice}, C.~J. 2003, \apjs, 147, 1, \dodoi{10.1086/375001}

\bibitem[{{Cranmer}(2023)}]{Cranmer2023}
{Cranmer}, M. 2023, arXiv e-prints, arXiv:2305.01582,
  \dodoi{10.48550/arXiv.2305.01582}

\bibitem[{{Dalcanton} {et~al.}(1997{\natexlab{a}}){Dalcanton}, {Spergel},
  {Gunn}, {Schmidt}, \& {Schneider}}]{Dalcanton1997a}
{Dalcanton}, J.~J., {Spergel}, D.~N., {Gunn}, J.~E., {Schmidt}, M., \&
  {Schneider}, D.~P. 1997{\natexlab{a}}, \aj, 114, 635, \dodoi{10.1086/118499}

\bibitem[{{Dalcanton} {et~al.}(1997{\natexlab{b}}){Dalcanton}, {Spergel},
  {Gunn}, {Schmidt}, \& {Schneider}}]{Dalcanton1997}
---. 1997{\natexlab{b}}, \aj, 114, 635, \dodoi{10.1086/118499}

\bibitem[{{Danieli} \& {van Dokkum}(2019)}]{Danieli2019}
{Danieli}, S., \& {van Dokkum}, P. 2019, \apj, 875, 155,
  \dodoi{10.3847/1538-4357/ab14f3}

\bibitem[{{Danieli} {et~al.}(2019){Danieli}, {van Dokkum}, {Conroy}, {Abraham},
  \& {Romanowsky}}]{Danieli2019DF2}
{Danieli}, S., {van Dokkum}, P., {Conroy}, C., {Abraham}, R., \& {Romanowsky},
  A.~J. 2019, \apjl, 874, L12, \dodoi{10.3847/2041-8213/ab0e8c}

\bibitem[{{Danieli} {et~al.}(2020){Danieli}, {Lokhorst}, {Zhang}, {Merritt},
  {van Dokkum}, {Abraham}, {Conroy}, {Gilhuly}, {Greco}, {Janssens}, {Li},
  {Liu}, {Miller}, \& {Mowla}}]{Danieli2020}
{Danieli}, S., {Lokhorst}, D., {Zhang}, J., {et~al.} 2020, \apj, 894, 119,
  \dodoi{10.3847/1538-4357/ab88a8}

\bibitem[{{Danieli} {et~al.}(2022){Danieli}, {van Dokkum}, {Trujillo-Gomez},
  {Kruijssen}, {Romanowsky}, {Carlsten}, {Shen}, {Li}, {Abraham}, {Brodie},
  {Conroy}, {Gannon}, \& {Greco}}]{Danieli2022}
{Danieli}, S., {van Dokkum}, P., {Trujillo-Gomez}, S., {et~al.} 2022, \apjl,
  927, L28, \dodoi{10.3847/2041-8213/ac590a}

\bibitem[{{de Vaucouleurs}(1948)}]{deVaucouleurs1948}
{de Vaucouleurs}, G. 1948, Annales d'Astrophysique, 11, 247

\bibitem[{{Dey} {et~al.}(2019){Dey}, {Schlegel}, {Lang}, {Blum}, {Burleigh},
  {Fan}, {Findlay}, {Finkbeiner}, {Herrera}, {Juneau}, {Landriau}, {Levi},
  {McGreer}, {Meisner}, {Myers}, {Moustakas}, {Nugent}, {Patej}, {Schlafly},
  {Walker}, {Valdes}, {Weaver}, {Y{\`e}che}, {Zou}, {Zhou}, {Abareshi},
  {Abbott}, {Abolfathi}, {Aguilera}, {Alam}, {Allen}, {Alvarez}, {Annis},
  {Ansarinejad}, {Aubert}, {Beechert}, {Bell}, {BenZvi}, {Beutler}, {Bielby},
  {Bolton}, {Brice{\~n}o}, {Buckley-Geer}, {Butler}, {Calamida}, {Carlberg},
  {Carter}, {Casas}, {Castander}, {Choi}, {Comparat}, {Cukanovaite}, {Delubac},
  {DeVries}, {Dey}, {Dhungana}, {Dickinson}, {Ding}, {Donaldson}, {Duan},
  {Duckworth}, {Eftekharzadeh}, {Eisenstein}, {Etourneau}, {Fagrelius},
  {Farihi}, {Fitzpatrick}, {Font-Ribera}, {Fulmer}, {G{\"a}nsicke},
  {Gaztanaga}, {George}, {Gerdes}, {Gontcho}, {Gorgoni}, {Green}, {Guy},
  {Harmer}, {Hernandez}, {Honscheid}, {Huang}, {James}, {Jannuzi}, {Jiang},
  {Joyce}, {Karcher}, {Karkar}, {Kehoe}, {Kneib}, {Kueter-Young}, {Lan},
  {Lauer}, {Le Guillou}, {Le Van Suu}, {Lee}, {Lesser}, {Perreault Levasseur},
  {Li}, {Mann}, {Marshall}, {Mart{\'\i}nez-V{\'a}zquez}, {Martini}, {du Mas des
  Bourboux}, {McManus}, {Meier}, {M{\'e}nard}, {Metcalfe},
  {Mu{\~n}oz-Guti{\'e}rrez}, {Najita}, {Napier}, {Narayan}, {Newman}, {Nie},
  {Nord}, {Norman}, {Olsen}, {Paat}, {Palanque-Delabrouille}, {Peng},
  {Poppett}, {Poremba}, {Prakash}, {Rabinowitz}, {Raichoor}, {Rezaie},
  {Robertson}, {Roe}, {Ross}, {Ross}, {Rudnick}, {Safonova}, {Saha},
  {S{\'a}nchez}, {Savary}, {Schweiker}, {Scott}, {Seo}, {Shan}, {Silva},
  {Slepian}, {Soto}, {Sprayberry}, {Staten}, {Stillman}, {Stupak}, {Summers},
  {Sien Tie}, {Tirado}, {Vargas-Maga{\~n}a}, {Vivas}, {Wechsler}, {Williams},
  {Yang}, {Yang}, {Yapici}, {Zaritsky}, {Zenteno}, {Zhang}, {Zhang}, {Zhou}, \&
  {Zhou}}]{Dey2019}
{Dey}, A., {Schlegel}, D.~J., {Lang}, D., {et~al.} 2019, \aj, 157, 168,
  \dodoi{10.3847/1538-3881/ab089d}

\bibitem[{{Di Cintio} {et~al.}(2017){Di Cintio}, {Brook}, {Dutton},
  {Macci{\`o}}, {Obreja}, \& {Dekel}}]{DiCintio2017}
{Di Cintio}, A., {Brook}, C.~B., {Dutton}, A.~A., {et~al.} 2017, \mnras, 466,
  L1, \dodoi{10.1093/mnrasl/slw210}

\bibitem[{{Diemer}(2018)}]{Colossus}
{Diemer}, B. 2018, \apjs, 239, 35, \dodoi{10.3847/1538-4365/aaee8c}

\bibitem[{{Du} {et~al.}(2020){Du}, {Cheng}, {Zheng}, \& {Wu}}]{Du2020}
{Du}, W., {Cheng}, C., {Zheng}, Z., \& {Wu}, H. 2020, \aj, 159, 138,
  \dodoi{10.3847/1538-3881/ab6efb}

\bibitem[{{Eigenthaler} {et~al.}(2018){Eigenthaler}, {Puzia}, {Taylor},
  {Ordenes-Brice{\~n}o}, {Mu{\~n}oz}, {Ribbeck}, {Alamo-Mart{\'\i}nez},
  {Zhang}, {{\'A}ngel}, {Capaccioli}, {C{\^o}t{\'e}}, {Ferrarese}, {Galaz},
  {Grebel}, {Hempel}, {Hilker}, {Lan{\c{c}}on}, {Mieske}, {Miller}, {Paolillo},
  {Powalka}, {Richtler}, {Roediger}, {Rong}, {S{\'a}nchez-Janssen}, \&
  {Spengler}}]{Eigenthaler2018}
{Eigenthaler}, P., {Puzia}, T.~H., {Taylor}, M.~A., {et~al.} 2018, \apj, 855,
  142, \dodoi{10.3847/1538-4357/aaab60}

\bibitem[{{Erwin}(2015)}]{imfit}
{Erwin}, P. 2015, \apj, 799, 226, \dodoi{10.1088/0004-637X/799/2/226}

\bibitem[{{Ferrarese} {et~al.}(2020){Ferrarese}, {C{\^o}t{\'e}}, {MacArthur},
  {Durrell}, {Gwyn}, {Duc}, {S{\'a}nchez-Janssen}, {Santos}, {Blakeslee},
  {Boselli}, {Boyer}, {Cantiello}, {Courteau}, {Cuillandre}, {Emsellem},
  {Erben}, {Gavazzi}, {Guhathakurta}, {Huertas-Company}, {Jord{\'a}n},
  {Lan{\c{c}}on}, {Liu}, {Mei}, {Mihos}, {Peng}, {Puzia}, {Roediger}, {Schade},
  {Taylor}, {Toloba}, \& {Zhang}}]{Ferrarese2020}
{Ferrarese}, L., {C{\^o}t{\'e}}, P., {MacArthur}, L.~A., {et~al.} 2020, \apj,
  890, 128, \dodoi{10.3847/1538-4357/ab339f}

\bibitem[{{Ferr{\'e}-Mateu} {et~al.}(2018){Ferr{\'e}-Mateu}, {Alabi}, {Forbes},
  {Romanowsky}, {Brodie}, {Pandya}, {Mart{\'\i}n-Navarro}, {Bellstedt},
  {Wasserman}, {Stone}, \& {Okabe}}]{Ferre-Mateu2018}
{Ferr{\'e}-Mateu}, A., {Alabi}, A., {Forbes}, D.~A., {et~al.} 2018, \mnras,
  479, 4891, \dodoi{10.1093/mnras/sty1597}

\bibitem[{{Fliri} \& {Trujillo}(2016)}]{Fliri2016}
{Fliri}, J., \& {Trujillo}, I. 2016, \mnras, 456, 1359,
  \dodoi{10.1093/mnras/stv2686}

\bibitem[{{Font} {et~al.}(2022){Font}, {McCarthy}, {Belokurov}, {Brown}, \&
  {Stafford}}]{Font2022}
{Font}, A.~S., {McCarthy}, I.~G., {Belokurov}, V., {Brown}, S.~T., \&
  {Stafford}, S.~G. 2022, \mnras, 511, 1544, \dodoi{10.1093/mnras/stac183}

\bibitem[{{Forbes} {et~al.}(2020){Forbes}, {Alabi}, {Romanowsky}, {Brodie}, \&
  {Arimoto}}]{Forbes2020}
{Forbes}, D.~A., {Alabi}, A., {Romanowsky}, A.~J., {Brodie}, J.~P., \&
  {Arimoto}, N. 2020, \mnras, 492, 4874, \dodoi{10.1093/mnras/staa180}

\bibitem[{{Gaia Collaboration} {et~al.}(2016){Gaia Collaboration}, {Prusti},
  {de Bruijne}, {Brown}, {Vallenari}, {Babusiaux}, {Bailer-Jones}, {Bastian},
  {Biermann}, {Evans}, {Eyer}, {Jansen}, {Jordi}, {Klioner}, {Lammers},
  {Lindegren}, {Luri}, {Mignard}, {Milligan}, {Panem}, {Poinsignon},
  {Pourbaix}, {Randich}, {Sarri}, {Sartoretti}, {Siddiqui}, {Soubiran},
  {Valette}, {van Leeuwen}, {Walton}, {Aerts}, {Arenou}, {Cropper}, {Drimmel},
  {H{\o}g}, {Katz}, {Lattanzi}, {O'Mullane}, {Grebel}, {Holland}, {Huc},
  {Passot}, {Bramante}, {Cacciari}, {Casta{\~n}eda}, {Chaoul}, {Cheek}, {De
  Angeli}, {Fabricius}, {Guerra}, {Hern{\'a}ndez}, {Jean-Antoine-Piccolo},
  {Masana}, {Messineo}, {Mowlavi}, {Nienartowicz}, {Ord{\'o}{\~n}ez-Blanco},
  {Panuzzo}, {Portell}, {Richards}, {Riello}, {Seabroke}, {Tanga},
  {Th{\'e}venin}, {Torra}, {Els}, {Gracia-Abril}, {Comoretto},
  {Garcia-Reinaldos}, {Lock}, {Mercier}, {Altmann}, {Andrae}, {Astraatmadja},
  {Bellas-Velidis}, {Benson}, {Berthier}, {Blomme}, {Busso}, {Carry},
  {Cellino}, {Clementini}, {Cowell}, {Creevey}, {Cuypers}, {Davidson}, {De
  Ridder}, {de Torres}, {Delchambre}, {Dell'Oro}, {Ducourant}, {Fr{\'e}mat},
  {Garc{\'\i}a-Torres}, {Gosset}, {Halbwachs}, {Hambly}, {Harrison}, {Hauser},
  {Hestroffer}, {Hodgkin}, {Huckle}, {Hutton}, {Jasniewicz}, {Jordan},
  {Kontizas}, {Korn}, {Lanzafame}, {Manteiga}, {Moitinho}, {Muinonen},
  {Osinde}, {Pancino}, {Pauwels}, {Petit}, {Recio-Blanco}, {Robin}, {Sarro},
  {Siopis}, {Smith}, {Smith}, {Sozzetti}, {Thuillot}, {van Reeven}, {Viala},
  {Abbas}, {Abreu Aramburu}, {Accart}, {Aguado}, {Allan}, {Allasia},
  {Altavilla}, {{\'A}lvarez}, {Alves}, {Anderson}, {Andrei}, {Anglada Varela},
  {Antiche}, {Antoja}, {Ant{\'o}n}, {Arcay}, {Atzei}, {Ayache}, {Bach},
  {Baker}, {Balaguer-N{\'u}{\~n}ez}, {Barache}, {Barata}, {Barbier}, {Barblan},
  {Baroni}, {Barrado y Navascu{\'e}s}, {Barros}, {Barstow}, {Becciani},
  {Bellazzini}, {Bellei}, {Bello Garc{\'\i}a}, {Belokurov}, {Bendjoya},
  {Berihuete}, {Bianchi}, {Bienaym{\'e}}, {Billebaud}, {Blagorodnova},
  {Blanco-Cuaresma}, {Boch}, {Bombrun}, {Borrachero}, {Bouquillon}, {Bourda},
  {Bouy}, {Bragaglia}, {Breddels}, {Brouillet}, {Br{\"u}semeister},
  {Bucciarelli}, {Budnik}, {Burgess}, {Burgon}, {Burlacu}, {Busonero}, {Buzzi},
  {Caffau}, {Cambras}, {Campbell}, {Cancelliere}, {Cantat-Gaudin}, {Carlucci},
  {Carrasco}, {Castellani}, {Charlot}, {Charnas}, {Charvet}, {Chassat},
  {Chiavassa}, {Clotet}, {Cocozza}, {Collins}, {Collins}, {Costigan}, {Crifo},
  {Cross}, {Crosta}, {Crowley}, {Dafonte}, {Damerdji}, {Dapergolas}, {David},
  {David}, {De Cat}, {de Felice}, {de Laverny}, {De Luise}, {De March}, {de
  Martino}, {de Souza}, {Debosscher}, {del Pozo}, {Delbo}, {Delgado},
  {Delgado}, {di Marco}, {Di Matteo}, {Diakite}, {Distefano}, {Dolding}, {Dos
  Anjos}, {Drazinos}, {Dur{\'a}n}, {Dzigan}, {Ecale}, {Edvardsson}, {Enke},
  {Erdmann}, {Escolar}, {Espina}, {Evans}, {Eynard Bontemps}, {Fabre},
  {Fabrizio}, {Faigler}, {Falc{\~a}o}, {Farr{\`a}s Casas}, {Faye}, {Federici},
  {Fedorets}, {Fern{\'a}ndez-Hern{\'a}ndez}, {Fernique}, {Fienga}, {Figueras},
  {Filippi}, {Findeisen}, {Fonti}, {Fouesneau}, {Fraile}, {Fraser}, {Fuchs},
  {Furnell}, {Gai}, {Galleti}, {Galluccio}, {Garabato}, {Garc{\'\i}a-Sedano},
  {Gar{\'e}}, {Garofalo}, {Garralda}, {Gavras}, {Gerssen}, {Geyer}, {Gilmore},
  {Girona}, {Giuffrida}, {Gomes}, {Gonz{\'a}lez-Marcos},
  {Gonz{\'a}lez-N{\'u}{\~n}ez}, {Gonz{\'a}lez-Vidal}, {Granvik}, {Guerrier},
  {Guillout}, {Guiraud}, {G{\'u}rpide}, {Guti{\'e}rrez-S{\'a}nchez}, {Guy},
  {Haigron}, {Hatzidimitriou}, {Haywood}, {Heiter}, {Helmi}, {Hobbs},
  {Hofmann}, {Holl}, {Holland}, {Hunt}, {Hypki}, {Icardi}, {Irwin}, {Jevardat
  de Fombelle}, {Jofr{\'e}}, {Jonker}, {Jorissen}, {Julbe}, {Karampelas},
  {Kochoska}, {Kohley}, {Kolenberg}, {Kontizas}, {Koposov}, {Kordopatis},
  {Koubsky}, {Kowalczyk}, {Krone-Martins}, {Kudryashova}, {Kull}, {Bachchan},
  {Lacoste-Seris}, {Lanza}, {Lavigne}, {Le Poncin-Lafitte}, {Lebreton},
  {Lebzelter}, {Leccia}, {Leclerc}, {Lecoeur-Taibi}, {Lemaitre}, {Lenhardt},
  {Leroux}, {Liao}, {Licata}, {Lindstr{\o}m}, {Lister}, {Livanou}, {Lobel},
  {L{\"o}ffler}, {L{\'o}pez}, {Lopez-Lozano}, {Lorenz}, {Loureiro},
  {MacDonald}, {Magalh{\~a}es Fernandes}, {Managau}, {Mann}, {Mantelet},
  {Marchal}, {Marchant}, {Marconi}, {Marie}, {Marinoni}, {Marrese},
  {Marschalk{\'o}}, {Marshall}, {Mart{\'\i}n-Fleitas}, {Martino}, {Mary},
  {Matijevi{\v{c}}}, {Mazeh}, {McMillan}, {Messina}, {Mestre}, {Michalik},
  {Millar}, {Miranda}, {Molina}, {Molinaro}, {Molinaro}, {Moln{\'a}r},
  {Moniez}, {Montegriffo}, {Monteiro}, {Mor}, {Mora}, {Morbidelli}, {Morel},
  {Morgenthaler}, {Morley}, {Morris}, {Mulone}, {Muraveva}, {Musella},
  {Narbonne}, {Nelemans}, {Nicastro}, {Noval}, {Ord{\'e}novic},
  {Ordieres-Mer{\'e}}, {Osborne}, {Pagani}, {Pagano}, {Pailler}, {Palacin},
  {Palaversa}, {Parsons}, {Paulsen}, {Pecoraro}, {Pedrosa}, {Pentik{\"a}inen},
  {Pereira}, {Pichon}, {Piersimoni}, {Pineau}, {Plachy}, {Plum}, {Poujoulet},
  {Pr{\v{s}}a}, {Pulone}, {Ragaini}, {Rago}, {Rambaux}, {Ramos-Lerate},
  {Ranalli}, {Rauw}, {Read}, {Regibo}, {Renk}, {Reyl{\'e}}, {Ribeiro},
  {Rimoldini}, {Ripepi}, {Riva}, {Rixon}, {Roelens}, {Romero-G{\'o}mez},
  {Rowell}, {Royer}, {Rudolph}, {Ruiz-Dern}, {Sadowski}, {Sagrist{\`a}
  Sell{\'e}s}, {Sahlmann}, {Salgado}, {Salguero}, {Sarasso}, {Savietto},
  {Schnorhk}, {Schultheis}, {Sciacca}, {Segol}, {Segovia}, {Segransan},
  {Serpell}, {Shih}, {Smareglia}, {Smart}, {Smith}, {Solano}, {Solitro},
  {Sordo}, {Soria Nieto}, {Souchay}, {Spagna}, {Spoto}, {Stampa}, {Steele},
  {Steidelm{\"u}ller}, {Stephenson}, {Stoev}, {Suess}, {S{\"u}veges}, {Surdej},
  {Szabados}, {Szegedi-Elek}, {Tapiador}, {Taris}, {Tauran}, {Taylor},
  {Teixeira}, {Terrett}, {Tingley}, {Trager}, {Turon}, {Ulla}, {Utrilla},
  {Valentini}, {van Elteren}, {Van Hemelryck}, {van Leeuwen}, {Varadi},
  {Vecchiato}, {Veljanoski}, {Via}, {Vicente}, {Vogt}, {Voss}, {Votruba},
  {Voutsinas}, {Walmsley}, {Weiler}, {Weingrill}, {Werner}, {Wevers},
  {Whitehead}, {Wyrzykowski}, {Yoldas}, {{\v{Z}}erjal}, {Zucker}, {Zurbach},
  {Zwitter}, {Alecu}, {Allen}, {Allende Prieto}, {Amorim},
  {Anglada-Escud{\'e}}, {Arsenijevic}, {Azaz}, {Balm}, {Beck}, {Bernstein},
  {Bigot}, {Bijaoui}, {Blasco}, {Bonfigli}, {Bono}, {Boudreault}, {Bressan},
  {Brown}, {Brunet}, {Bunclark}, {Buonanno}, {Butkevich}, {Carret}, {Carrion},
  {Chemin}, {Ch{\'e}reau}, {Corcione}, {Darmigny}, {de Boer}, {de Teodoro}, {de
  Zeeuw}, {Delle Luche}, {Domingues}, {Dubath}, {Fodor}, {Fr{\'e}zouls},
  {Fries}, {Fustes}, {Fyfe}, {Gallardo}, {Gallegos}, {Gardiol}, {Gebran},
  {Gomboc}, {G{\'o}mez}, {Grux}, {Gueguen}, {Heyrovsky}, {Hoar}, {Iannicola},
  {Isasi Parache}, {Janotto}, {Joliet}, {Jonckheere}, {Keil}, {Kim},
  {Klagyivik}, {Klar}, {Knude}, {Kochukhov}, {Kolka}, {Kos}, {Kutka}, {Lainey},
  {LeBouquin}, {Liu}, {Loreggia}, {Makarov}, {Marseille}, {Martayan},
  {Martinez-Rubi}, {Massart}, {Meynadier}, {Mignot}, {Munari}, {Nguyen},
  {Nordlander}, {Ocvirk}, {O'Flaherty}, {Olias Sanz}, {Ortiz}, {Osorio},
  {Oszkiewicz}, {Ouzounis}, {Palmer}, {Park}, {Pasquato}, {Peltzer}, {Peralta},
  {P{\'e}turaud}, {Pieniluoma}, {Pigozzi}, {Poels}, {Prat}, {Prod'homme},
  {Raison}, {Rebordao}, {Risquez}, {Rocca-Volmerange}, {Rosen}, {Ruiz-Fuertes},
  {Russo}, {Sembay}, {Serraller Vizcaino}, {Short}, {Siebert}, {Silva},
  {Sinachopoulos}, {Slezak}, {Soffel}, {Sosnowska}, {Strai{\v{z}}ys}, {ter
  Linden}, {Terrell}, {Theil}, {Tiede}, {Troisi}, {Tsalmantza}, {Tur},
  {Vaccari}, {Vachier}, {Valles}, {Van Hamme}, {Veltz}, {Virtanen}, {Wallut},
  {Wichmann}, {Wilkinson}, {Ziaeepour}, \& {Zschocke}}]{GAIA2016}
{Gaia Collaboration}, {Prusti}, T., {de Bruijne}, J.~H.~J., {et~al.} 2016,
  \aap, 595, A1, \dodoi{10.1051/0004-6361/201629272}

\bibitem[{{Gaia Collaboration} {et~al.}(2018){Gaia Collaboration}, {Brown},
  {Vallenari}, {Prusti}, {de Bruijne}, {Babusiaux}, {Bailer-Jones}, {Biermann},
  {Evans}, {Eyer}, {Jansen}, {Jordi}, {Klioner}, {Lammers}, {Lindegren},
  {Luri}, {Mignard}, {Panem}, {Pourbaix}, {Randich}, {Sartoretti}, {Siddiqui},
  {Soubiran}, {van Leeuwen}, {Walton}, {Arenou}, {Bastian}, {Cropper},
  {Drimmel}, {Katz}, {Lattanzi}, {Bakker}, {Cacciari}, {Casta{\~n}eda},
  {Chaoul}, {Cheek}, {De Angeli}, {Fabricius}, {Guerra}, {Holl}, {Masana},
  {Messineo}, {Mowlavi}, {Nienartowicz}, {Panuzzo}, {Portell}, {Riello},
  {Seabroke}, {Tanga}, {Th{\'e}venin}, {Gracia-Abril}, {Comoretto},
  {Garcia-Reinaldos}, {Teyssier}, {Altmann}, {Andrae}, {Audard},
  {Bellas-Velidis}, {Benson}, {Berthier}, {Blomme}, {Burgess}, {Busso},
  {Carry}, {Cellino}, {Clementini}, {Clotet}, {Creevey}, {Davidson}, {De
  Ridder}, {Delchambre}, {Dell'Oro}, {Ducourant},
  {Fern{\'a}ndez-Hern{\'a}ndez}, {Fouesneau}, {Fr{\'e}mat}, {Galluccio},
  {Garc{\'\i}a-Torres}, {Gonz{\'a}lez-N{\'u}{\~n}ez}, {Gonz{\'a}lez-Vidal},
  {Gosset}, {Guy}, {Halbwachs}, {Hambly}, {Harrison}, {Hern{\'a}ndez},
  {Hestroffer}, {Hodgkin}, {Hutton}, {Jasniewicz}, {Jean-Antoine-Piccolo},
  {Jordan}, {Korn}, {Krone-Martins}, {Lanzafame}, {Lebzelter}, {L{\"o}ffler},
  {Manteiga}, {Marrese}, {Mart{\'\i}n-Fleitas}, {Moitinho}, {Mora}, {Muinonen},
  {Osinde}, {Pancino}, {Pauwels}, {Petit}, {Recio-Blanco}, {Richards},
  {Rimoldini}, {Robin}, {Sarro}, {Siopis}, {Smith}, {Sozzetti}, {S{\"u}veges},
  {Torra}, {van Reeven}, {Abbas}, {Abreu Aramburu}, {Accart}, {Aerts},
  {Altavilla}, {{\'A}lvarez}, {Alvarez}, {Alves}, {Anderson}, {Andrei},
  {Anglada Varela}, {Antiche}, {Antoja}, {Arcay}, {Astraatmadja}, {Bach},
  {Baker}, {Balaguer-N{\'u}{\~n}ez}, {Balm}, {Barache}, {Barata}, {Barbato},
  {Barblan}, {Barklem}, {Barrado}, {Barros}, {Barstow}, {Bartholom{\'e}
  Mu{\~n}oz}, {Bassilana}, {Becciani}, {Bellazzini}, {Berihuete}, {Bertone},
  {Bianchi}, {Bienaym{\'e}}, {Blanco-Cuaresma}, {Boch}, {Boeche}, {Bombrun},
  {Borrachero}, {Bossini}, {Bouquillon}, {Bourda}, {Bragaglia}, {Bramante},
  {Breddels}, {Bressan}, {Brouillet}, {Br{\"u}semeister}, {Brugaletta},
  {Bucciarelli}, {Burlacu}, {Busonero}, {Butkevich}, {Buzzi}, {Caffau},
  {Cancelliere}, {Cannizzaro}, {Cantat-Gaudin}, {Carballo}, {Carlucci},
  {Carrasco}, {Casamiquela}, {Castellani}, {Castro-Ginard}, {Charlot},
  {Chemin}, {Chiavassa}, {Cocozza}, {Costigan}, {Cowell}, {Crifo}, {Crosta},
  {Crowley}, {Cuypers}, {Dafonte}, {Damerdji}, {Dapergolas}, {David}, {David},
  {de Laverny}, {De Luise}, {De March}, {de Martino}, {de Souza}, {de Torres},
  {Debosscher}, {del Pozo}, {Delbo}, {Delgado}, {Delgado}, {Di Matteo},
  {Diakite}, {Diener}, {Distefano}, {Dolding}, {Drazinos}, {Dur{\'a}n},
  {Edvardsson}, {Enke}, {Eriksson}, {Esquej}, {Eynard Bontemps}, {Fabre},
  {Fabrizio}, {Faigler}, {Falc{\~a}o}, {Farr{\`a}s Casas}, {Federici},
  {Fedorets}, {Fernique}, {Figueras}, {Filippi}, {Findeisen}, {Fonti},
  {Fraile}, {Fraser}, {Fr{\'e}zouls}, {Gai}, {Galleti}, {Garabato},
  {Garc{\'\i}a-Sedano}, {Garofalo}, {Garralda}, {Gavel}, {Gavras}, {Gerssen},
  {Geyer}, {Giacobbe}, {Gilmore}, {Girona}, {Giuffrida}, {Glass}, {Gomes},
  {Granvik}, {Gueguen}, {Guerrier}, {Guiraud}, {Guti{\'e}rrez-S{\'a}nchez},
  {Haigron}, {Hatzidimitriou}, {Hauser}, {Haywood}, {Heiter}, {Helmi}, {Heu},
  {Hilger}, {Hobbs}, {Hofmann}, {Holland}, {Huckle}, {Hypki}, {Icardi},
  {Jan{\ss}en}, {Jevardat de Fombelle}, {Jonker}, {Juh{\'a}sz}, {Julbe},
  {Karampelas}, {Kewley}, {Klar}, {Kochoska}, {Kohley}, {Kolenberg},
  {Kontizas}, {Kontizas}, {Koposov}, {Kordopatis}, {Kostrzewa-Rutkowska},
  {Koubsky}, {Lambert}, {Lanza}, {Lasne}, {Lavigne}, {Le Fustec}, {Le
  Poncin-Lafitte}, {Lebreton}, {Leccia}, {Leclerc}, {Lecoeur-Taibi},
  {Lenhardt}, {Leroux}, {Liao}, {Licata}, {Lindstr{\o}m}, {Lister}, {Livanou},
  {Lobel}, {L{\'o}pez}, {Managau}, {Mann}, {Mantelet}, {Marchal}, {Marchant},
  {Marconi}, {Marinoni}, {Marschalk{\'o}}, {Marshall}, {Martino}, {Marton},
  {Mary}, {Massari}, {Matijevi{\v{c}}}, {Mazeh}, {McMillan}, {Messina},
  {Michalik}, {Millar}, {Molina}, {Molinaro}, {Moln{\'a}r}, {Montegriffo},
  {Mor}, {Morbidelli}, {Morel}, {Morris}, {Mulone}, {Muraveva}, {Musella},
  {Nelemans}, {Nicastro}, {Noval}, {O'Mullane}, {Ord{\'e}novic},
  {Ord{\'o}{\~n}ez-Blanco}, {Osborne}, {Pagani}, {Pagano}, {Pailler},
  {Palacin}, {Palaversa}, {Panahi}, {Pawlak}, {Piersimoni}, {Pineau}, {Plachy},
  {Plum}, {Poggio}, {Poujoulet}, {Pr{\v{s}}a}, {Pulone}, {Racero}, {Ragaini},
  {Rambaux}, {Ramos-Lerate}, {Regibo}, {Reyl{\'e}}, {Riclet}, {Ripepi}, {Riva},
  {Rivard}, {Rixon}, {Roegiers}, {Roelens}, {Romero-G{\'o}mez}, {Rowell},
  {Royer}, {Ruiz-Dern}, {Sadowski}, {Sagrist{\`a} Sell{\'e}s}, {Sahlmann},
  {Salgado}, {Salguero}, {Sanna}, {Santana-Ros}, {Sarasso}, {Savietto},
  {Schultheis}, {Sciacca}, {Segol}, {Segovia}, {S{\'e}gransan}, {Shih},
  {Siltala}, {Silva}, {Smart}, {Smith}, {Solano}, {Solitro}, {Sordo}, {Soria
  Nieto}, {Souchay}, {Spagna}, {Spoto}, {Stampa}, {Steele},
  {Steidelm{\"u}ller}, {Stephenson}, {Stoev}, {Suess}, {Surdej}, {Szabados},
  {Szegedi-Elek}, {Tapiador}, {Taris}, {Tauran}, {Taylor}, {Teixeira},
  {Terrett}, {Teyssandier}, {Thuillot}, {Titarenko}, {Torra Clotet}, {Turon},
  {Ulla}, {Utrilla}, {Uzzi}, {Vaillant}, {Valentini}, {Valette}, {van Elteren},
  {Van Hemelryck}, {van Leeuwen}, {Vaschetto}, {Vecchiato}, {Veljanoski},
  {Viala}, {Vicente}, {Vogt}, {von Essen}, {Voss}, {Votruba}, {Voutsinas},
  {Walmsley}, {Weiler}, {Wertz}, {Wevers}, {Wyrzykowski}, {Yoldas},
  {{\v{Z}}erjal}, {Ziaeepour}, {Zorec}, {Zschocke}, {Zucker}, {Zurbach}, \&
  {Zwitter}}]{GAIA2018}
{Gaia Collaboration}, {Brown}, A.~G.~A., {Vallenari}, A., {et~al.} 2018, \aap,
  616, A1, \dodoi{10.1051/0004-6361/201833051}

\bibitem[{{Gannon} {et~al.}(2022){Gannon}, {Forbes}, {Romanowsky},
  {Ferr{\'e}-Mateu}, {Couch}, {Brodie}, {Huang}, {Janssens}, \&
  {Okabe}}]{Gannon2022}
{Gannon}, J.~S., {Forbes}, D.~A., {Romanowsky}, A.~J., {et~al.} 2022, \mnras,
  510, 946, \dodoi{10.1093/mnras/stab3297}

\bibitem[{{Geha} {et~al.}(2017){Geha}, {Wechsler}, {Mao}, {Tollerud}, {Weiner},
  {Bernstein}, {Hoyle}, {Marchi}, {Marshall}, {Mu{\~n}oz}, \& {Lu}}]{SAGA-I}
{Geha}, M., {Wechsler}, R.~H., {Mao}, Y.-Y., {et~al.} 2017, \apj, 847, 4,
  \dodoi{10.3847/1538-4357/aa8626}

\bibitem[{{Graham} \& {Driver}(2005)}]{Graham2005}
{Graham}, A.~W., \& {Driver}, S.~P. 2005, \pasa, 22, 118,
  \dodoi{10.1071/AS05001}

\bibitem[{{Graham} \& {Guzm{\'a}n}(2003)}]{Graham2003}
{Graham}, A.~W., \& {Guzm{\'a}n}, R. 2003, \aj, 125, 2936,
  \dodoi{10.1086/374992}

\bibitem[{{Greco} {et~al.}(2018){Greco}, {Greene}, {Strauss}, {Macarthur},
  {Flowers}, {Goulding}, {Huang}, {Kim}, {Komiyama}, {Leauthaud}, {Leisman},
  {Lupton}, {Sif{\'o}n}, \& {Wang}}]{Greco2018}
{Greco}, J.~P., {Greene}, J.~E., {Strauss}, M.~A., {et~al.} 2018, \apj, 857,
  104, \dodoi{10.3847/1538-4357/aab842}

\bibitem[{{Greene} {et~al.}(2022){Greene}, {Greco}, {Goulding}, {Huang},
  {Kado-Fong}, {Danieli}, {Li}, {Kim}, {Komiyama}, {Leauthaud}, {MacArthur}, \&
  {Sifon}}]{Greene2022}
{Greene}, J.~E., {Greco}, J.~P., {Goulding}, A.~D., {et~al.} 2022, arXiv
  e-prints, arXiv:2204.11883.
\newblock \doarXiv{2204.11883}

\bibitem[{{Gu} {et~al.}(2018){Gu}, {Conroy}, {Law}, {van Dokkum}, {Yan},
  {Wake}, {Bundy}, {Merritt}, {Abraham}, {Zhang}, {Bershady}, {Bizyaev},
  {Brinkmann}, {Drory}, {Grabowski}, {Masters}, {Pan}, {Parejko}, {Weijmans},
  \& {Zhang}}]{Gu2018}
{Gu}, M., {Conroy}, C., {Law}, D., {et~al.} 2018, \apj, 859, 37,
  \dodoi{10.3847/1538-4357/aabbae}

\bibitem[{Harris {et~al.}(2020)Harris, Millman, van~der Walt, Gommers,
  Virtanen, Cournapeau, Wieser, Taylor, Berg, Smith, Kern, Picus, Hoyer, van
  Kerkwijk, Brett, Haldane, del R{'{\i}}o, Wiebe, Peterson,
  G{'{e}}rard-Marchant, Sheppard, Reddy, Weckesser, Abbasi, Gohlke, \&
  Oliphant}]{Numpy}
Harris, C.~R., Millman, K.~J., van~der Walt, S.~J., {et~al.} 2020, Nature, 585,
  357, \dodoi{10.1038/s41586-020-2649-2}

\bibitem[{{H{\"a}ussler} {et~al.}(2007){H{\"a}ussler}, {McIntosh}, {Barden},
  {Bell}, {Rix}, {Borch}, {Beckwith}, {Caldwell}, {Heymans}, {Jahnke}, {Jogee},
  {Koposov}, {Meisenheimer}, {S{\'a}nchez}, {Somerville}, {Wisotzki}, \&
  {Wolf}}]{Haussler2007}
{H{\"a}ussler}, B., {McIntosh}, D.~H., {Barden}, M., {et~al.} 2007, \apjs, 172,
  615, \dodoi{10.1086/518836}

\bibitem[{{Hearin} {et~al.}(2017){Hearin}, {Campbell}, {Tollerud}, {Behroozi},
  {Diemer}, {Goldbaum}, {Jennings}, {Leauthaud}, {Mao}, {More}, {Parejko},
  {Sinha}, {Sip{\"o}cz}, \& {Zentner}}]{Hearin2017}
{Hearin}, A.~P., {Campbell}, D., {Tollerud}, E., {et~al.} 2017, \aj, 154, 190,
  \dodoi{10.3847/1538-3881/aa859f}

\bibitem[{{Herrmann} {et~al.}(2016){Herrmann}, {Hunter}, {Zhang}, \&
  {Elmegreen}}]{Herrmann2016}
{Herrmann}, K.~A., {Hunter}, D.~A., {Zhang}, H.-X., \& {Elmegreen}, B.~G. 2016,
  \aj, 152, 177, \dodoi{10.3847/0004-6256/152/6/177}

\bibitem[{{Huang} {et~al.}(2018{\natexlab{a}}){Huang}, {Leauthaud}, {Greene},
  {Bundy}, {Lin}, {Tanaka}, {Miyazaki}, \& {Komiyama}}]{Huang2018}
{Huang}, S., {Leauthaud}, A., {Greene}, J.~E., {et~al.} 2018{\natexlab{a}},
  \mnras, 475, 3348, \dodoi{10.1093/mnras/stx3200}

\bibitem[{{Huang} {et~al.}(2018{\natexlab{b}}){Huang}, {Leauthaud}, {Murata},
  {Bosch}, {Price}, {Lupton}, {Mandelbaum}, {Lackner}, {Bickerton}, {Miyazaki},
  {Coupon}, \& {Tanaka}}]{Huang2018synpipe}
{Huang}, S., {Leauthaud}, A., {Murata}, R., {et~al.} 2018{\natexlab{b}}, \pasj,
  70, S6, \dodoi{10.1093/pasj/psx126}

\bibitem[{{Hunter}(2007)}]{matplotlib}
{Hunter}, J.~D. 2007, Computing in Science Engineering, 9, 90,
  \dodoi{10.1109/MCSE.2007.55}

\bibitem[{{Impey} {et~al.}(1988){Impey}, {Bothun}, \& {Malin}}]{Impey1988}
{Impey}, C., {Bothun}, G., \& {Malin}, D. 1988, \apj, 330, 634,
  \dodoi{10.1086/166500}

\bibitem[{{Into} \& {Portinari}(2013)}]{Into2013}
{Into}, T., \& {Portinari}, L. 2013, \mnras, 430, 2715,
  \dodoi{10.1093/mnras/stt071}

\bibitem[{{Irwin}(1985)}]{Irwin1985}
{Irwin}, M.~J. 1985, \mnras, 214, 575, \dodoi{10.1093/mnras/214.4.575}

\bibitem[{{Ivezi{\'c}} {et~al.}(2019){Ivezi{\'c}}, {Kahn}, {Tyson}, {Abel},
  {Acosta}, {Allsman}, {Alonso}, {AlSayyad}, {Anderson}, {Andrew}, {Angel},
  {Angeli}, {Ansari}, {Antilogus}, {Araujo}, {Armstrong}, {Arndt}, {Astier},
  {Aubourg}, {Auza}, {Axelrod}, {Bard}, {Barr}, {Barrau}, {Bartlett}, {Bauer},
  {Bauman}, {Baumont}, {Bechtol}, {Bechtol}, {Becker}, {Becla}, {Beldica},
  {Bellavia}, {Bianco}, {Biswas}, {Blanc}, {Blazek}, {Blandford}, {Bloom},
  {Bogart}, {Bond}, {Booth}, {Borgland}, {Borne}, {Bosch}, {Boutigny},
  {Brackett}, {Bradshaw}, {Brandt}, {Brown}, {Bullock}, {Burchat}, {Burke},
  {Cagnoli}, {Calabrese}, {Callahan}, {Callen}, {Carlin}, {Carlson},
  {Chandrasekharan}, {Charles-Emerson}, {Chesley}, {Cheu}, {Chiang}, {Chiang},
  {Chirino}, {Chow}, {Ciardi}, {Claver}, {Cohen-Tanugi}, {Cockrum}, {Coles},
  {Connolly}, {Cook}, {Cooray}, {Covey}, {Cribbs}, {Cui}, {Cutri}, {Daly},
  {Daniel}, {Daruich}, {Daubard}, {Daues}, {Dawson}, {Delgado}, {Dellapenna},
  {de Peyster}, {de Val-Borro}, {Digel}, {Doherty}, {Dubois},
  {Dubois-Felsmann}, {Durech}, {Economou}, {Eifler}, {Eracleous}, {Emmons},
  {Fausti Neto}, {Ferguson}, {Figueroa}, {Fisher-Levine}, {Focke}, {Foss},
  {Frank}, {Freemon}, {Gangler}, {Gawiser}, {Geary}, {Gee}, {Geha}, {Gessner},
  {Gibson}, {Gilmore}, {Glanzman}, {Glick}, {Goldina}, {Goldstein}, {Goodenow},
  {Graham}, {Gressler}, {Gris}, {Guy}, {Guyonnet}, {Haller}, {Harris},
  {Hascall}, {Haupt}, {Hernandez}, {Herrmann}, {Hileman}, {Hoblitt}, {Hodgson},
  {Hogan}, {Howard}, {Huang}, {Huffer}, {Ingraham}, {Innes}, {Jacoby}, {Jain},
  {Jammes}, {Jee}, {Jenness}, {Jernigan}, {Jevremovi{\'c}}, {Johns}, {Johnson},
  {Johnson}, {Jones}, {Juramy-Gilles}, {Juri{\'c}}, {Kalirai}, {Kallivayalil},
  {Kalmbach}, {Kantor}, {Karst}, {Kasliwal}, {Kelly}, {Kessler}, {Kinnison},
  {Kirkby}, {Knox}, {Kotov}, {Krabbendam}, {Krughoff}, {Kub{\'a}nek},
  {Kuczewski}, {Kulkarni}, {Ku}, {Kurita}, {Lage}, {Lambert}, {Lange},
  {Langton}, {Le Guillou}, {Levine}, {Liang}, {Lim}, {Lintott}, {Long},
  {Lopez}, {Lotz}, {Lupton}, {Lust}, {MacArthur}, {Mahabal}, {Mandelbaum},
  {Markiewicz}, {Marsh}, {Marshall}, {Marshall}, {May}, {McKercher}, {McQueen},
  {Meyers}, {Migliore}, {Miller}, {Mills}, {Miraval}, {Moeyens}, {Moolekamp},
  {Monet}, {Moniez}, {Monkewitz}, {Montgomery}, {Morrison}, {Mueller},
  {Muller}, {Mu{\~n}oz Arancibia}, {Neill}, {Newbry}, {Nief}, {Nomerotski},
  {Nordby}, {O'Connor}, {Oliver}, {Olivier}, {Olsen}, {O'Mullane}, {Ortiz},
  {Osier}, {Owen}, {Pain}, {Palecek}, {Parejko}, {Parsons}, {Pease},
  {Peterson}, {Peterson}, {Petravick}, {Libby Petrick}, {Petry},
  {Pierfederici}, {Pietrowicz}, {Pike}, {Pinto}, {Plante}, {Plate}, {Plutchak},
  {Price}, {Prouza}, {Radeka}, {Rajagopal}, {Rasmussen}, {Regnault}, {Reil},
  {Reiss}, {Reuter}, {Ridgway}, {Riot}, {Ritz}, {Robinson}, {Roby}, {Roodman},
  {Rosing}, {Roucelle}, {Rumore}, {Russo}, {Saha}, {Sassolas}, {Schalk},
  {Schellart}, {Schindler}, {Schmidt}, {Schneider}, {Schneider}, {Schoening},
  {Schumacher}, {Schwamb}, {Sebag}, {Selvy}, {Sembroski}, {Seppala}, {Serio},
  {Serrano}, {Shaw}, {Shipsey}, {Sick}, {Silvestri}, {Slater}, {Smith},
  {Smith}, {Sobhani}, {Soldahl}, {Storrie-Lombardi}, {Stover}, {Strauss},
  {Street}, {Stubbs}, {Sullivan}, {Sweeney}, {Swinbank}, {Szalay}, {Takacs},
  {Tether}, {Thaler}, {Thayer}, {Thomas}, {Thornton}, {Thukral}, {Tice},
  {Trilling}, {Turri}, {Van Berg}, {Vanden Berk}, {Vetter}, {Virieux},
  {Vucina}, {Wahl}, {Walkowicz}, {Walsh}, {Walter}, {Wang}, {Wang}, {Warner},
  {Wiecha}, {Willman}, {Winters}, {Wittman}, {Wolff}, {Wood-Vasey}, {Wu},
  {Xin}, {Yoachim}, \& {Zhan}}]{LSST2019}
{Ivezi{\'c}}, {\v{Z}}., {Kahn}, S.~M., {Tyson}, J.~A., {et~al.} 2019, \apj,
  873, 111, \dodoi{10.3847/1538-4357/ab042c}

\bibitem[{{Janssens} {et~al.}(2017){Janssens}, {Abraham}, {Brodie}, {Forbes},
  {Romanowsky}, \& {van Dokkum}}]{Janssens2017}
{Janssens}, S., {Abraham}, R., {Brodie}, J., {et~al.} 2017, \apjl, 839, L17,
  \dodoi{10.3847/2041-8213/aa667d}

\bibitem[{{Jiang} {et~al.}(2019){Jiang}, {Dekel}, {Freundlich}, {Romanowsky},
  {Dutton}, {Macci{\`o}}, \& {Di Cintio}}]{Jiang2019}
{Jiang}, F., {Dekel}, A., {Freundlich}, J., {et~al.} 2019, \mnras, 487, 5272,
  \dodoi{10.1093/mnras/stz1499}

\bibitem[{Jones {et~al.}(2001)Jones, Oliphant, Peterson, {et~al.}}]{scipy}
Jones, E., Oliphant, T., Peterson, P., {et~al.} 2001, {SciPy}: Open source
  scientific tools for {Python}.
\newblock \url{http://www.scipy.org/}

\bibitem[{{Juri{\'c}} {et~al.}(2017){Juri{\'c}}, {Kantor}, {Lim}, {Lupton},
  {Dubois-Felsmann}, {Jenness}, {Axelrod}, {Aleksi{\'c}}, {Allsman},
  {AlSayyad}, {Alt}, {Armstrong}, {Basney}, {Becker}, {Becla}, {Biswas},
  {Bosch}, {Boutigny}, {Kind}, {Ciardi}, {Connolly}, {Daniel}, {Daues},
  {Economou}, {Chiang}, {Fausti}, {Fisher-Levine}, {Freemon}, {Gris},
  {Hernandez}, {Hoblitt}, {Ivezi{\'c}}, {Jammes}, {Jevremovi{\'c}}, {Jones},
  {Kalmbach}, {Kasliwal}, {Krughoff}, {Lurie}, {Lust}, {MacArthur}, {Melchior},
  {Moeyens}, {Nidever}, {Owen}, {Parejko}, {Peterson}, {Petravick},
  {Pietrowicz}, {Price}, {Reiss}, {Shaw}, {Sick}, {Slater}, {Strauss},
  {Sullivan}, {Swinbank}, {Van Dyk}, {Vuj{\v{c}}i{\'c}}, {Withers}, \&
  {Yoachim}}]{LSST-pipeline}
{Juri{\'c}}, M., {Kantor}, J., {Lim}, K.~T., {et~al.} 2017, in Astronomical
  Society of the Pacific Conference Series, Vol. 512, Astronomical Data
  Analysis Software and Systems XXV, ed. N.~P.~F. {Lorente}, K.~{Shortridge},
  \& R.~{Wayth}, 279.
\newblock \doarXiv{1512.07914}

\bibitem[{{Kado-Fong} {et~al.}(2022){Kado-Fong}, {Greene}, {Huang}, \&
  {Goulding}}]{Kado-Fong2022}
{Kado-Fong}, E., {Greene}, J.~E., {Huang}, S., \& {Goulding}, A. 2022, \apj,
  941, 11, \dodoi{10.3847/1538-4357/ac9964}

\bibitem[{{Kado-Fong} {et~al.}(2018){Kado-Fong}, {Greene}, {Hendel},
  {Price-Whelan}, {Greco}, {Goulding}, {Huang}, {Johnston}, {Komiyama}, {Lee},
  {Lust}, {Strauss}, \& {Tanaka}}]{Kado-Fong2018}
{Kado-Fong}, E., {Greene}, J.~E., {Hendel}, D., {et~al.} 2018, \apj, 866, 103,
  \dodoi{10.3847/1538-4357/aae0f0}

\bibitem[{{Kado-Fong} {et~al.}(2021){Kado-Fong}, {Petrescu}, {Mohammad},
  {Greco}, {Greene}, {Adams}, {Huang}, {Leisman}, {Munshi}, {Tanoglidis}, \&
  {Van Nest}}]{Kado-Fong2021}
{Kado-Fong}, E., {Petrescu}, M., {Mohammad}, M., {et~al.} 2021, \apj, 920, 72,
  \dodoi{10.3847/1538-4357/ac15f0}

\bibitem[{{Kadowaki} {et~al.}(2021){Kadowaki}, {Zaritsky}, {Donnerstein}, {RS},
  {Karunakaran}, \& {Spekkens}}]{Kadowaki2021}
{Kadowaki}, J., {Zaritsky}, D., {Donnerstein}, R.~L., {et~al.} 2021, \apj, 923,
  257, \dodoi{10.3847/1538-4357/ac2948}

\bibitem[{{Karachentsev} {et~al.}(2013){Karachentsev}, {Makarov}, \&
  {Kaisina}}]{Karachentsev2013}
{Karachentsev}, I.~D., {Makarov}, D.~I., \& {Kaisina}, E.~I. 2013, \aj, 145,
  101, \dodoi{10.1088/0004-6256/145/4/101}

\bibitem[{{Karunakaran} \& {Zaritsky}(2022)}]{Karunakaran2022b}
{Karunakaran}, A., \& {Zaritsky}, D. 2022, arXiv e-prints, arXiv:2210.00009.
\newblock \doarXiv{2210.00009}

\bibitem[{{Keim} {et~al.}(2022){Keim}, {Dokkum}, {Danieli}, {Lokhorst}, {Li},
  {Shen}, {Abraham}, {Chen}, {Gilhuly}, {Liu}, {Merritt}, {Miller}, {Pasha}, \&
  {Polzin}}]{Keim2022}
{Keim}, M.~A., {Dokkum}, P.~v., {Danieli}, S., {et~al.} 2022, \apj, 935, 160,
  \dodoi{10.3847/1538-4357/ac7dab}

\bibitem[{{Koda} {et~al.}(2015){Koda}, {Yagi}, {Yamanoi}, \&
  {Komiyama}}]{Koda2015}
{Koda}, J., {Yagi}, M., {Yamanoi}, H., \& {Komiyama}, Y. 2015, \apjl, 807, L2,
  \dodoi{10.1088/2041-8205/807/1/L2}

\bibitem[{{Kong} {et~al.}(2022){Kong}, {Kaplinghat}, {Yu}, {Fraternali}, \&
  {Mancera Pi{\~n}a}}]{Kong2022}
{Kong}, D., {Kaplinghat}, M., {Yu}, H.-B., {Fraternali}, F., \& {Mancera
  Pi{\~n}a}, P.~E. 2022, \apj, 936, 166, \dodoi{10.3847/1538-4357/ac8875}

\bibitem[{{La Marca} {et~al.}(2022){La Marca}, {Iodice}, {Cantiello}, {Forbes},
  {Rejkuba}, {Hilker}, {Arnaboldi}, {Greggio}, {Spiniello}, {Mieske},
  {Venhola}, {Spavone}, {D'Ago}, {Raj}, {Ragusa}, {Mirabile}, {Rampazzo},
  {Peletier}, {Paolillo}, {Challapa}, \& {Schipani}}]{LaMarca2022}
{La Marca}, A., {Iodice}, E., {Cantiello}, M., {et~al.} 2022, \aap, 665, A105,
  \dodoi{10.1051/0004-6361/202142367}

\bibitem[{{Lang} {et~al.}(2016){Lang}, {Hogg}, \& {Mykytyn}}]{Lang2016}
{Lang}, D., {Hogg}, D.~W., \& {Mykytyn}, D. 2016, {The Tractor: Probabilistic
  astronomical source detection and measurement}, Astrophysics Source Code
  Library, record ascl:1604.008.
\newblock \doeprint{1604.008}

\bibitem[{{Lange} {et~al.}(2015){Lange}, {Driver}, {Robotham}, {Kelvin},
  {Graham}, {Alpaslan}, {Andrews}, {Baldry}, {Bamford}, {Bland-Hawthorn},
  {Brough}, {Cluver}, {Conselice}, {Davies}, {Haeussler}, {Konstantopoulos},
  {Loveday}, {Moffett}, {Norberg}, {Phillipps}, {Taylor},
  {L{\'o}pez-S{\'a}nchez}, \& {Wilkins}}]{Lange2015}
{Lange}, R., {Driver}, S.~P., {Robotham}, A. S.~G., {et~al.} 2015, \mnras, 447,
  2603, \dodoi{10.1093/mnras/stu2467}

\bibitem[{{Lee} {et~al.}(2017){Lee}, {Kang}, {Lee}, \& {Jang}}]{Lee2017}
{Lee}, M.~G., {Kang}, J., {Lee}, J.~H., \& {Jang}, I.~S. 2017, \apj, 844, 157,
  \dodoi{10.3847/1538-4357/aa78fb}

\bibitem[{{Leisman} {et~al.}(2017){Leisman}, {Haynes}, {Janowiecki},
  {Hallenbeck}, {J{\'o}zsa}, {Giovanelli}, {Adams}, {Bernal Neira}, {Cannon},
  {Janesh}, {Rhode}, \& {Salzer}}]{Leisman2017}
{Leisman}, L., {Haynes}, M.~P., {Janowiecki}, S., {et~al.} 2017, \apj, 842,
  133, \dodoi{10.3847/1538-4357/aa7575}

\bibitem[{{Li} {et~al.}(2023){Li}, {Greene}, {Greco}, {Beaton}, {Danieli},
  {Goulding}, {Huang}, \& {Kado-Fong}}]{Li2023}
{Li}, J., {Greene}, J.~E., {Greco}, J., {et~al.} 2023, arXiv e-prints,
  arXiv:2302.14108, \dodoi{10.48550/arXiv.2302.14108}

\bibitem[{{Li} {et~al.}(2022){Li}, {Huang}, {Leauthaud}, {Moustakas},
  {Danieli}, {Greene}, {Abraham}, {Ardila}, {Kado-Fong}, {Lokhorst}, {Lupton},
  \& {Price}}]{Li2021}
{Li}, J., {Huang}, S., {Leauthaud}, A., {et~al.} 2022, \mnras, 515, 5335,
  \dodoi{10.1093/mnras/stac2121}

\bibitem[{{Liao} {et~al.}(2019){Liao}, {Gao}, {Frenk}, {Grand}, {Guo},
  {G{\'o}mez}, {Marinacci}, {Pakmor}, {Shao}, \& {Springel}}]{Liao2019}
{Liao}, S., {Gao}, L., {Frenk}, C.~S., {et~al.} 2019, \mnras, 490, 5182,
  \dodoi{10.1093/mnras/stz2969}

\bibitem[{{Licquia} \& {Newman}(2015)}]{Licquia2015}
{Licquia}, T.~C., \& {Newman}, J.~A. 2015, \apj, 806, 96,
  \dodoi{10.1088/0004-637X/806/1/96}

\bibitem[{{Lim} {et~al.}(2020){Lim}, {C{\^o}t{\'e}}, {Peng}, {Ferrarese},
  {Roediger}, {Durrell}, {Mihos}, {Wang}, {Gwyn}, {Cuillandre}, {Liu},
  {S{\'a}nchez-Janssen}, {Toloba}, {Sales}, {Guhathakurta}, {Lan{\c{c}}on}, \&
  {Puzia}}]{Lim2020}
{Lim}, S., {C{\^o}t{\'e}}, P., {Peng}, E.~W., {et~al.} 2020, \apj, 899, 69,
  \dodoi{10.3847/1538-4357/aba433}

\bibitem[{{Liu} {et~al.}(2022){Liu}, {Abraham}, {Gilhuly}, {van Dokkum},
  {Martin}, {Li}, {Greco}, {Lokhorst}, {Chen}, {Danieli}, {Keim}, {Merritt},
  {Miller}, {Pasha}, {Polzin}, {Shen}, \& {Zhang}}]{Liu2022}
{Liu}, Q., {Abraham}, R., {Gilhuly}, C., {et~al.} 2022, \apj, 925, 219,
  \dodoi{10.3847/1538-4357/ac32c6}

\bibitem[{{Lotz} {et~al.}(2004){Lotz}, {Primack}, \& {Madau}}]{Lotz2004}
{Lotz}, J.~M., {Primack}, J., \& {Madau}, P. 2004, \aj, 128, 163,
  \dodoi{10.1086/421849}

\bibitem[{{Lotz} {et~al.}(2008){Lotz}, {Davis}, {Faber}, {Guhathakurta},
  {Gwyn}, {Huang}, {Koo}, {Le Floc'h}, {Lin}, {Newman}, {Noeske}, {Papovich},
  {Willmer}, {Coil}, {Conselice}, {Cooper}, {Hopkins}, {Metevier}, {Primack},
  {Rieke}, \& {Weiner}}]{Lotz2008}
{Lotz}, J.~M., {Davis}, M., {Faber}, S.~M., {et~al.} 2008, \apj, 672, 177,
  \dodoi{10.1086/523659}

\bibitem[{{LSST Science Collaboration} {et~al.}(2009){LSST Science
  Collaboration}, {Abell}, {Allison}, {Anderson}, {Andrew}, {Angel}, {Armus},
  {Arnett}, {Asztalos}, {Axelrod}, {Bailey}, {Ballantyne}, {Bankert},
  {Barkhouse}, {Barr}, {Barrientos}, {Barth}, {Bartlett}, {Becker}, {Becla},
  {Beers}, {Bernstein}, {Biswas}, {Blanton}, {Bloom}, {Bochanski}, {Boeshaar},
  {Borne}, {Bradac}, {Brandt}, {Bridge}, {Brown}, {Brunner}, {Bullock},
  {Burgasser}, {Burge}, {Burke}, {Cargile}, {Chandrasekharan}, {Chartas},
  {Chesley}, {Chu}, {Cinabro}, {Claire}, {Claver}, {Clowe}, {Connolly}, {Cook},
  {Cooke}, {Cooray}, {Covey}, {Culliton}, {de Jong}, {de Vries}, {Debattista},
  {Delgado}, {Dell'Antonio}, {Dhital}, {Di Stefano}, {Dickinson}, {Dilday},
  {Djorgovski}, {Dobler}, {Donalek}, {Dubois-Felsmann}, {Durech},
  {Eliasdottir}, {Eracleous}, {Eyer}, {Falco}, {Fan}, {Fassnacht}, {Ferguson},
  {Fernandez}, {Fields}, {Finkbeiner}, {Figueroa}, {Fox}, {Francke}, {Frank},
  {Frieman}, {Fromenteau}, {Furqan}, {Galaz}, {Gal-Yam}, {Garnavich},
  {Gawiser}, {Geary}, {Gee}, {Gibson}, {Gilmore}, {Grace}, {Green}, {Gressler},
  {Grillmair}, {Habib}, {Haggerty}, {Hamuy}, {Harris}, {Hawley}, {Heavens},
  {Hebb}, {Henry}, {Hileman}, {Hilton}, {Hoadley}, {Holberg}, {Holman},
  {Howell}, {Infante}, {Ivezic}, {Jacoby}, {Jain}, {R}, {Jedicke}, {Jee},
  {Garrett Jernigan}, {Jha}, {Johnston}, {Jones}, {Juric}, {Kaasalainen},
  {Styliani}, {Kafka}, {Kahn}, {Kaib}, {Kalirai}, {Kantor}, {Kasliwal},
  {Keeton}, {Kessler}, {Knezevic}, {Kowalski}, {Krabbendam}, {Krughoff},
  {Kulkarni}, {Kuhlman}, {Lacy}, {Lepine}, {Liang}, {Lien}, {Lira}, {Long},
  {Lorenz}, {Lotz}, {Lupton}, {Lutz}, {Macri}, {Mahabal}, {Mandelbaum},
  {Marshall}, {May}, {McGehee}, {Meadows}, {Meert}, {Milani}, {Miller},
  {Miller}, {Mills}, {Minniti}, {Monet}, {Mukadam}, {Nakar}, {Neill}, {Newman},
  {Nikolaev}, {Nordby}, {O'Connor}, {Oguri}, {Oliver}, {Olivier}, {Olsen},
  {Olsen}, {Olszewski}, {Oluseyi}, {Padilla}, {Parker}, {Pepper}, {Peterson},
  {Petry}, {Pinto}, {Pizagno}, {Popescu}, {Prsa}, {Radcka}, {Raddick},
  {Rasmussen}, {Rau}, {Rho}, {Rhoads}, {Richards}, {Ridgway}, {Robertson},
  {Roskar}, {Saha}, {Sarajedini}, {Scannapieco}, {Schalk}, {Schindler},
  {Schmidt}, {Schmidt}, {Schneider}, {Schumacher}, {Scranton}, {Sebag},
  {Seppala}, {Shemmer}, {Simon}, {Sivertz}, {Smith}, {Allyn Smith}, {Smith},
  {Spitz}, {Stanford}, {Stassun}, {Strader}, {Strauss}, {Stubbs}, {Sweeney},
  {Szalay}, {Szkody}, {Takada}, {Thorman}, {Trilling}, {Trimble}, {Tyson}, {Van
  Berg}, {Vanden Berk}, {VanderPlas}, {Verde}, {Vrsnak}, {Walkowicz},
  {Wandelt}, {Wang}, {Wang}, {Warner}, {Wechsler}, {West}, {Wiecha},
  {Williams}, {Willman}, {Wittman}, {Wolff}, {Wood-Vasey}, {Wozniak}, {Young},
  {Zentner}, \& {Zhan}}]{lsst2009}
{LSST Science Collaboration}, {Abell}, P.~A., {Allison}, J., {et~al.} 2009,
  arXiv e-prints, arXiv:0912.0201.
\newblock \doarXiv{0912.0201}

\bibitem[{{Mancera Pi{\~n}a} {et~al.}(2019{\natexlab{a}}){Mancera Pi{\~n}a},
  {Aguerri}, {Peletier}, {Venhola}, {Trager}, \& {Choque
  Challapa}}]{ManceraPina2019a}
{Mancera Pi{\~n}a}, P.~E., {Aguerri}, J.~A.~L., {Peletier}, R.~F., {et~al.}
  2019{\natexlab{a}}, \mnras, 485, 1036, \dodoi{10.1093/mnras/stz238}

\bibitem[{{Mancera Pi{\~n}a} {et~al.}(2022{\natexlab{a}}){Mancera Pi{\~n}a},
  {Fraternali}, {Oosterloo}, {Adams}, {di Teodoro}, {Bacchini}, \&
  {Iorio}}]{Mancera2022b}
{Mancera Pi{\~n}a}, P.~E., {Fraternali}, F., {Oosterloo}, T., {et~al.}
  2022{\natexlab{a}}, \mnras, 514, 3329, \dodoi{10.1093/mnras/stac1508}

\bibitem[{{Mancera Pi{\~n}a} {et~al.}(2022{\natexlab{b}}){Mancera Pi{\~n}a},
  {Fraternali}, {Oosterloo}, {Adams}, {Oman}, \& {Leisman}}]{ManceraPina2022}
---. 2022{\natexlab{b}}, \mnras, 512, 3230, \dodoi{10.1093/mnras/stab3491}

\bibitem[{{Mancera Pi{\~n}a} {et~al.}(2018){Mancera Pi{\~n}a}, {Peletier},
  {Aguerri}, {Venhola}, {Trager}, \& {Choque Challapa}}]{ManceraPina2018}
{Mancera Pi{\~n}a}, P.~E., {Peletier}, R.~F., {Aguerri}, J.~A.~L., {et~al.}
  2018, \mnras, 481, 4381, \dodoi{10.1093/mnras/sty2574}

\bibitem[{{Mancera Pi{\~n}a} {et~al.}(2019{\natexlab{b}}){Mancera Pi{\~n}a},
  {Fraternali}, {Adams}, {Marasco}, {Oosterloo}, {Oman}, {Leisman}, {di
  Teodoro}, {Posti}, {Battipaglia}, {Cannon}, {Gault}, {Haynes}, {Janowiecki},
  {McAllan}, {Pagel}, {Reiter}, {Rhode}, {Salzer}, \&
  {Smith}}]{ManceraPina2019b}
{Mancera Pi{\~n}a}, P.~E., {Fraternali}, F., {Adams}, E. A.~K., {et~al.}
  2019{\natexlab{b}}, \apjl, 883, L33, \dodoi{10.3847/2041-8213/ab40c7}

\bibitem[{{Mancera Pi{\~n}a} {et~al.}(2020){Mancera Pi{\~n}a}, {Fraternali},
  {Oman}, {Adams}, {Bacchini}, {Marasco}, {Oosterloo}, {Pezzulli}, {Posti},
  {Leisman}, {Cannon}, {di Teodoro}, {Gault}, {Haynes}, {Reiter}, {Rhode},
  {Salzer}, \& {Smith}}]{ManceraPina2020}
{Mancera Pi{\~n}a}, P.~E., {Fraternali}, F., {Oman}, K.~A., {et~al.} 2020,
  \mnras, 495, 3636, \dodoi{10.1093/mnras/staa1256}

\bibitem[{{Mao} {et~al.}(2021){Mao}, {Geha}, {Wechsler}, {Weiner}, {Tollerud},
  {Nadler}, \& {Kallivayalil}}]{SAGA-II}
{Mao}, Y.-Y., {Geha}, M., {Wechsler}, R.~H., {et~al.} 2021, \apj, 907, 85,
  \dodoi{10.3847/1538-4357/abce58}

\bibitem[{{Martin} {et~al.}(2019){Martin}, {Kaviraj}, {Laigle}, {Devriendt},
  {Jackson}, {Peirani}, {Dubois}, {Pichon}, \& {Slyz}}]{Martin2019}
{Martin}, G., {Kaviraj}, S., {Laigle}, C., {et~al.} 2019, \mnras, 485, 796,
  \dodoi{10.1093/mnras/stz356}

\bibitem[{{McConnachie}(2012)}]{McConnachie2012}
{McConnachie}, A.~W. 2012, \aj, 144, 4, \dodoi{10.1088/0004-6256/144/1/4}

\bibitem[{{McGaugh} {et~al.}(1995){McGaugh}, {Bothun}, \&
  {Schombert}}]{McGaugh1995}
{McGaugh}, S.~S., {Bothun}, G.~D., \& {Schombert}, J.~M. 1995, \aj, 110, 573,
  \dodoi{10.1086/117543}

\bibitem[{{Melchior} {et~al.}(2019){Melchior}, {Joseph}, \&
  {Moolekamp}}]{Melchior2019}
{Melchior}, P., {Joseph}, R., \& {Moolekamp}, F. 2019, arXiv e-prints,
  arXiv:1910.10094.
\newblock \doarXiv{1910.10094}

\bibitem[{{Melchior} {et~al.}(2021){Melchior}, {Joseph}, {Sanchez}, {MacCrann},
  \& {Gruen}}]{Melchior2021}
{Melchior}, P., {Joseph}, R., {Sanchez}, J., {MacCrann}, N., \& {Gruen}, D.
  2021, Nature Reviews Physics, 3, 712, \dodoi{10.1038/s42254-021-00353-y}

\bibitem[{{Melchior} {et~al.}(2018){Melchior}, {Moolekamp}, {Jerdee},
  {Armstrong}, {Sun}, {Bosch}, \& {Lupton}}]{Melchior2018}
{Melchior}, P., {Moolekamp}, F., {Jerdee}, M., {et~al.} 2018, Astronomy and
  Computing, 24, 129, \dodoi{10.1016/j.ascom.2018.07.001}

\bibitem[{{Mihos} {et~al.}(2015){Mihos}, {Durrell}, {Ferrarese}, {Feldmeier},
  {C{\^o}t{\'e}}, {Peng}, {Harding}, {Liu}, {Gwyn}, \&
  {Cuillandre}}]{Mihos2015}
{Mihos}, J.~C., {Durrell}, P.~R., {Ferrarese}, L., {et~al.} 2015, \apjl, 809,
  L21, \dodoi{10.1088/2041-8205/809/2/L21}

\bibitem[{{Miller} {et~al.}(2019){Miller}, {van Dokkum}, {Mowla}, \& {van der
  Wel}}]{Miller2019}
{Miller}, T.~B., {van Dokkum}, P., {Mowla}, L., \& {van der Wel}, A. 2019,
  \apjl, 872, L14, \dodoi{10.3847/2041-8213/ab0380}

\bibitem[{{Misgeld} \& {Hilker}(2011)}]{Misgeld2011}
{Misgeld}, I., \& {Hilker}, M. 2011, \mnras, 414, 3699,
  \dodoi{10.1111/j.1365-2966.2011.18669.x}

\bibitem[{{Misgeld} {et~al.}(2008){Misgeld}, {Mieske}, \&
  {Hilker}}]{Misgeld2008}
{Misgeld}, I., {Mieske}, S., \& {Hilker}, M. 2008, \aap, 486, 697,
  \dodoi{10.1051/0004-6361:200810014}

\bibitem[{{Miyazaki} {et~al.}(2012){Miyazaki}, Komiyama, {Nakaya}, {Kamata},
  {Doi}, Takashi, {Karoji}, {Furusawa}, {Kawanomoto}, {Morokuma}, {Ishizuka},
  {Nariai}, \& et~al.}]{Miyazaki2012}
{Miyazaki}, S., Komiyama, Y., {Nakaya}, H., {et~al.} 2012, Proc.SPIE, 8446,
  8446 , \dodoi{10.1117/12.926844}

\bibitem[{{Miyazaki} {et~al.}(2018){Miyazaki}, {Komiyama}, {Kawanomoto}, {Doi},
  {Furusawa}, {Hamana}, {Hayashi}, {Ikeda}, {Kamata}, {Karoji}, {Koike},
  {Kurakami}, {Miyama}, {Morokuma}, {Nakata}, {Namikawa}, {Nakaya}, {Nariai},
  {Obuchi}, {Oishi}, {Okada}, {Okura}, {Tait}, {Takata}, {Tanaka}, {Tanaka},
  {Terai}, {Tomono}, {Uraguchi}, {Usuda}, {Utsumi}, {Yamada}, {Yamanoi},
  {Aihara}, {Fujimori}, {Mineo}, {Miyatake}, {Oguri}, {Uchida}, {Tanaka},
  {Yasuda}, {Takada}, {Murayama}, {Nishizawa}, {Sugiyama}, {Chiba}, {Futamase},
  {Wang}, {Chen}, {Ho}, {Liaw}, {Chiu}, {Ho}, {Lai}, {Lee}, {Jeng}, {Iwamura},
  {Armstrong}, {Bickerton}, {Bosch}, {Gunn}, {Lupton}, {Loomis}, {Price},
  {Smith}, {Strauss}, {Turner}, {Suzuki}, {Miyazaki}, {Muramatsu}, {Yamamoto},
  {Endo}, {Ezaki}, {Ito}, {Kawaguchi}, {Sofuku}, {Taniike}, {Akutsu}, {Dojo},
  {Kasumi}, {Matsuda}, {Imoto}, {Miwa}, {Suzuki}, {Takeshi}, \&
  {Yokota}}]{Miyazaki2018}
{Miyazaki}, S., {Komiyama}, Y., {Kawanomoto}, S., {et~al.} 2018, \pasj, 70, S1,
  \dodoi{10.1093/pasj/psx063}

\bibitem[{{Moster} {et~al.}(2013){Moster}, {Naab}, \& {White}}]{Moster2013}
{Moster}, B.~P., {Naab}, T., \& {White}, S. D.~M. 2013, \mnras, 428, 3121,
  \dodoi{10.1093/mnras/sts261}

\bibitem[{{Mowla} {et~al.}(2019){Mowla}, {van der Wel}, {van Dokkum}, \&
  {Miller}}]{Mowla2019}
{Mowla}, L., {van der Wel}, A., {van Dokkum}, P., \& {Miller}, T.~B. 2019,
  \apjl, 872, L13, \dodoi{10.3847/2041-8213/ab0379}

\bibitem[{{Mowla} {et~al.}(2017){Mowla}, {van Dokkum}, {Merritt}, {Abraham},
  {Yagi}, \& {Koda}}]{Mowla2017}
{Mowla}, L., {van Dokkum}, P., {Merritt}, A., {et~al.} 2017, \apj, 851, 27,
  \dodoi{10.3847/1538-4357/aa961b}

\bibitem[{{Mu{\~n}oz} {et~al.}(2015){Mu{\~n}oz}, {Eigenthaler}, {Puzia},
  {Taylor}, {Ordenes-Brice{\~n}o}, {Alamo-Mart{\'\i}nez}, {Ribbeck},
  {{\'A}ngel}, {Capaccioli}, {C{\^o}t{\'e}}, {Ferrarese}, {Galaz}, {Hempel},
  {Hilker}, {Jord{\'a}n}, {Lan{\c{c}}on}, {Mieske}, {Paolillo}, {Richtler},
  {S{\'a}nchez-Janssen}, \& {Zhang}}]{Munoz2015}
{Mu{\~n}oz}, R.~P., {Eigenthaler}, P., {Puzia}, T.~H., {et~al.} 2015, \apjl,
  813, L15, \dodoi{10.1088/2041-8205/813/1/L15}

\bibitem[{{Nashimoto} {et~al.}(2022){Nashimoto}, {Tanaka}, {Chiba}, {Hayashi},
  {Komiyama}, \& {Okamoto}}]{Nashimoto2022}
{Nashimoto}, M., {Tanaka}, M., {Chiba}, M., {et~al.} 2022, arXiv e-prints,
  arXiv:2207.11992.
\newblock \doarXiv{2207.11992}

\bibitem[{{Neumayer} {et~al.}(2020){Neumayer}, {Seth}, \&
  {B{\"o}ker}}]{Neumayer2020}
{Neumayer}, N., {Seth}, A., \& {B{\"o}ker}, T. 2020, \aapr, 28, 4,
  \dodoi{10.1007/s00159-020-00125-0}

\bibitem[{{Oke} \& {Gunn}(1983)}]{Oke1983}
{Oke}, J.~B., \& {Gunn}, J.~E. 1983, \apj, 266, 713, \dodoi{10.1086/160817}

\bibitem[{{Pandya} {et~al.}(2018){Pandya}, {Romanowsky}, {Laine}, {Brodie},
  {Johnson}, {Glaccum}, {Villaume}, {Cuillandre}, {Gwyn}, {Krick}, {Lasker},
  {Mart{\'\i}n-Navarro}, {Martinez-Delgado}, \& {van Dokkum}}]{Pandya2018}
{Pandya}, V., {Romanowsky}, A.~J., {Laine}, S., {et~al.} 2018, \apj, 858, 29,
  \dodoi{10.3847/1538-4357/aab498}

\bibitem[{{Planck Collaboration} {et~al.}(2016){Planck Collaboration}, {Ade},
  {Aghanim}, {Arnaud}, {Ashdown}, {Aumont}, {Baccigalupi}, {Banday},
  {Barreiro}, {Bartlett}, {Bartolo}, {Battaner}, {Battye}, {Benabed},
  {Beno{\^\i}t}, {Benoit-L{\'e}vy}, {Bernard}, {Bersanelli}, {Bielewicz},
  {Bock}, {Bonaldi}, {Bonavera}, {Bond}, {Borrill}, {Bouchet}, {Boulanger},
  {Bucher}, {Burigana}, {Butler}, {Calabrese}, {Cardoso}, {Catalano},
  {Challinor}, {Chamballu}, {Chary}, {Chiang}, {Chluba}, {Christensen},
  {Church}, {Clements}, {Colombi}, {Colombo}, {Combet}, {Coulais}, {Crill},
  {Curto}, {Cuttaia}, {Danese}, {Davies}, {Davis}, {de Bernardis}, {de Rosa},
  {de Zotti}, {Delabrouille}, {D{\'e}sert}, {Di Valentino}, {Dickinson},
  {Diego}, {Dolag}, {Dole}, {Donzelli}, {Dor{\'e}}, {Douspis}, {Ducout},
  {Dunkley}, {Dupac}, {Efstathiou}, {Elsner}, {En{\ss}lin}, {Eriksen},
  {Farhang}, {Fergusson}, {Finelli}, {Forni}, {Frailis}, {Fraisse},
  {Franceschi}, {Frejsel}, {Galeotta}, {Galli}, {Ganga}, {Gauthier}, {Gerbino},
  {Ghosh}, {Giard}, {Giraud-H{\'e}raud}, {Giusarma}, {Gjerl{\o}w},
  {Gonz{\'a}lez-Nuevo}, {G{\'o}rski}, {Gratton}, {Gregorio}, {Gruppuso},
  {Gudmundsson}, {Hamann}, {Hansen}, {Hanson}, {Harrison}, {Helou},
  {Henrot-Versill{\'e}}, {Hern{\'a}ndez-Monteagudo}, {Herranz}, {Hildebrandt},
  {Hivon}, {Hobson}, {Holmes}, {Hornstrup}, {Hovest}, {Huang}, {Huffenberger},
  {Hurier}, {Jaffe}, {Jaffe}, {Jones}, {Juvela}, {Keih{\"a}nen}, {Keskitalo},
  {Kisner}, {Kneissl}, {Knoche}, {Knox}, {Kunz}, {Kurki-Suonio}, {Lagache},
  {L{\"a}hteenm{\"a}ki}, {Lamarre}, {Lasenby}, {Lattanzi}, {Lawrence}, {Leahy},
  {Leonardi}, {Lesgourgues}, {Levrier}, {Lewis}, {Liguori}, {Lilje},
  {Linden-V{\o}rnle}, {L{\'o}pez-Caniego}, {Lubin}, {Mac{\'\i}as-P{\'e}rez},
  {Maggio}, {Maino}, {Mandolesi}, {Mangilli}, {Marchini}, {Maris}, {Martin},
  {Martinelli}, {Mart{\'\i}nez-Gonz{\'a}lez}, {Masi}, {Matarrese}, {McGehee},
  {Meinhold}, {Melchiorri}, {Melin}, {Mendes}, {Mennella}, {Migliaccio},
  {Millea}, {Mitra}, {Miville-Desch{\^e}nes}, {Moneti}, {Montier}, {Morgante},
  {Mortlock}, {Moss}, {Munshi}, {Murphy}, {Naselsky}, {Nati}, {Natoli},
  {Netterfield}, {N{\o}rgaard-Nielsen}, {Noviello}, {Novikov}, {Novikov},
  {Oxborrow}, {Paci}, {Pagano}, {Pajot}, {Paladini}, {Paoletti}, {Partridge},
  {Pasian}, {Patanchon}, {Pearson}, {Perdereau}, {Perotto}, {Perrotta},
  {Pettorino}, {Piacentini}, {Piat}, {Pierpaoli}, {Pietrobon}, {Plaszczynski},
  {Pointecouteau}, {Polenta}, {Popa}, {Pratt}, {Pr{\'e}zeau}, {Prunet},
  {Puget}, {Rachen}, {Reach}, {Rebolo}, {Reinecke}, {Remazeilles}, {Renault},
  {Renzi}, {Ristorcelli}, {Rocha}, {Rosset}, {Rossetti}, {Roudier},
  {Rouill{\'e} d'Orfeuil}, {Rowan-Robinson}, {Rubi{\~n}o-Mart{\'\i}n},
  {Rusholme}, {Said}, {Salvatelli}, {Salvati}, {Sandri}, {Santos},
  {Savelainen}, {Savini}, {Scott}, {Seiffert}, {Serra}, {Shellard}, {Spencer},
  {Spinelli}, {Stolyarov}, {Stompor}, {Sudiwala}, {Sunyaev}, {Sutton},
  {Suur-Uski}, {Sygnet}, {Tauber}, {Terenzi}, {Toffolatti}, {Tomasi},
  {Tristram}, {Trombetti}, {Tucci}, {Tuovinen}, {T{\"u}rler}, {Umana},
  {Valenziano}, {Valiviita}, {Van Tent}, {Vielva}, {Villa}, {Wade}, {Wandelt},
  {Wehus}, {White}, {White}, {Wilkinson}, {Yvon}, {Zacchei}, \&
  {Zonca}}]{Planck15}
{Planck Collaboration}, {Ade}, P.~A.~R., {Aghanim}, N., {et~al.} 2016, \aap,
  594, A13, \dodoi{10.1051/0004-6361/201525830}

\bibitem[{{Posti} {et~al.}(2019){Posti}, {Fraternali}, \&
  {Marasco}}]{Posti2019}
{Posti}, L., {Fraternali}, F., \& {Marasco}, A. 2019, \aap, 626, A56,
  \dodoi{10.1051/0004-6361/201935553}

\bibitem[{{Prole} {et~al.}(2019){Prole}, {van der Burg}, {Hilker}, \&
  {Davies}}]{Prole2019}
{Prole}, D.~J., {van der Burg}, R.~F.~J., {Hilker}, M., \& {Davies}, J.~I.
  2019, \mnras, 488, 2143, \dodoi{10.1093/mnras/stz1843}

\bibitem[{{Rodriguez-Gomez} {et~al.}(2019){Rodriguez-Gomez}, {Snyder}, {Lotz},
  {Nelson}, {Pillepich}, {Springel}, {Genel}, {Weinberger}, {Tacchella},
  {Pakmor}, {Torrey}, {Marinacci}, {Vogelsberger}, {Hernquist}, \&
  {Thilker}}]{statmorph}
{Rodriguez-Gomez}, V., {Snyder}, G.~F., {Lotz}, J.~M., {et~al.} 2019, \mnras,
  483, 4140, \dodoi{10.1093/mnras/sty3345}

\bibitem[{{Rom{\'a}n} {et~al.}(2019){Rom{\'a}n}, {Beasley}, {Ruiz-Lara}, \&
  {Valls-Gabaud}}]{Roman2019}
{Rom{\'a}n}, J., {Beasley}, M.~A., {Ruiz-Lara}, T., \& {Valls-Gabaud}, D. 2019,
  \mnras, 486, 823, \dodoi{10.1093/mnras/stz835}

\bibitem[{{Rom{\'a}n} {et~al.}(2021){Rom{\'a}n}, {Castilla}, \&
  {Pascual-Granado}}]{Roman2021}
{Rom{\'a}n}, J., {Castilla}, A., \& {Pascual-Granado}, J. 2021, \aap, 656, A44,
  \dodoi{10.1051/0004-6361/202142161}

\bibitem[{{Rom{\'a}n} \& {Trujillo}(2017{\natexlab{a}})}]{Roman2017b}
{Rom{\'a}n}, J., \& {Trujillo}, I. 2017{\natexlab{a}}, \mnras, 468, 4039,
  \dodoi{10.1093/mnras/stx694}

\bibitem[{{Rom{\'a}n} \& {Trujillo}(2017{\natexlab{b}})}]{Roman2017a}
---. 2017{\natexlab{b}}, \mnras, 468, 703, \dodoi{10.1093/mnras/stx438}

\bibitem[{{Rowe} {et~al.}(2015){Rowe}, {Jarvis}, {Mandelbaum}, {Bernstein},
  {Bosch}, {Simet}, {Meyers}, {Kacprzak}, {Nakajima}, {Zuntz}, {Miyatake},
  {Dietrich}, {Armstrong}, {Melchior}, \& {Gill}}]{Rowe2015}
{Rowe}, B.~T.~P., {Jarvis}, M., {Mandelbaum}, R., {et~al.} 2015, Astronomy and
  Computing, 10, 121, \dodoi{10.1016/j.ascom.2015.02.002}

\bibitem[{{Sandage} \& {Binggeli}(1984)}]{Sandage1984}
{Sandage}, A., \& {Binggeli}, B. 1984, \aj, 89, 919, \dodoi{10.1086/113588}

\bibitem[{{Schlafly} \& {Finkbeiner}(2011)}]{Schlafly2011}
{Schlafly}, E.~F., \& {Finkbeiner}, D.~P. 2011, \apj, 737, 103,
  \dodoi{10.1088/0004-637X/737/2/103}

\bibitem[{{Schlegel} {et~al.}(1998){Schlegel}, {Finkbeiner}, \&
  {Davis}}]{SFD1998}
{Schlegel}, D.~J., {Finkbeiner}, D.~P., \& {Davis}, M. 1998, \apj, 500, 525,
  \dodoi{10.1086/305772}

\bibitem[{{S{\'e}rsic}(1963)}]{Sersic1963}
{S{\'e}rsic}, J.~L. 1963, Boletin de la Asociacion Argentina de Astronomia La
  Plata Argentina, 6, 41

\bibitem[{{Simon}(2019)}]{Simon2019}
{Simon}, J.~D. 2019, \araa, 57, 375,
  \dodoi{10.1146/annurev-astro-091918-104453}

\bibitem[{{Smith Castelli} {et~al.}(2008){Smith Castelli}, {Bassino},
  {Richtler}, {Cellone}, {Aruta}, \& {Infante}}]{SmithCastelli2008}
{Smith Castelli}, A.~V., {Bassino}, L.~P., {Richtler}, T., {et~al.} 2008,
  \mnras, 386, 2311, \dodoi{10.1111/j.1365-2966.2008.13211.x}

\bibitem[{{Somalwar} {et~al.}(2020){Somalwar}, {Greene}, {Greco}, {Huang},
  {Beaton}, {Goulding}, \& {Lancaster}}]{Somalwar2020}
{Somalwar}, J.~J., {Greene}, J.~E., {Greco}, J.~P., {et~al.} 2020, \apj, 902,
  45, \dodoi{10.3847/1538-4357/abb1b2}

\bibitem[{{Spergel}(2010)}]{Spergel2010}
{Spergel}, D.~N. 2010, \apjs, 191, 58, \dodoi{10.1088/0067-0049/191/1/58}

\bibitem[{Starck {et~al.}(2015)Starck, Murtagh, \& Bertero}]{Starck2015}
Starck, J.-L., Murtagh, F., \& Bertero, M. 2015, Starlet Transform in
  Astronomical Data Processing, ed. O.~Scherzer (New York, NY: Springer New
  York), 2053--2098, \dodoi{10.1007/978-1-4939-0790-8_34}

\bibitem[{{Tanoglidis} {et~al.}(2022){Tanoglidis}, {{\'C}iprijanovi{\'c}}, \&
  {Drlica-Wagner}}]{Tanoglidis2022ICML}
{Tanoglidis}, D., {{\'C}iprijanovi{\'c}}, A., \& {Drlica-Wagner}, A. 2022,
  arXiv e-prints, arXiv:2207.03471.
\newblock \doarXiv{2207.03471}

\bibitem[{{Tanoglidis} {et~al.}(2021){Tanoglidis}, {Drlica-Wagner}, {Wei},
  {Li}, {S{\'a}nchez}, {Zhang}, {Peter}, {Feldmeier-Krause}, {Prat}, {Casey},
  {Palmese}, {S{\'a}nchez}, {DeRose}, {Conselice}, {Gagnon}, {Abbott},
  {Aguena}, {Allam}, {Avila}, {Bechtol}, {Bertin}, {Bhargava}, {Brooks},
  {Burke}, {Rosell}, {Kind}, {Carretero}, {Chang}, {Costanzi}, {da Costa}, {De
  Vicente}, {Desai}, {Diehl}, {Doel}, {Eifler}, {Everett}, {Evrard},
  {Flaugher}, {Frieman}, {Garc{\'\i}a-Bellido}, {Gerdes}, {Gruendl},
  {Gschwend}, {Gutierrez}, {Hartley}, {Hollowood}, {Huterer}, {James},
  {Krause}, {Kuehn}, {Kuropatkin}, {Maia}, {March}, {Marshall}, {Menanteau},
  {Miquel}, {Ogando}, {Paz-Chinch{\'o}n}, {Romer}, {Roodman}, {Sanchez},
  {Scarpine}, {Serrano}, {Sevilla-Noarbe}, {Smith}, {Suchyta}, {Tarle},
  {Thomas}, {Tucker}, {Walker}, \& {DES Collaboration}}]{Tanoglidis2021}
{Tanoglidis}, D., {Drlica-Wagner}, A., {Wei}, K., {et~al.} 2021, \apjs, 252,
  18, \dodoi{10.3847/1538-4365/abca89}

\bibitem[{{Tremmel} {et~al.}(2020){Tremmel}, {Wright}, {Brooks}, {Munshi},
  {Nagai}, \& {Quinn}}]{Tremmel2020}
{Tremmel}, M., {Wright}, A.~C., {Brooks}, A.~M., {et~al.} 2020, \mnras, 497,
  2786, \dodoi{10.1093/mnras/staa2015}

\bibitem[{{Trujillo} {et~al.}(2020){Trujillo}, {Chamba}, \&
  {Knapen}}]{Trujillo2020}
{Trujillo}, I., {Chamba}, N., \& {Knapen}, J.~H. 2020, \mnras, 493, 87,
  \dodoi{10.1093/mnras/staa236}

\bibitem[{{Trujillo} {et~al.}(2007){Trujillo}, {Conselice}, {Bundy}, {Cooper},
  {Eisenhardt}, \& {Ellis}}]{Trujillo2007}
{Trujillo}, I., {Conselice}, C.~J., {Bundy}, K., {et~al.} 2007, \mnras, 382,
  109, \dodoi{10.1111/j.1365-2966.2007.12388.x}

\bibitem[{{Trujillo} {et~al.}(2017){Trujillo}, {Roman}, {Filho}, \&
  {S{\'a}nchez Almeida}}]{Trujillo2017}
{Trujillo}, I., {Roman}, J., {Filho}, M., \& {S{\'a}nchez Almeida}, J. 2017,
  \apj, 836, 191, \dodoi{10.3847/1538-4357/aa5cbb}

\bibitem[{{van der Burg} {et~al.}(2016){van der Burg}, {Muzzin}, \&
  {Hoekstra}}]{vdBurg2016}
{van der Burg}, R. F.~J., {Muzzin}, A., \& {Hoekstra}, H. 2016, \aap, 590, A20,
  \dodoi{10.1051/0004-6361/201628222}

\bibitem[{{van der Burg} {et~al.}(2017){van der Burg}, {Hoekstra}, {Muzzin},
  {Sif{\'o}n}, {Viola}, {Bremer}, {Brough}, {Driver}, {Erben}, {Heymans},
  {Hildebrandt}, {Holwerda}, {Klaes}, {Kuijken}, {McGee}, {Nakajima},
  {Napolitano}, {Norberg}, {Taylor}, \& {Valentijn}}]{vdBurg2017}
{van der Burg}, R. F.~J., {Hoekstra}, H., {Muzzin}, A., {et~al.} 2017, \aap,
  607, A79, \dodoi{10.1051/0004-6361/201731335}

\bibitem[{{van Dokkum} {et~al.}(2019){van Dokkum}, {Danieli}, {Abraham},
  {Conroy}, \& {Romanowsky}}]{vanDokkum2019}
{van Dokkum}, P., {Danieli}, S., {Abraham}, R., {Conroy}, C., \& {Romanowsky},
  A.~J. 2019, \apjl, 874, L5, \dodoi{10.3847/2041-8213/ab0d92}

\bibitem[{{van Dokkum} {et~al.}(2017){van Dokkum}, {Abraham}, {Romanowsky},
  {Brodie}, {Conroy}, {Danieli}, {Lokhorst}, {Merritt}, {Mowla}, \&
  {Zhang}}]{vanDokkum2017}
{van Dokkum}, P., {Abraham}, R., {Romanowsky}, A.~J., {et~al.} 2017, \apjl,
  844, L11, \dodoi{10.3847/2041-8213/aa7ca2}

\bibitem[{{van Dokkum} {et~al.}(2018){van Dokkum}, {Danieli}, {Cohen},
  {Merritt}, {Romanowsky}, {Abraham}, {Brodie}, {Conroy}, {Lokhorst}, {Mowla},
  {O'Sullivan}, \& {Zhang}}]{vanDokkum2018}
{van Dokkum}, P., {Danieli}, S., {Cohen}, Y., {et~al.} 2018, \nat, 555, 629,
  \dodoi{10.1038/nature25767}

\bibitem[{{van Dokkum} {et~al.}(2022{\natexlab{a}}){van Dokkum}, {Shen},
  {Romanowsky}, {Abraham}, {Conroy}, {Danieli}, {Dutta Chowdhury}, {Keim},
  {Kruijssen}, {Leja}, \& {Trujillo-Gomez}}]{vanDokkum2022GC}
{van Dokkum}, P., {Shen}, Z., {Romanowsky}, A.~J., {et~al.} 2022{\natexlab{a}},
  arXiv e-prints, arXiv:2207.07129.
\newblock \doarXiv{2207.07129}

\bibitem[{{van Dokkum} {et~al.}(2022{\natexlab{b}}){van Dokkum}, {Shen},
  {Keim}, {Trujillo-Gomez}, {Danieli}, {Dutta Chowdhury}, {Abraham}, {Conroy},
  {Kruijssen}, {Nagai}, \& {Romanowsky}}]{vandokkum2022Nat}
{van Dokkum}, P., {Shen}, Z., {Keim}, M.~A., {et~al.} 2022{\natexlab{b}}, \nat,
  605, 435, \dodoi{10.1038/s41586-022-04665-6}

\bibitem[{{van Dokkum} {et~al.}(2015){van Dokkum}, {Abraham}, {Merritt},
  {Zhang}, {Geha}, \& {Conroy}}]{vanDokkum2015}
{van Dokkum}, P.~G., {Abraham}, R., {Merritt}, A., {et~al.} 2015, \apjl, 798,
  L45, \dodoi{10.1088/2041-8205/798/2/L45}

\bibitem[{{van Dokkum} {et~al.}(2013){van Dokkum}, {Leja}, {Nelson}, {Patel},
  {Skelton}, {Momcheva}, {Brammer}, {Whitaker}, {Lundgren}, {Fumagalli},
  {Conroy}, {F{\"o}rster Schreiber}, {Franx}, {Kriek}, {Labb{\'e}},
  {Marchesini}, {Rix}, {van der Wel}, \& {Wuyts}}]{vanDokkum2013}
{van Dokkum}, P.~G., {Leja}, J., {Nelson}, E.~J., {et~al.} 2013, \apjl, 771,
  L35, \dodoi{10.1088/2041-8205/771/2/L35}

\bibitem[{{Van Nest} {et~al.}(2022){Van Nest}, {Munshi}, {Wright}, {Tremmel},
  {Brooks}, {Nagai}, \& {Quinn}}]{vanNest2022}
{Van Nest}, J.~D., {Munshi}, F., {Wright}, A.~C., {et~al.} 2022, \apj, 926, 92,
  \dodoi{10.3847/1538-4357/ac43b7}

\bibitem[{{Venhola} {et~al.}(2022){Venhola}, {Peletier}, {Salo}, {Laurikainen},
  {Janz}, {Haigh}, {Wilkinson}, {Iodice}, {Hilker}, {Mieske}, {Cantiello}, \&
  {Spavone}}]{Venhola2022}
{Venhola}, A., {Peletier}, R.~F., {Salo}, H., {et~al.} 2022, \aap, 662, A43,
  \dodoi{10.1051/0004-6361/202141756}

\bibitem[{{Villaume} {et~al.}(2022){Villaume}, {Romanowsky}, {Brodie}, {van
  Dokkum}, {Conroy}, {Forbes}, {Danieli}, {Martin}, \&
  {Matuszewski}}]{Villaume2022}
{Villaume}, A., {Romanowsky}, A.~J., {Brodie}, J., {et~al.} 2022, \apj, 924,
  32, \dodoi{10.3847/1538-4357/ac341e}

\bibitem[{{Wang} {et~al.}(2021){Wang}, {Takada}, {Li}, {Carlsten}, {Lan},
  {Shi}, {Miyatake}, {More}, {Beaton}, {Lupton}, {Lin}, {Qiu}, \&
  {Luo}}]{Wang2021}
{Wang}, W., {Takada}, M., {Li}, X., {et~al.} 2021, \mnras, 500, 3776,
  \dodoi{10.1093/mnras/staa3495}

\bibitem[{{Wasserman} {et~al.}(2019){Wasserman}, {van Dokkum}, {Romanowsky},
  {Brodie}, {Danieli}, {Forbes}, {Abraham}, {Martin}, {Matuszewski},
  {Villaume}, {Tamanas}, \& {Profumo}}]{Wasserman2019}
{Wasserman}, A., {van Dokkum}, P., {Romanowsky}, A.~J., {et~al.} 2019, \apj,
  885, 155, \dodoi{10.3847/1538-4357/ab3eb9}

\bibitem[{{Willmer}(2018)}]{Willmer2018}
{Willmer}, C. N.~A. 2018, \apjs, 236, 47, \dodoi{10.3847/1538-4365/aabfdf}

\bibitem[{{Wright} {et~al.}(2021){Wright}, {Tremmel}, {Brooks}, {Munshi},
  {Nagai}, {Sharma}, \& {Quinn}}]{Wright2021}
{Wright}, A.~C., {Tremmel}, M., {Brooks}, A.~M., {et~al.} 2021, \mnras, 502,
  5370, \dodoi{10.1093/mnras/stab081}

\bibitem[{{Wu} {et~al.}(2022){Wu}, {Peek}, {Tollerud}, {Mao}, {Nadler}, {Geha},
  {Wechsler}, {Kallivayalil}, \& {Weiner}}]{xSAGA-I}
{Wu}, J.~F., {Peek}, J.~E.~G., {Tollerud}, E.~J., {et~al.} 2022, \apj, 927,
  121, \dodoi{10.3847/1538-4357/ac4eea}

\bibitem[{{Yagi} {et~al.}(2016){Yagi}, {Koda}, {Komiyama}, \&
  {Yamanoi}}]{Yagi2016}
{Yagi}, M., {Koda}, J., {Komiyama}, Y., \& {Yamanoi}, H. 2016, \apjs, 225, 11,
  \dodoi{10.3847/0067-0049/225/1/11}

\bibitem[{{York} {et~al.}(2000){York}, {Adelman}, {Anderson}, {Anderson},
  {Annis}, {Bahcall}, {Bakken}, {Barkhouser}, {Bastian}, {Berman}, {Boroski},
  {Bracker}, {Briegel}, {Briggs}, {Brinkmann}, {Brunner}, {Burles}, {Carey},
  {Carr}, {Castander}, {Chen}, {Colestock}, {Connolly}, {Crocker}, {Csabai},
  {Czarapata}, {Davis}, {Doi}, {Dombeck}, {Eisenstein}, {Ellman}, {Elms},
  {Evans}, {Fan}, {Federwitz}, {Fiscelli}, {Friedman}, {Frieman}, {Fukugita},
  {Gillespie}, {Gunn}, {Gurbani}, {de Haas}, {Haldeman}, {Harris}, {Hayes},
  {Heckman}, {Hennessy}, {Hindsley}, {Holm}, {Holmgren}, {Huang}, {Hull},
  {Husby}, {Ichikawa}, {Ichikawa}, {Ivezi{\'c}}, {Kent}, {Kim}, {Kinney},
  {Klaene}, {Kleinman}, {Kleinman}, {Knapp}, {Korienek}, {Kron}, {Kunszt},
  {Lamb}, {Lee}, {Leger}, {Limmongkol}, {Lindenmeyer}, {Long}, {Loomis},
  {Loveday}, {Lucinio}, {Lupton}, {MacKinnon}, {Mannery}, {Mantsch}, {Margon},
  {McGehee}, {McKay}, {Meiksin}, {Merelli}, {Monet}, {Munn}, {Narayanan},
  {Nash}, {Neilsen}, {Neswold}, {Newberg}, {Nichol}, {Nicinski}, {Nonino},
  {Okada}, {Okamura}, {Ostriker}, {Owen}, {Pauls}, {Peoples}, {Peterson},
  {Petravick}, {Pier}, {Pope}, {Pordes}, {Prosapio}, {Rechenmacher}, {Quinn},
  {Richards}, {Richmond}, {Rivetta}, {Rockosi}, {Ruthmansdorfer}, {Sandford},
  {Schlegel}, {Schneider}, {Sekiguchi}, {Sergey}, {Shimasaku}, {Siegmund},
  {Smee}, {Smith}, {Snedden}, {Stone}, {Stoughton}, {Strauss}, {Stubbs},
  {SubbaRao}, {Szalay}, {Szapudi}, {Szokoly}, {Thakar}, {Tremonti}, {Tucker},
  {Uomoto}, {Vanden Berk}, {Vogeley}, {Waddell}, {Wang}, {Watanabe},
  {Weinberg}, {Yanny}, {Yasuda}, \& {SDSS Collaboration}}]{York2000}
{York}, D.~G., {Adelman}, J., {Anderson}, John~E., J., {et~al.} 2000, \aj, 120,
  1579, \dodoi{10.1086/301513}

\bibitem[{{Zaritsky} {et~al.}(2021){Zaritsky}, {Donnerstein}, {Karunakaran},
  {Barbosa}, {Dey}, {Kadowaki}, {Spekkens}, \& {Zhang}}]{Zaritsky2021}
{Zaritsky}, D., {Donnerstein}, R., {Karunakaran}, A., {et~al.} 2021, \apjs,
  257, 60, \dodoi{10.3847/1538-4365/ac2607}

\bibitem[{{Zaritsky} {et~al.}(2022){Zaritsky}, {Donnerstein}, {Karunakaran},
  {Barbosa}, {Dey}, {Kadowaki}, {Spekkens}, \& {Zhang}}]{Zaritsky2022}
---. 2022, \apjs, 261, 11, \dodoi{10.3847/1538-4365/ac6ceb}

\bibitem[{{Zaritsky} {et~al.}(2019){Zaritsky}, {Donnerstein}, {Dey},
  {Kadowaki}, {Zhang}, {Karunakaran}, {Mart{\'\i}nez-Delgado}, {Rahman}, \&
  {Spekkens}}]{Zaritsky2019}
{Zaritsky}, D., {Donnerstein}, R., {Dey}, A., {et~al.} 2019, \apjs, 240, 1,
  \dodoi{10.3847/1538-4365/aaefe9}

\end{thebibliography}
\bibliographystyle{aasjournal}

\newpage
\appendix

\section{LSBG Detection}\label{ap:detection}
In this appendix, we provide a more detailed description of our LSBG search method and how it is different from \citetalias{Greco2018}.

\subsection{Bright Source Removal}
Bright sources and their associated LSB outskirts can mimic objects of interest and obstruct the detection of LSBGs. In this step, we replace pixels related to bright sources and small compact sources with sky noise. The bright sources and their diffuse outskirts are detected in the $i$ band by applying a high thresholding and a low thresholding to the image, respectively. We associate a diffuse light component with a bright source if more than 15\% of its pixels are above the high threshold. In this way, we generate a footprint of bright sources and their associated LSB components. Then for the $gri$ bands, we replace every pixel within this footprint with Gaussian noise, where the noise level is determined by masking out all detected sources in the image. 
    
In \citetalias{Greco2018}, the thresholds are set based on the noise level of the local sky. Thanks to the global sky subtraction in HSC PDR2, LSB features are well conserved after subtracting the sky and the sky noise level varies less than in prior reductions. We therefore set the thresholds based on the characteristic surface brightness instead of a certain sigma value above the sky background. In this work, we set the high threshold to $\mu_{\rm high} = 22\ \sbunit$ to capture all bright sources above this surface brightness, and the low threshold to $\mu_{\rm low} = 24.5\ \sbunit$ to capture the associated diffuse light. Unlike in \citetalias{Greco2018}, we find it unnecessary to smooth the image prior to the thresholding based on our completeness tests (see \S \ref{sec:comp_meas},\S\ref{ap:comp_meas_unc}).   

After this step, there are still a number of small and compact LSB objects, which are typically marginally resolved galaxies, blended sources, or just pixels standing out because of noise fluctuation. We add an extra step to detect and remove them as our main goal is to detect extended LSBGs. We run \code{sep} on the ``cleaned'' images as described above. Based on the segmentation map, we generate a mask for sources smaller than $r_{\rm min} = 2\arcsec$, and pixels within this mask are also replaced by sky noise. This step significantly improves the purity of LSBG search. The values of ($\mu_{\rm high},\ \mu_{\rm low},\ r_{\rm min})$ are chosen by trial and error, but guided by the completeness tests. 
    
\subsection{Source Extraction}
We use \code{SourceExtractor} to detect sources on the ``cleaned'' images where the bright sources and small compact sources are replaced by sky noise. This step remains largely the same as in \citetalias{Greco2018}. The images are convolved with a Gaussian kernel of FWHM=$1\arcsec$ to enhance the contrast between LSBGs and sky background \citep[e.g.,][]{Irwin1985,Akhlaghi2015,Greco2018}. We take a mesh size of $43\arcsec$ (double the mesh size used in \citetalias{Greco2018} to allow bigger galaxies in our sample) to measure the local background and detect objects that are 0.7$\sigma$ per pixel above the local sky background. We also require the object to contain at least 100 contiguous pixels (equivalent to a square box with $1\farcs7$ on a side) to further remove small compact objects. We perform the detection in the $g$ band, but require that all sources are also detected in the $r$ band to exclude spurious detection and artifacts.
    
\subsection{Initial Sample Selection} 
We take the output catalog from \code{SourceExtractor} and remove those objects that are not likely to be LSBGs based on their sizes and colors. To be specific, we require objects to have $g$-band half-light radii (measured by \code{SourceExtractor}) greater than $r_{\rm min} = 2.0\arcsec$. We also require the measured colors to satisfy $-0.1 < g-i < 1.4$ and $|(g-r) - 0.7\cdot (g-i)| < 0.4$. This color box is relatively conservative with respect to the color distribution of discovered LSBGs \citep[e.g.,][]{SAGA-I,Greco2018,Zaritsky2019,Tanoglidis2021}. 

Compared with \citetalias{Greco2018}, we add a new metric based on the morphology of sources to further remove compact point-like sources. For each source, we compute the ratio between the 2D autocorrelation function (ACT) within an aperture of 5 pixels and the ACT within an annulus between 5 pixels and 9 pixels from the center. The ACT ratio essentially characterizes the peakiness of the source. We expect extended LSBG to have a strong correlation across a large number of pixels, making the ACT less peaky than point-like sources. Thus we apply a cut on the ACT ratio $a_{\rm corr} < 2.5$ to remove point sources. This step helps to improve the purity of our sample.


\section{Deblending details}\label{ap:deblending}
In \S\ref{sec:deblending}, we briefly introduce how we use \code{scarlet} to model the LSBG candidates in a nonparametric fashion and filter out false positives based on the structural and morphological parameters. In this appendix, we describe details of the implementation of vanilla \code{scarlet} to help interested readers better understand the technique.

\subsection{Peak detection}\label{sec:peak}
We generate cutout images with a size of $1\arcmin$ in the $griz$ bands for each LSBG candidate in our initial sample. We then construct a detection image by taking an average of the four bands weighted by the inverse variance of each band. This detection image is considered to be deeper than the single-band images. 
Next, we run \code{sep} on the detection image using a threshold of 4$\sigma$ above the sky, a mesh size of 48 pixels ($8\arcsec$), and a kernel size of 3 pixels. This step identifies extended sources in the detection image. However, there are still faint and compact peaks not detected. We apply a wavelet decomposition to the detection image \citep{Starck2015} and only keep the high spatial frequency components (also see \citealt{Zaritsky2019} for an example of using wavelet filtering in the context of LSBGs). Another round of \code{sep} is run on this high-frequency image using a detection threshold of 2.5$\sigma$, a mesh size of 24 pixels, and a kernel size of 3 pixels. This step detects many compact sources that were not included in the previous step. In the end, we combine the two detection catalogs and remove duplicates. 

\subsection{Model Initialization and Optimization}
After the peak detection step, we need to decide which peaks to model. It is not necessary to model all detected peaks because the deblending step is designed to model the sources only in the vicinity of the target LSBG candidate. Therefore, the size of the target object determines which peaks are relevant. Nevertheless, it is also important to choose the appropriate model for each source and initialize them as best as possible. In practice, we draw a square bounding box around the target object and model all the peaks inside this box. The bounding box is in turn determined as we initialize the model for the target object. We describe the procedures as follows. 

First, in vanilla \code{scarlet}, objects can be modeled with different types of models\footnote{\url{https://pmelchior.github.io/scarlet/1-concepts.html\#Source}} including point source, single-extended source, multi-extended source, compact source, and flat-sky source. The morphology image of a point source is simply the normalized PSF model. The single-extended source has a morphology image that follows positivity and monotonicity constraints. The multi-extended source is a combination of two or more co-centered single-extended sources, making it possible to model galaxies with more complex structures and capture any color gradients. The compact source is a single-extended source initialized using the morphology image of a point source, which encourages the model to be compact, but still allows the morphology image to be extended. The flat-sky source has uniform color and morphology within the bounding box.

For modeling the LSBG candidates, we use an extended source model with two components such that the model is able to capture the galaxy structure and color gradient. The target source is initialized in the following way. We convolve the detection image with a circular Gaussian kernel with $\sigma=1.5$ pixels to boost the contrast between the signal and the sky noise. After smoothing, the LSB outskirts of the target galaxy become more prominent, which helps in determining the bounding box and initializing the morphology image. We take the smoothed image and threshold it with $0.1\sigma$ to remove the sky noise. The initial morphology image $S_i$ is determined by constructing a symmetric and monotonic approximation to the smoothed image around the target object. The initial SED vector $A_i$ is set to be the sum of the morphology matrices $S_i$ in each band. As a byproduct of initialization, we get a bounding box from $S_i$ which indicates the extent of the target object. Consequently, the size of the bounding box depends on the smoothing kernel and the threshold. We choose the above values by trial and error such that the box is not so large as to include many irrelevant peaks, but not so small as to lose a significant fraction of LSB outskirts of the target galaxy. 

Then we take all peaks within the bounding box of the target galaxy and initialize them in the same way as described above. Recall that we have two rounds of peak detection (Sec \ref{sec:peak}), focusing on extended and compact sources respectively. Extended objects detected in the first peak detection step are modeled as single-extended sources. Compact objects that are only detected in the second peak detection are modeled as point sources if their FWHM $<$ 5 pixels, otherwise as compact-extended sources. We also add a flat-sky source to model the local sky around the target. This is helpful for situations where an object overlaps with the LSB outskirts of a bright galaxy (e.g., the bottom panel in Figure \ref{fig:vanilla_scarlet_demo}). We note that adding a flat-sky source could significantly change the size and total magnitude of the \code{scarlet} model.

Although we only model peaks within the bounding box of the target, the scattered light from nearby bright stars and galaxies could bias the modeling of the sources within the bounding box (e.g., the yellow star in the top left panel of Figure \ref{fig:vanilla_scarlet_demo}). We match our field with the Gaia catalog \citep{GAIA2016,GAIA2018} and mask out stars outside of the bounding box. As we have already detected peaks throughout the whole cutout image, we also generate a mask for objects outside the bounding box to reduce the impact of scattered light from bright galaxies on the modeling of the flat-sky source. 

The optimization process uses the adaptive proximal gradient method \citep{Melchior2019}, which is a robust method for optimization with constraints. The model is considered to be converged when the relative changes of parameters are smaller than \code{e\_rel\,=\,2e-4}. Typically, convergence is achieved after $\sim 50$ steps of optimization, and the whole modeling process takes about 40s for a typical LSBG. It takes longer when the target galaxy has a large angular size as more peaks are modeled. We note that \code{scarlet} only finds the maximum likelihood estimation of the model instead of deriving the full posteriors, thus a good initialization becomes especially important for fast convergence.


\section{Spergel profiles}\label{ap:spergel}
\begin{figure*}[htbp!]
	\vbox{ 
		\centering
		\includegraphics[width=0.85\linewidth]{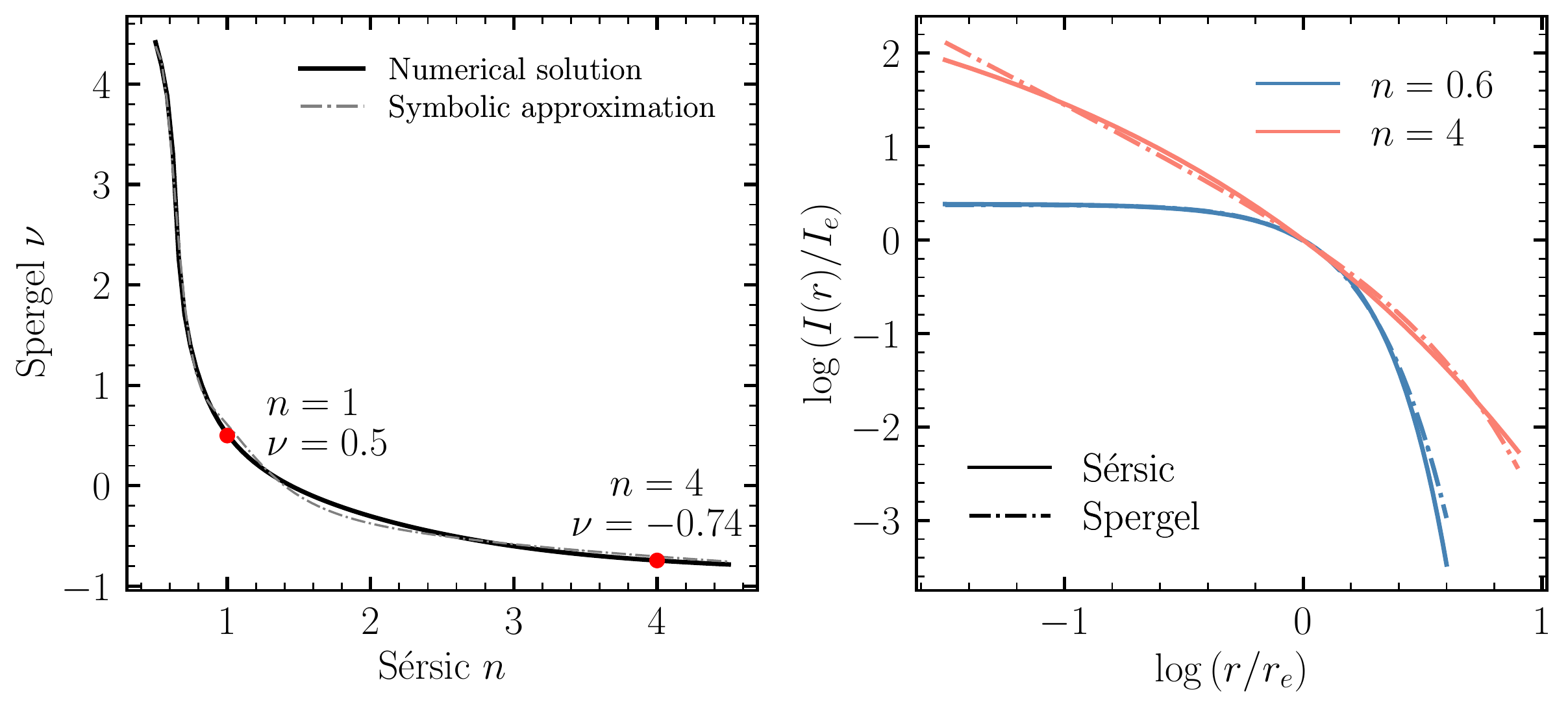}
	}
    \caption{Correspondence between the \sersic{} profile \eqref{eq:sersic} and the Spergel profile \eqref{eq:spergel}. We fit a Spergel profile to each \sersic{} while fixing the total luminosity and half-light radius. The black line in the left panel shows the best-fit Spergel index $\nu$ as a function of the \sersic{} index $n$. We also provide a symbolic approximation to this relation in \eqref{eq:symbolic}, shown as the gray dashed-dotted line. In the right panel, two \sersic{} profiles (solid) and their best-fit Spergel profiles (dashed-dotted) are shown. The Spergel profile approximates the \sersic{} profile well at the small \sersic{} index.  
    }
    \label{fig:spgl_calib}
\end{figure*}

\begin{figure*}
	\vbox{ 
		\centering
		\includegraphics[width=0.9\linewidth]{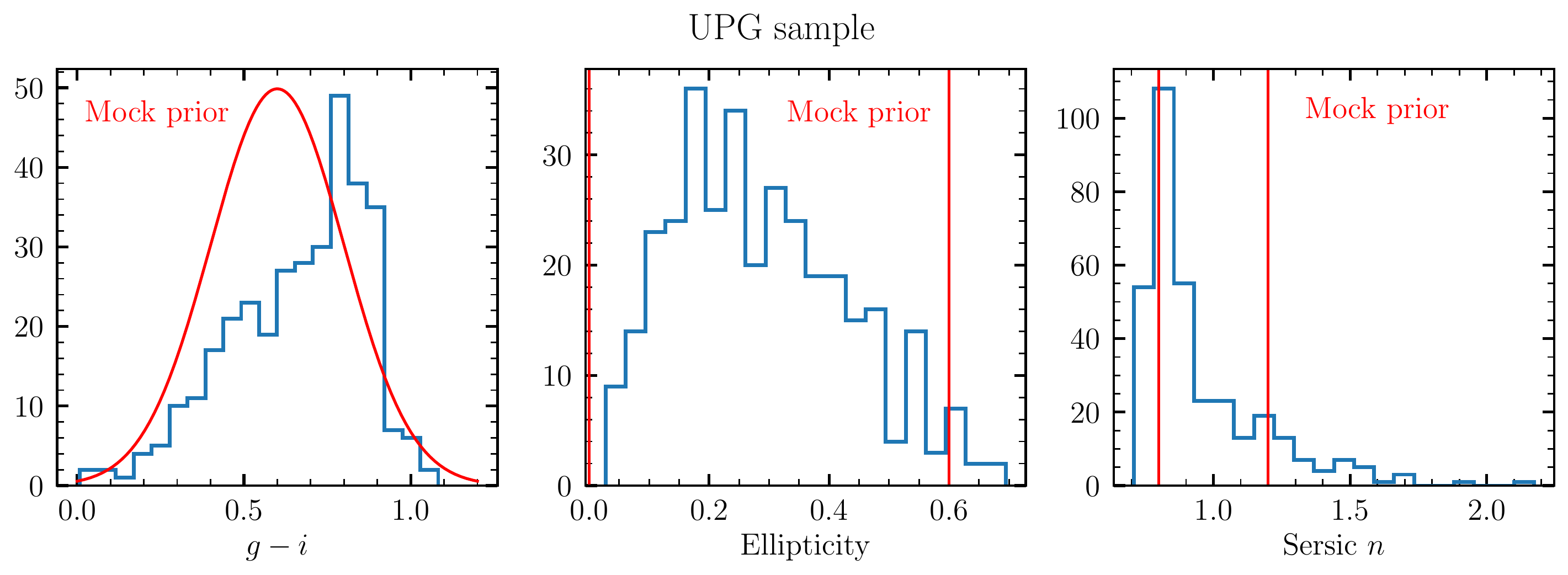}
	}
    \caption{Distribution of UPGs (blue) and mock galaxies (red) in $g-i$ color, ellipticity, and \sersic{} index $n$. We translate the Spergel index to the \sersic{} index using Equation \eqref{eq:symbolic}. The mock galaxy population represents the real UDG/UPG population well.
    }
    \label{fig:mock_params}
\end{figure*}

In this appendix, we demonstrate that the Spergel profile can approximate the \sersic{} profile and provide a lookup table for the correspondence between the \sersic{} index $n$ and Spergel index $\nu$.

The surface brightness of a Spergel profile has the form \citep{Spergel2010}:
\begin{equation}
    \label{eq:spergel}
    I_\nu(r) = \frac{c_{\nu}^{2} L_{0}}{2\pi r_{0}^{2}} f_{\nu}\left(\frac{c_{\nu} r}{r_{0}}\right),
\end{equation}
where 
\begin{equation}
    f_{\nu}(u)=\left(\frac{u}{2}\right)^{\nu} \frac{K_{\nu}(u)}{\Gamma(\nu+1)},
\end{equation}
and $K_\nu(u)$ is the Modified Bessel function of the second kind. The half-light radius is $r_0$, the total luminosity is $L_0$, and $c_\nu$ satisfies the equation $(1 + \nu)f_{\nu + 1}(c_\nu) = 1/4$. The Spergel profile has a simple analytical expression in Fourier space, making it easy to convolve with a PSF.

The surface brightness of a \sersic{} profile follows \citep{Sersic1963,Graham2005}:
\begin{equation}\label{eq:sersic}
    I(r)=I_{\mathrm{e}} \exp \left\{-b_{n}\left[\left(\frac{r}{r_{\mathrm{e}}}\right)^{1 / n}-1\right]\right\},
\end{equation}
where $r_e$ is the half-light radius, $I_e$ is the surface brightness at $r=r_e$, $n$ is the \sersic{} index. The value of $b_n$ satisfies $\Gamma(2 n)=2 \gamma\left(2 n, b_{n}\right)$, where $\gamma(a, x)$ is the incomplete gamma function. According to \citet{Graham2005}, the total luminosity of a \sersic{} profile is given by 
\begin{equation}\label{eq:sersic_lum}
    L_0 = I_{e} r_{e}^{2}\, 2 \pi n\, e^{b_{n}} \left(b_{n}\right)^{-2 n} \Gamma(2 n).
\end{equation}

To study the correspondence between \sersic{} and Spergel profiles, we generate \sersic{} 1D profiles with different \sersic{} indices using \code{astropy}\footnote{\url{https://docs.astropy.org/en/stable/api/astropy.modeling.functional_models.Sersic1D.html}}, and try to fit the \sersic{} profiles using the Spergel model. The \sersic{} index ranges from $n=0.5$ to $n=4.5$, and both profiles are normalized at the half-light radius (see the right panel in Figure \ref{fig:spgl_calib}). For each \sersic{} profile, we calculate the total luminosity $L_0$ according to \eqref{eq:sersic_lum} and plug it into the Spergel profile \eqref{eq:spergel} as a fixed value. Therefore, only the Spergel index $\nu$ is allowed to vary during the fitting. The fitting uses the least squares fitting with Levenberg--Marquardt algorithm implemented in \code{astropy.modeling.fitting}\footnote{\url{https://docs.astropy.org/en/stable/api/astropy.modeling.fitting.LevMarLSQFitter.html\#astropy.modeling.fitting.LevMarLSQFitter}}, which is a commonly used robust algorithm to solve nonlinear least square fitting problems. The best-fit Spergel index $\nu$ as a function of the \sersic{} index $n$ is shown in the left panel of Figure \ref{fig:spgl_calib}. A Spergel profile with $\nu=0.5$ is exactly an exponential profile with $n=1$. The de Vaucouleurs profile \citep{deVaucouleurs1948} with $n=4$ can be approximated by a Spergel profile with $\nu=-0.74$. As two examples, we show two \sersic{} profiles (solid) and their best-fit Spergel profiles (dashed-dotted) in the right panel. Overall, the Spergel profile provides a good approximation for the \sersic{} profile.

As shown in Figure \ref{fig:spgl_calib}, a \sersic{} profile with a small \sersic{} index ($0.5 < n < 1.5$) can be well-approximated by a Spergel profile, although the Spergel profile seems to be more extended than \sersic{} in the outskirts. For a \sersic{} profile with a high \sersic{} index, the approximation gets worse at both small and large radii. Overall, the Spergel profile is a good approximation to \sersic{} from $n\approx 0.5$ to $n\approx 4.5$. Using the symbolic regression package \code{PySR} \citep{Cranmer2023}, the best-fit relation between $n$ and $\nu$ for $0.5<n<4.5$ is given as follows:
\begin{equation}\label{eq:symbolic}
    \nu = 1.06\,n^{-2.5} \cos(\sin(3.74/n))^{1/2} - 0.365\,n^{1/2}
\end{equation}

It is well-known that the light profiles of low-mass galaxies are quite flat and can be described using \sersic{} profiles with $0.5 < n < 1.5$ \citep[e.g.,][]{vanDokkum2015,Lange2015,Greco2018,Zaritsky2021,ELVES-I}. It is thus reasonable to use Spergel profiles to model the LSBGs (\S\ref{sec:modeling}) and enjoy its convenience in Fourier space. 

\section{Completeness and Measurement Uncertainty}\label{ap:comp_meas_unc}
\subsection{Completeness}
The total completeness is a multiplication between the detection completeness and deblending completeness. Below we describe how we generate mock galaxies and derive the recovered fraction. 

To estimate the detection completeness, we inject $\sim 700,000$ mock galaxies with single-\sersic{} light profiles into the co-added images. The mock galaxies are generated using \texttt{GalSim} \citep{Rowe2015} and are injected after being convolved with HSC PSF generated by \code{hscPipe}. We directly inject the mock galaxies into the co-added images without adding extra noise. The galaxies are placed randomly on the image but we require all mock galaxies to be separated by at least 80\arcsec{} from each other to minimize the impact of mock galaxies on the image noise properties. We have also done extensive tests on injecting mock galaxies into the raw images, adding Poisson noise, and going through the entire data reduction pipeline using \code{SynPipe} \citep{Huang2018synpipe}. This is very expensive in terms of both CPU time and disk space as we must run the full \code{hscPipe}. However, we find no noticeable difference in the completeness between this method and the direct injection to co-added images.

To resemble the real LSBG population in \citetalias{Greco2018}, we generate mock galaxies following uniform distributions in size ($2\arcsec \leqslant r_{e} \leqslant 21\arcsec$), surface brightness ($23 \leqslant \overline{\mu}_{\rm eff}(g) \leqslant 28.5\ \mathrm{mag\ arcsec^{-2}}$), \sersic{} index ($0.8 < n < 1.2$), and ellipticity ($0 < \varepsilon < 0.6$). They are randomly assigned to have a blue ($g-i=0.47,\ g-r=0.32$), medium ($g-i=0.64,\ g-r=0.43$), and red ($g-i=0.82,\ g-r=0.56$) color with equal chance. The ranges spanned by these colors and \sersic{} indices cover most of the observed LSBGs in \citetalias{Greco2018}. Then we run the detection step described in \S \ref{sec:detection} and cross-match the detection catalog with the input mock galaxy catalog to calculate completeness. We split the size and surface brightness range into 15 bins with $\Delta r_e = 0\farcs86$, $\Delta \sbeff = 0.33\ \sbunit$, and interpolate over bins using an isotropic Gaussian kernel with $\sigma = 0.5$. We note that the size and surface brightness shown in Figure \ref{fig:completeness} are all the intrinsic values for the mock galaxies, which are different from the measured ones. We find negligible dependence of detection completeness on the \sersic{} index, color, and ellipticity. Therefore, we neglect the dependence of detection completeness on parameters other than size and surface brightness hereafter.\footnote{We performed a smaller set of image simulations where we extended the ellipticity range and also simulated galaxies with additional structures (such as star-forming clumps). We find the completeness declines for $\varepsilon > 0.6$, suggesting that edge-on disk galaxies may be missing from our sample. Please see \citet{Greene2022} for more details.}

For the deblending completeness, we inject 5000 mock \sersic{} galaxies into the co-added images and run vanilla \code{scarlet} on them. The mock galaxies follow the same uniform distribution in size, surface brightness, ellipticity, \sersic{} index as for deriving the detection completeness, but follow a Gaussian distribution in color: $g-i \sim \mathcal{N}(0.6, 0.2^2),\ g-r = 0.7 \cdot (g-i) + \mathcal{N}(0, 0.03^2)$. We show the distribution of mock galaxies and UPGs in Figure \ref{fig:mock_params}. This mock galaxy population represents the real UDG/UPG population well. To be specific, the Gaussian color prior covers 86\% of the UPGs within $1.5\sigma$, the ellipticity prior covers 98\% of them, and the \sersic{} index prior covers 60\% of them.
We measure the size and surface brightness on the scarlet models of the mock galaxies, then we apply the deblending cuts. We also find that the deblending cut mainly depends on the size and surface brightness.

\subsection{Measurement Bias and Uncertainty}\label{sec:meas_unc}

Due to the low-surface-brightness nature of LSBGs, their size, magnitude, surface brightness, and shape are hard to characterize and are typically associated with large uncertainties. \citet{Haussler2007} used mock single-\sersic{} galaxies and parametric fitting codes to demonstrate that the estimated size and total magnitude are sensitive to local sky estimation and masking of neighboring objects. In the low-surface-brightness regime, the measured size and total magnitude can be quite biased and have large uncertainties. Therefore, measurement bias and uncertainty must be considered when studying the properties of LSBGs. \citet{Zaritsky2021,Zaritsky2022} characterized their measurement bias and uncertainty by injecting mock \sersic{} galaxies into co-added images and comparing the recovered properties with the truth. \citet{Tanoglidis2022ICML} recently proposed a new method to estimate the measurement error using a Bayesian neural network. In this paper, we simply take the former method because of its simplicity. 

In order to test how well we recover the photometric and structural parameters in our measurements (Sec \ref{sec:modeling}), we take the 5000 mock \sersic{} galaxies used for computing the deblending completeness and model them using the Spergel light profile. We model the bias and uncertainty in size, surface brightness, total magnitude, and color as a function of other parameters including size, surface brightness, and shape. However, we do not find any significant dependence of the bias on color, ellipticity, and the Spergel index. Thus we just model the bias and uncertainty as a function of the \textit{measured} angular size $r_e$ and surface brightness $\sbeff$.

Given the size and surface brightness of our simulated galaxies, we set the range of the observed size and surface brightness to be $r_e\in[1\arcsec, 15\arcsec],\ \sbeff\in[23, 29]\,\sbunit$. We then split the observed $r_e-\overline{\mu}_{\rm eff}(g)$ plane using an $8\times 8$ grid, and calculate the mean bias $\Delta X = X_{\rm truth} - X_{\rm meas}$ within each bin, where $X=\{\sbeff,\ g-i,\ g-r\}$. For $r_e$, we calculate the relative bias $\Delta r_e / r_e$ instead because it is less dependent on the angular size. Then we interpolate over the grid using a multi-quadratic kernel\footnote{\url{https://docs.scipy.org/doc/scipy/reference/generated/scipy.interpolate.RBFInterpolator.html}} with \code{epsilon=0.5, smooth=1}. Unlike \citet{Zaritsky2021} where they fit models in 4-D space, we find that interpolation works well enough in 2D and we do not use polynomial fitting to avoid meaningless results outside of the fitting range. We emphasize that the bias and uncertainty are modeled to be functions of the measured properties, not intrinsic ones, such that we can correct for bias based on our measurements. 

For LSBGs in our sample, we apply corrections for the bias using the interpolated bias terms. We first correct for the bias in size, $g$-band average surface brightness, and $g-r$ and $g-i$ colors. Then we calculate the $g$-band total magnitude following $m_g = \overline{\mu}_{\rm eff}(g) - 2.5\log(2\pi r_e^2)$. The magnitudes and surface brightnesses in other bands are derived using $g$-band magnitude, surface brightness, and colors. In this way, we apply a self-consistent correction for the measurement biases to the data. 
The measurement uncertainty consists of a statistical uncertainty, which is determined by the shape of the likelihood (posterior) surface, and a systematic uncertainty, which is related to various factors including sky subtraction, neighbor contamination, etc. Unlike other parametric modeling codes such as \code{imfit} \citep{imfit} and \code{the Tractor} \citep{Lang2016}, \code{scarlet} does not explore the full posterior space but rather finds one optimal solution. Thus we have no access to the statistical error on the measured properties from \code{scarlet}. Fortunately, by comparing the recovered properties of mock galaxies with the truth, we can empirically estimate the measurement uncertainty without knowing the impact of each factor. Following the same method as that of calculating the bias, in each bin we compute the $1\sigma$ standard deviation of the difference between the truth and the bias-corrected measurement, then we interpolate over the grid. The measurement errors $\sigma(X)$ are shown in Figure \ref{fig:meas_err} as contours, and they have the same units as the biases. We set minimum uncertainties to be $\sigma(r_e) \geqslant 0.3,\ \sigma(\overline{\mu}_{\rm eff}) \geqslant 0.05,\ \sigma(g-i) \geqslant 0.05$ to avoid meaningless uncertainty due to small statistics.

\end{CJK*}
\end{document}